\documentclass[prb,twocolumn,aps,eqsecnum,epsf,floatfix]{revtex4}
\usepackage{epsfig}
\preprint{}
\newcommand{\ket}[1]{{\left | {#1} \right\rangle}}
\newcommand{\bra}[1]{{\left\langle {#1} \right |}}
\newcommand{\av}[1]{{\left\langle {#1} \right\rangle}}
\begin{document}

\title{\bf
Superconducting Quantum Circuits, Qubits and Computing}
\author{
G. Wendin and V.S. Shumeiko}
\affiliation{Department of Microtechnology and Nanoscience - MC2,
Chalmers University of Technology, \\SE-41296 Gothenburg, Sweden}

\date{\today}

\begin{abstract}

This paper gives an introduction to the physics and principles of operation
of quantized superconducting electrical circuits for quantum information
processing.

\end{abstract}
\pacs{PACS: 00}

\maketitle
\section*{Table of contents}
\noindent
{\bf I. Introduction}\\
{\bf II. Nanotechnology, computers and qubits}\\
{\bf III. Basics of quantum computation}\\
(a) Conditions for quantum information processing\\
(b) Qubits and entanglement\\
(c) Operations and gates\\
(d) Readout and state preparation\\
{\bf IV. Dynamics of two-level systems}\\
(a) The two-level state \\
(b) State evolution on the Bloch sphere \\
(c) dc-pulses, sudden switching and precession \\
(d) Adiabatic switching \\
(e) Harmonic perturbation and Rabi oscillation \\
(g) Decoherence of qubit systems\\
{\bf V. Classical superconducting circuits}\\
(a) Current biased Josephson junction\\
(b) rf-SQUID \\
(c) dc-SQUID \\
(d) Single Cooper pair Box \\
{\bf VI. Quantum superconducting circuits}\\
{\bf VII. Basic qubits}\\
(a) Josephson junction (JJ) qubit \\
(b) Charge qubits \\
    Single Cooper pair Box (SCB) \\
    Single Cooper pair Transistor (SCT) \\
(d) Flux qubits \\
    rf-SQUID \\
    3-junction SQUID - persistent current qubit (PCQ) \\
(e) Potential qubits \\
{\bf VIII. Qubit read-out and measurement of quantum information}\\
(a) Readout: why, when and how? \\
(b) Direct qubit measurement \\
(c) Measurement of charge qubit with SET \\
(d) Measurement via coupled oscillator \\
(e) Threshold detection \\
{\bf IX. Physical coupling schemes for two qubits}\\
(a) General principles \\
(b) Inductive coupling of flux qubits \\
(c) Capacitive coupling of single JJ qubits \\
(d) JJ coupling of charge qubits \\
(f) Coupling via oscillators \\
Coupling of charge qubits \\
Phase coupling of SCT qubits \\
(g) Variable-coupling schemes \\
Variable inductive coupling \\
Variable Josephson coupling \\
Variable phase coupling \\
Variable capacitive coupling \\
 (h) Two qubits coupled via a resonator \\
{\bf X. Dynamics of multi-qubit systems}\\
(a) General N-qubit formulation \\
(b) Two qubits, Ising-type transverse zz coupling \\
Biasing far away from the degeneracy point \\
Biasing at the degeneracy point \\
(c) Two qubits, transverse xx coupling  \\
(d) Two qubits, yy coupling  \\
(e) Effects of the environment: noise and decoherence \\
{\bf XI. Experiments with single qubits and readout devices} \\
(a) Readout detectors \\
(b) Operation and measurement procedures \\
(c) NIST current-biased Josephson junction qubit \\
(d) Flux qubits \\
(e) Charge-phase qubit \\
{\bf XII. Experiments with qubits coupled to quantum oscillators} \\
(a) General discussion \\
(b) Delft persistent current flux qubit coupled to a quantum oscillator \\
(c) Yale charge-phase qubit coupled to a strip-line resonator \\
(d) Comparison of the Delft and Yale approaches \\
{\bf XIII. Experimental manipulation of coupled two-qubit systems} \\
(a) Capacitively coupled charge qubits\\
(b) Inductively coupled charge qubits\\
(c) Capacitively coupled JJ phase qubits\\
{\bf XIV. Quantum state engineering with multi-qubit JJ systems}\\
(a) Bell measurements\\
(b) Teleportation\\
(c) Qubit buses and entanglement transfer \\
(d) Qubit encoding and quantum error correction \\
{\bf XV. Conclusion and perspectives}\\
{\bf Glossary}\\
{\bf References}\\

%\newpage
\section{Introduction}

The first demonstration of oscillation of a superconducting qubit by Nakamura et al. in 1999 \cite{Nakamura1999} can be said to represent the "tip of the iceberg": it rests on a huge volume of advanced research on Josephson junctions (JJ) and circuits developed during the last 25 years. Some of this work has concerned fundamental research on Josephson junctions and superconducting quantum interferometers (SQUIDs) aimed at understanding macroscopic quantum coherence (MQC) \cite{CaldeiraLeggett1981,Leggett1987,Devoret1985,Clarke1988}, providing the foundation of the persistent current flux qubit \cite{Mooij1999,vdWal2000,Friedman2000}. However, there has also been intense research aimed at developing superconducting flux-based digital electronics and computers \cite{Likharev1986,Likharev2001,Gaj1997}. Moreover, in the 1990's the single-Cooper-pair box/transistor (SCB, SCT) \cite{LikharevZorin1985}, was developed experimentally and used to demonstrate the quantization of Cooper pairs on a small superconducting island \cite{Bouchiat1998}, which is the foundation of the charge qubit \cite{Nakamura1999,Shnirman1997}.

Since then there has been a steady development \cite{Wendin_RS2003,Wendin_PW2003,DevoretMartinis2004,DevWallMart2004,EsteveVion2004}, with observation of microwave-induced Rabi oscillation of the two-level populations in charge \cite{Vion2002,Vion2003,Collin2004}, and flux \cite{Chiorescu2003,Chiorescu2004,Bertet2004,Ilichev2003} qubits and dc-pulse driven oscillation of charge qubits with rf-SET detection \cite{Duty2004}. An important step is the development of the charge-phase qubit, a hybrid version of the charge qubit consisting of an SCT in a superconducting loop \cite{Vion2002,Vion2003}, demonstrating Rabi oscillations with very long coherence time, of the order of 1 $\mu$s, allowing a large set of basic and advanced ("NMR-like") one-qubit operations (gates) to be performed \cite{Collin2004}. In addition, coherent oscillations have been demonstrated in the "simplest" JJ qubits of them all, namely a single Josephson junction \cite{Martinis2002,Martinis2003,Simmonds2004,Cooper2004}, or a two-JJ dc-SQUID \cite{Claudon2004}, where the qubit is formed by the two lowest states in the periodic potential of the JJ itself.

Although a powerful JJ-based quantum computer with hundreds of qubits remains a distant goal, systems with 5-10 qubits will be built and tested by, say, 2010.
Pairwise coupling of qubits for two-qubit gate operations is then an essential task, and a few experiments with coupled JJ-qubits with fixed capacitive or inductive couplings have been reported
\cite{Pashkin2003,Yamamoto2003,MajerPhD2002,Majer2003,Izmalkov2004,Berkley2003}, in particular the first realization of a controlled-NOT gate with two coupled SCBs \cite{Yamamoto2003}, used together with a one-qubit Hadamard gate to generate an entangled two-qubit state.

For scalability, and simple operation, the ability to control qubit couplings, e.g. switching them on and off, will be essential.
So far, experiments on coupled JJ qubits have been performed without direct physical control of the qubit coupling, but there are many proposed schemes for
two(multi)-qubit gates based on fixed or controllable physical qubit-qubit couplings or tunings of qubits and bus resonators.

All of the JJ-circuit devices introduced above are based on nanoscale science and technology and represent emerging technologies for quantum engineering and, at best, information processing. One may debate the importance of quantum computers on any time scale, but there is no doubt that the research will be a powerful driver of the development of solid-state quantum state engineering and quantum technology, e.g. performing measurements "at the edge of the impossible".

This article aims at describing the inner workings of superconducting Josephson junction (JJ) circuits, how these can form two-level systems acting as qubits, and how they can be coupled together to multi-qubit networks. Since the field of experimental qubit applications is only five years old, it is not even clear if the field represents an emerging technology for computers. Nevertheless, the JJ-technology is presently the only example of a working solid state qubit with long coherence time, with demonstrated two-qubit gate operation and readout, and with potential for scalability. This makes it worthwile to describe this system in some detail.

It needs to be said, however, that much of the basic theory for coupled JJ-qubits was worked out well ahead of experiment \cite{Shnirman1997,Makhlin1999,MakhlinRMP2001}, defining and elaborating basic operation and coupling schemes. We recommend the reader to take a good look at the excellent research and review paper by Makhlin et al. \cite{MakhlinRMP2001} which describes the basic principles of a multi-JJ-qubit information processor, including essential schemes for qubit-qubit coupling.
The ambition of the present article is to provide a both introductory and in-depth overview of essential Josephson junction quantum circuits, discuss basic issues of readout and measurement, and connect to the recent experimental progress with JJ-based qubits for quantum information processing (JJ-QIP).

\section{Nanotechnology, computers and qubits}

The scaling down of microelectronics into the nanometer range will
inevitably make quantum effects like tunneling and wave propagation
important. This will eventually impede the functioning of classical
transistor components, but will also open up new opportunities for
multi-terminal components and logic circuits built on e.g. resonant
tunneling, ballistic transport, single electronics, etc.

There are two main branches of fundamentally different computer architectures,
namely logically irreversible and logically reversible. Ordinary computers
are irreversible because they dissipate both energy and information. Even if
CMOS circuits in principle only draw current when switching, this is
nevertheless the source of intense local heat generation, threatening to burn
up future processor chips. The energy dissipation can in principle be reduced
by going to single-electron devices, superconducting electronics and quantum
devices, but this does not alter the fact that the information processing is
logically irreversible. In the simplest case of an AND gate, two incoming bit
lines only result in a single bit output, which means that one bit is in
practice erased (initialized to zero) and the heat irreversibly dissipated to
the environment. The use of quantum-effect devices does not change the fact
that we are dealing with computers where each gate is logically irreversible
and where discarded information constantly is erased and turned into heat. A
computer with quantum device components therefore does not make a quantum
computer.

A quantum information processor has to be built on fundamentally reversible
schemes with reversible gates where no information is discarded, and where
all internal processes in the components are elastic. This issue is connected
with the problem of the minimum energy needed for performing a calculation
\cite{Landauer1961} (connected with the entropy change created by erasing the
final result, i.e. reading the result and then clearing the register). A
reversible information processor can in principle be built by classical means
using adiabatically switched networks of different kinds. The principles were
investigated in the 1980's and form a background for much of the work on
quantum computation
\cite{FredkinToffoli1982,Likharev1982,Likharev1985,Bennett1988,Feynman1996,NielsenChuang2000,Gruska1999}.
Recently there has been some very interesting development of reversible
computers (see \cite{Gershenfeld1996,Frank2002,Frank2003,Semenov2003} and
references therein). However, a reversible computer still does not make a
quantum computer. What is characteristic for a quantum computer is that it is
reversible and {\em quantum coherent}, meaning that one can build entangled
non-classical multi-qubit states.

One can broadly distinguish between microscopic, mesoscopic and macroscopic
qubits. Microscopic effective two-level systems are localized systems
confined by natural or artificial constraining potentials. Natural systems
then typically are atomic or molecular impurities utilizing electronic
charge, or electronic or nuclear spins. These systems may be implanted
\cite{Clark2003,Schofield2003} or naturally occurring due to the material
growth process \cite{Leuenberger2003}. A related type of impurity qubit
involves endohedral fullerenes, i.e. atoms implanted into C60 or similar
cages \cite{Hahrneit2003}, to be placed at specific positions on a surface
prepared for control and readout.

Mesoscopic qubit systems typically involve geometrically defined confining
potentials like quantum dots. Quantum dots (QD) for qubits
\cite{LossDiVincenzo1998} are usually made in semiconductor materials. One
type of QD is a small natural or artificial semiconductor grain with
quantized electronic levels. The electronic excitations may be excitonic (excitons, \cite{Zrenner2002,KamadaGotoh2004} or biexcitons, \cite{Li2003}),
charge-like \cite{Hayashi2003},  or spin-like (e.g. singlet-triplet)
\cite{vdWiel2003,Elzerman2005}. Another type of QD is geometrically
defined in semiconductor 2DEG by electrostatic split-gate arrangements.
Although the host materials are epitaxially layered semiconducting materials
(e.g.GaAs/AlGaAs), the ungated 2DEG electronic system is metallic. A
split-gate arrangement can then define a voltage-controlled system of quantum
dots coupled to metallic reservoir electrodes, creating a system for electron
charge and spin transport through quantum point contacts and (effective)
two-level quantum dots
\cite{vdWiel2003,Hanson2003,Elzerman2004,Elzerman2005}.

There is also a potential qubit based on liquid He superfluid technology,
namely "Electrons on Helium", EoH \cite{PlatzmanDykman1999}. This is really
an atomic-like microscopic qubit: a thin film of liquid He is made to cover a
Si surface, and electrons are bound by the image force above the He surface,
forming an electronic two-level system. Qubits are laterally defined by
electrostatic gate patterns in the Si substrate, which also defines circuits
for qubit control, qubit coupling and readout.

Macroscopic superconducting qubits - the subject of this article - are based
on electrical circuits containing Josephson junctions (JJ). Looking at two
extreme examples, the principle is actually very simple. In one limit, the
qubit is simply represented by the two rotation directions of the persistent
supercurrent of Cooper pairs in a superconducting ring containing Josephson
tunnel junctions (rf-SQUID)(flux qubit)
\cite{Mooij1999,vdWal2000,Friedman2000}, shown below in Fig. \ref{3J_vdWal}
for the Delft 3-junction flux qubit,

\begin{figure}[t]
\centerline{\epsfxsize=0.35\textwidth\epsfbox{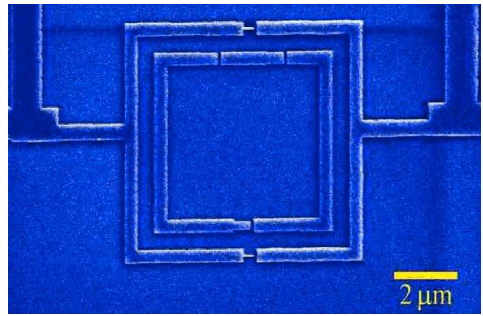}}
\caption{3-junction persistent current flux qubit (PCQ) (inner loop) surrounded by a 2-junction SQUID.
{\em Courtesy of J.E. Mooij, TU Delft.}
}
\label{3J_vdWal}
\end{figure}

\begin{figure}[t]
\centerline{\epsfxsize=0.40\textwidth\epsfbox{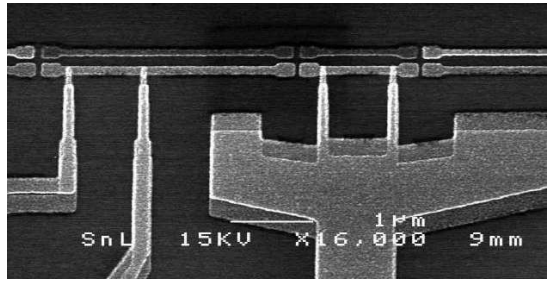}}
\caption{Single Cooper pair box (SCB) (right) coupled to a single-electron transistor (SET) (left) for readout.
{\em Courtesy of P. Delsing, Chalmers.}
}
\label{SCB_Delsing}
\end{figure}

In another limit, the qubit is represented by the presence or absence of a
Cooper pair on a small superconducting island (Single Cooper pair Box, SCB,
or Transistor, SCT) (charge qubit)
\cite{Nakamura1999,Bouchiat1998,Duty2004,Aassime2001}, as illustrated in Fig.
\ref{SCB_Delsing}. Hybrid circuits  \cite{Vion2002,Cottet2002,Zorin2002} can
in principle be tuned between these limits by varing the relations between
the electrostatic charging energy $E_C$ and the Josephson tunneling energy
$E_J$.

All of these solid-state qubits have advantages and disadvantages, and only
systematic research and development of multi-qubit systems will show the
practical and ultimate limitations of various systems. The superconducting
systems have presently the undisputable advantage of acutally existing,
showing Rabi oscillations and responding to one- and two-qubit gate
operations. In fact, even an elementary SCB two-qubit entangling gate
creating Bell-type states has been demonstrated very recently
\cite{Yamamoto2003}. All of the non-superconducting qubits are so far,
promising but still potential qubits. Several of the impurity electron spin
qubits show impressive relaxation lifetimes in bulk measurements, but it
remains to demonstrate how to read out individual qubit spins.

\section{Basics of quantum computation}
\label{BasicsQC}

\subsection{Conditions for quantum information processing}
DiVincenzo \cite{DiVincenzo2000} has formulated a set of rules and conditions that need to be fulfilled in order for quantum computing to be possible:

1. Register of $2$-level systems (qubits), $n = 2^N$ states $\ket{101..01}$  ($N$ qubits)\\
2. Initialization of the qubit register: e.g. setting it to $\ket{000..00}$\\
3. Tools for manipulation: $1$- and $2$-qubit gates, e.g. Hadamard ($H$) gates to flip the spin to the equator, $U_H \ket{0} = (\ket{0}+\ket{1})/2$,  and Controlled-NOT ($CNOT$) gates to create entangled states, $U_{CNOT}U_H|00> =(\ket{00}+\ket{11})/2$   (Bell state)\\
4. Read-out of single qubits $\ket{\psi} = a\ket{0} + b e^{i\phi}\ket{1}$  $\rightarrow$ $a$, $b$ (spin projection; phase $\phi$ of qubit lost)\\
5. Long decoherence times: $>10^4$ 2-qubit gate operations needed for error correction to maintain coherence "forever".\\
6. Transport qubits and to transfer entanglement between different coherent systems (quantum-quantum interfaces).\\
7. Create classical-quantum interfaces for control, readout and information storage.

\subsection{Qubits and entanglement}
A qubit is a two-level quantum system caracterized by the state vector
\begin{equation}\label{angles}
|\psi\rangle =\cos{\theta\over2}|0\rangle + \sin{\theta\over2}e^{i\phi}|1\rangle
\end{equation}
Expressing $|0\rangle$ and $|1\rangle$ in terms of the eigenvectors of the Pauli matrix $\sigma_z$,
\begin{equation}
|0\rangle =  \left(
\begin{array}{c}
1\\ 0
\end{array}
\right), \;\;\; |1\rangle =  \left(
\begin{array}{c}
0\\ 1
\end{array}
\right).
\end{equation}
this can be described as a rotation from the north pole of the $|0\rangle$ state,
\begin{equation}\label{TLSEq}
|\psi\rangle =\left(\begin{array}{cc}
1 & 0 \\ 0  & e^{i\phi} \end{array}\right)
\left(\begin{array}{cc}
\cos{\theta\over2} & -\sin{\theta\over2}\\ \sin{\theta\over2} & \cos{\theta\over2} \end{array}\right)
\left( \begin{array}{c}1\\0\end{array}\right)
\end{equation}
can be characterised by a unit vector on the Bloch sphere:
\begin{figure}[t]
\centerline{\epsfxsize=0.40\textwidth\epsfbox{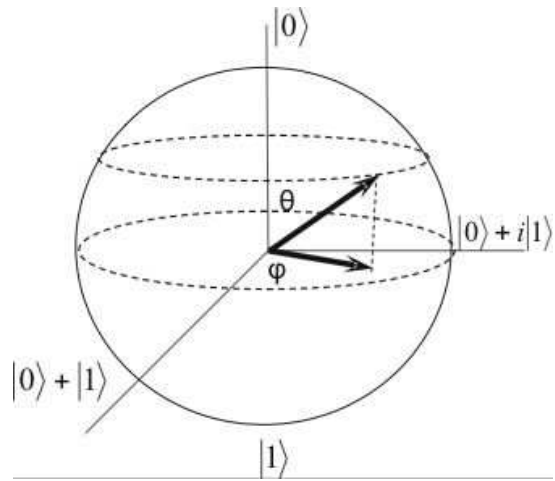}}
\caption{The Bloch sphere. Points on the sphere correspond to
the quantum states $\ket{\psi}$; in particular, the north and south poles correspond to the computational basis states $\ket{0}$ and $\ket{1}$;
superposition cat-states $|\psi\rangle = |0\rangle + e^{i\phi}|1\rangle $
are situated on the equator.}
\label{Blochsphere1}
\end{figure}

The state vector can be represented as a unitary vector on the Bloch sphere,
and general unitary (rotation) operations make it possible to reach every
point on the Bloch sphere. The qubit is therefore an analogue object with a
continuum of possible states. Only in the case of spin 1/2 systems do we have
a true two-level system. In the general case, the qubit is represented by the
lowest levels of a multi-level system, which means that the length of the
state vector may not be conserved due to transitions to other levels. The
first condition will therefore be to operate the qubit so that it stays on
the Bloch sphere (fidelity). Competing with normal operation, noise from the
environment may cause fluctuation of both qubit amplitude and phase, leading
to relaxation and decoherence. It is a delicate matter to isolate the qubit
from a perturbing environment, and desirable operation and unwanted
perturbation (noise) easily go hand in hand. It is a major issue to design
qubit control and read-out such that the necessary communication lines can be
blocked when not in use.

The state of N independent qubits can be represented as a product state,
\begin{equation}
|\psi\rangle =  |\psi_1\rangle |\psi_2\rangle .... |\psi_N\rangle
= |\psi_1\psi_2 .... \psi_N\rangle
\end{equation}
involving any one of all of the configurations $|00...0>$, $|00...1>$, ...., $|11...1>$.
A general state of an N-qubit memory register (i.e. a many-body system) can
then be written as a time-dependent superposition of many-particle configurations
\begin{eqnarray}
\ket{\psi(t)} =  c_1(t) \ket{0...00} + \;c_2(t) \ket{0...01} \\ \nonumber
+ \;c_3(t) \ket{0...10} + ....+ \;c_n(t) \ket{1...11}
\end{eqnarray}
where the amplitudes $c_i(t)$ are complex, providing phase information. This state represents a time-dependent superposition of $2N$ N-body configurations which in general cannot be written as a product of one-qubit states and then represents an entangled (quantum correlated) many-body state.

In the case of two qubits, the maximally  entangled states are the so-called Bell states,
\begin{eqnarray}
\label{psi_GHZ}
\ket{\psi} = (\ket{00} + \ket{11})/2\\
\ket{\psi} = (\ket{00} - \ket{11})/2\\
\ket{\psi} = (\ket{01} + \ket{10})/2\\
\ket{\psi} = (\ket{01} - \ket{10})/2
\end{eqnarray}
where the last one is the singlet state.
In the case of three qubits, the corresponding maximally entangled ("cat") states are the Greenberger-Horne-Zeilinger (GHZ) states \cite{GHZ1993}
\begin{eqnarray}
\ket{\psi} = (\ket{000} \pm \ket{111})/\sqrt{8}
\end{eqnarray}
Another interesting entangled three-qubit state appears in the teleportation process,
\begin{eqnarray}
\label{psi_teleport}
\ket{\psi} = [\ket{00}(a\ket{0}+b\ket{1})+ \ket{01}(a\ket{0}-b\ket{1}) \\ \nonumber
+\ket{10}(b\ket{0}+a\ket{1})+ \ket{11}(b\ket{0}-a\ket{1})]/\sqrt{8}
\end{eqnarray}

\subsection{Operations and gates}

Quantum computation basically means allowing the $N$-body state to develop in a fully coherent fashion through unitary transformations acting on all $N$ qubits. The time evolution of the many-body system of $N$ two-level subsystems can be described by the Schr\"odinger equation for the $N$-level state vector $\ket{\psi(t)}$,
\begin{equation}
i\hbar \partial_t \ket{\psi} = \hat H \ket{\psi}.
\end{equation}
in terms of the time-evolution operator characterizing by the time-dependent many-body Hamiltonian $\hat{H}(t)$ of the system determined by the external control operations and the perturbing noise from the environment,
\begin{equation}
\ket{\psi(t)} = U(t,t_0) \ket{\psi(t_0)}.
\end{equation}
The solution of Schr\"odinger equation for $U(t,t_0)$
\begin{equation}
i\hbar \partial_t U(t,t_0) = \hat H U(t,t_0)
\end{equation}
may be written as
\begin{equation}
U(t,t_0) = 1-\frac{i}{\hbar}\int_{t_0}^{t} \hat{H}(t')U(t',t_0) dt' \;,
\end{equation}
and finally, in terms of the time-ordering operator T, as
\begin{equation}
\label{UHTDgen}
U(t,t_0) = T \;e^{-\frac{i}{\hbar} \int_{t_0}^{t} \hat{H}(t') dt'} \;,
\end{equation}
describing the time evolution of the entire $N$-particle state in the interval $[t,t_0]$.
If the total Hamiltonian commutes with itself at different times, the time ordering can be omitted,
\begin{equation}
\label{UHTD}
U(t,t_0) = \;e^{-\frac{i}{\hbar} \int_{t_0}^{t} \hat{H}(t') dt'} \;.
\end{equation}
This describes the time-evolution controlled by a homogeneous time-dependent potential or electromagnetic field, e.g. dc or ac pulses with finite rise times but having no space-dependence. If the Hamiltonian is constant  in the interval $[t_0,t]$, then the evolution operator takes the simple form
\begin{equation}\label{UHconst}
U(t,t_0) = e^{-\frac{i}{\hbar}\hat{H}(t-t_0)} \;,
\end{equation}
describing the time-evolution controlled by square dc pulses.

The time-development will depend on how many terms are switched on in the Hamiltonian during this time interval. In the ideal case, usually not realizable, all terms are switched off except for those selected for the specific computational step. A single qubit gate operation then involves turning on a particular term in the Hamiltonian for a specific qubit, while a two-qubit gate involves turning on an interaction term between two specific qubits. In principle one can perform an $N$-qubit gate operation by turning on interactions for all $N$ qubits. In practical cases, many terms in the Hamiltonian are turned on all the time, leading to a "background" time development that has to be taken into account.

The basic model for a two-level qubit is the spin-$1/2$ in a magnetic field. A system of interacting qubits can then be modelled by a collection of interacting spins, described by the Heisenberg Hamiltonian
\begin{eqnarray}\label{HH}
\hat H(t) = \sum {\bf h}_i(t)\;{\bf S}_i
+ {1\over2}  \sum J_{ij}(t)\;{\bf S}_i\;{\bf S}_j
\end{eqnarray}
controlled by a time-dependent external magnetic field ${\bf h}_i(t)$ and by a time-dependent spin-spin coupling $J_{ij}(t)$.

Expressing the Hamiltonian in Cartesian components, ${\bf h}_i(t)= (h_x,h_y,h_z)$,  ${\bf S}_i(t)= (S_x,S_y,S_z)$
and introducing the Pauli $\sigma$-matrices, $(S_x,S_y,S_z) = {1\over2}\hbar(\sigma_x,\sigma_y,\sigma_z)$
we obtain a general N-qubit Hamiltonian with general qubit-qubit coupling:
\begin{eqnarray} \label{NqubitHam}
\hat H = -\frac{1}{2}\sum_i (\epsilon_i\sigma_{zi} + Re\Delta_i\sigma_{xi} + Im\Delta_i\sigma_{yi}) \; \\ \nonumber
+\;  \frac{1}{2} \sum_{ij;\nu} \lambda_{\nu,ij}(t)\;\sigma_{\nu i}\sigma_{\nu j} \\ \nonumber
+ \sum_i (f_i(t)\sigma_{zi} + g_{xi}(t)\sigma_{xi} + g_{yi}(t)\sigma_{yi})
\end{eqnarray}
We have here introduced time-independent components of the external field defining qubit energy level splittings $\epsilon_i$, $Re\Delta_i$ and $Im\Delta_i$ along the z and x,y axes, as well as time-dependent components $f_i(t)$ and $g_i(t)$  explicitly describing qubit operation and readout signals and noise.

Inserted into Eq.(\ref{UHTD}), this Hamiltonian determines the time evolution of the many-qubit state. In the ideal case one can turn on and off each individual term of the Hamiltonian, including the two-body interaction, giving complete control of the evolution of the state.

It has been shown that any unitary transformation can be achieved through a quantum network of sequential application of one- and two-qubit gates. Moreover, the size of the coherent workspace of the multi-qubit memory can be varied (in principle) by switching on and off qubit-qubit interactions.

In the common case of NMR applied to molecules \cite{NMR_Grover,NMR_order,NMR_Shor}, one has no control of the fixed, direct spin-spin coupling. With an external magnetic field one can control the qubit Zeeman level splittings (no control of individual qubits). Individual qubits can be addressed by external RF-fields since the qubits have different resonance frequencies (due to different chemical environments in a molecule). Two-qubit coupling can be induced by simultaneous resonant excitation of two qubits.

In the case of engineered solid state JJ-circuits, individual qubits can be
addressed by local gates, controlling the local electric or magnetic field.
Extensive single-qubit operation has recently been demonstrated by Collin et
al. \cite{Collin2004}. Regarding two-qubit coupling, the field is just
starting up, and different options are only beginning to be tested. The most
straightforward approaches involve fixed capacitive
\cite{Pashkin2003,Yamamoto2003,Berkley2003} or inductive
\cite{MajerPhD2002,Majer2003} coupling; such systems could be operated in NMR
style, by detuning specific qubit pairs into resonance. Moreover, there are
various solutions for controlling the direct physical qubit-qubit coupling
strength, as will be described in Section \ref{SectIX}.

\subsection{Readout and state preparation}
\label{Readout1}

External perturbations, described by $f_i(t)$ and $g_i(t)$  in
Eq.(\ref{NqubitHam}) can influence the two-level system in typically two
ways: (i) shifting the individual energy levels, which may change the
transition energy and the phase of the qubit; and (ii) inducing transitions
between the levels, changing the level populations. These effects arise both
from desirable control operations and from unwanted noise (see \cite{MakhlinRMP2001,Tian2003,Orlando2002,Wilhelm2003a} for a discussion of superconducting circuits, and Grangier et al. \cite{Grangier1998} for a discussion of state preparation and quantum non-demolition (QND) measurements).

To control a qubit register, the important thing is to control the
decoherence during qubit operation and readout, and in the memory state. In
the qubit memory state, the qubit must be isolated from the environment.
Operation and read-out devices should be decoupled from the qubit and,
ideally,  not cause any dephasing or relaxation. The resulting intrinsic
qubit life times should be long compared to the duration of the calculation.
In the operation and readout state, the qubit must be connected to the
operation fields. This also opens up the system to a noisy environment, which
puts great demands on signal-to-noise ratios.

The readout operation is a particularly critical step. The ultimate purpose
is to perform a "single-shot" quantum non-demolition (QND) measurement,
determining the state ($\ket{0}$ or $\ket{1}$) of the qubit in a single
measurement and then leaving the qubit in that very state. During the
measurement time, the back-action noise from the "meter" will cause
relaxation and mixing, changing the qubit state. The "meter" must then be
sensitive enough to detect the qubit state on a time scale shorter than the
induced relaxation time $T_1^\ast$ in the presence of the dissipative
back-action of the "meter" itself. Under these conditions it is possible to
detect the qubit projection in a single measurement (single shot read-out).
Performing a single-shot projective measurement in the qubit eigenbasis then
provides a QND measurement (note the "trivial" fact that the phase is irreversibly lost in the measurement process under any circumstances).

The readout/measurement processes described above can be related to the
Stern-Gerlach (SG) experiment \cite{Grangier1998}. A SG spin filter acts as a beam splitter for flying qubits, creating separate paths for spin-up and spin-down atoms, preparing for the measurement by making it possible in principle to
distinguish spatially between the two states of the qubit. The measurement is
then performed by particle counters: a click in, say, the spin-up path
collapses the atom to the spin-up state in a single shot. There is
essentially no decohering back-action from the detector until the atom is
detected. However, after detection of the spin-up atom the state is
thoroughly destroyed: this qubit has not only decohered and relaxed, but is
removed from the system. However, if it was entangled with other qubits,
these are left in a specific eigenstate. For example, if the qubit was part
of the Bell pair $\ket{\psi} = (\ket{00} + \ket{11})/2$, detection of spin up ($\ket{0}$) selects $\ket{00}$ and leaves the other qubit in state $\ket{0}$. If instead the qubit was part of the Bell pair $\ket{\psi} = (\ket{01} + \ket{10})/2$, detection of spin up ($\ket{0}$) selects $\ket{01}$ and leaves the other qubit in the spin-down state $\ket{1}$. Furthermore, for the 3-qubit GHZ state $\ket{\psi} = (\ket{000} \pm \ket{111})/\sqrt{8}$, detection of the first qubit in the spin up ($\ket{0}$) state leaves the remaining 2-qubit system in the $\ket{00}$ state.

Finally, in the case of the three-qubit entangled state in the teleportation
process, Eq.(\ref{psi_teleport}), detection of the first qubit in the spin up
state $\ket{0}$, leaves the remaining 2-qubit system in  the  entangled state
$[\ket{0}(a\ket{0}+b\ket{1})+ \ket{1}(a\ket{0}-b\ket{1})]/2$. Moreover,
detection of also the second qubit in the spin-up state $\ket{0}$ will leave
the third qubit in the state $(a\ket{0}+b\ket{1})/\sqrt{2}$. In this way,
{\em measurement can be used for state preparation}.

Applying this discussion to the measurement and readout of JJ-qubit circuits,
an obvious difference is that the JJ-qubits are not flying particles. The
detection can there not be turned on simply by the qubit flying into the
detector. Instead, the detector is part of the JJ-qubit circuitry, and must
be turned on to discriminate between the two qubit states. These are not
spatially separated and therefore sensitive to level mixing by detector noise
with frequency around the qubit transition energy. The sensitivity of the
detector determines the time scale $T_m$ of the measurement, and the
detector-on back-action determines the time scale of qubit relaxation
$T_1^*$. Clearly $T_m \ll T_1^*$ is needed for single-shot discrimination of
$\ket{0}$ and $\ket{1}$. This requires a detector signal-to-noise (S/N) ratio
$\gg 1$, in which case one will have the possibility also to turn off the
detector and leave the qubit in the determined eigenstate, now relaxing on
the much longer time scale $T_1$ of the "isolated" qubit. This would then be
the ultimate QND measurement.

Note that in the above discussion, the qubit dephasing time ("isolated"
qubit) does not enter because it is assumed to be much longer than the
measurement time, $T_m \ll T_{\phi}$. This condition is obviously essential
for utilizing the measurement process for state preparation and error
correction.

\section{Dynamics of two-level systems}

To perform computational tasks one must be able to put a qubit in an
arbitrary state. This is usually done in two steps. The first step,
initialization, consists of relaxation of the initial qubit state to the
equilibrium state due to interaction with environment. At low temperature,
this state is close to the ground state. During the next step, time dependent
dc- or rf-pulses are applied to the controlling gates: electrostatic gate in
the case of charge qubits, bias flux in the case of flux qubits, and bias
current in the case of the JJ qubit. Formally, the pulses enter as
time-dependent contributions to the Hamiltonian and the state evolves under
the action of the time-evolution operator. To study the dynamics of a single
qubit two-level system we therefore first describe the two-level state, and
then the evolution of this state under the influence of the control pulses
("perturbations").

\subsection{The two-level state}

The general 1-qubit Hamiltonian has the form
\begin{eqnarray}
\hat{H} = -{1\over2}({\epsilon}\;\sigma_z + \Delta \;\sigma_x)
\end{eqnarray}
The qubit eigenstates are to be found from the stationary Schr\"odinger equation
\begin{equation}
\hat{H} \ket{\psi} = E\ket{\psi}
\end{equation}
To solve the S-equation we expand the 1-qubit state in a complete basis, e.g.
the basis states of the $\sigma_z$ operator,
\begin{eqnarray}
\ket{\psi} = \sum_k a_k \;\ket{k} = c_0\ket{0} + c_1\ket{1}
\end{eqnarray}
and project onto the basis states
\begin{equation}
\hat{H} \sum_m\ket{m}\bra{m}\psi\rangle = E \ket{\psi}
\end{equation}
obtaining the ususal matrix equation
\begin{equation}
\sum_m \bra{k} \hat{H} \ket{m} a_m = E  a_k
\end{equation}
where $a_k = \bra{k}\psi\rangle$, and

\begin{eqnarray}
H_{qp}
= -\frac{\epsilon}{2}\bra{k}\sigma_z\ket{m}
-\frac{\Delta}{2}\bra{k}\sigma_x\ket{m}
\end{eqnarray}

giving the Hamiltonian matrix

\begin{equation}
{\hat H}= -\frac{1}{2}
\left(\begin{array}{cc}
\epsilon &  \Delta \\
  \Delta  &  -\epsilon \\
\end{array}\right)
\end{equation}
The Schr\"odinger equation is then given by
\begin{equation}
({\hat H}-E)\ket{\psi} = -\frac{1}{2}
\left(\begin{array}{cc}
\epsilon + 2E &  \Delta \\
  \Delta  &  -\epsilon + 2E \\
\end{array}\right)
\left(\begin{array}{c}
a_1 \\
a_2 \\
\end{array}\right) = 0
\end{equation}
The eigenvalues are determined by
\begin{equation}
det({\hat H}-E) = E^2- \frac{1}{4}(\epsilon^2+\Delta^2) = 0
\end{equation}
with the result
\begin{equation}
E_{1,2} = \pm \frac{1}{2}\sqrt{\epsilon^2+\Delta^2}
\end{equation}
The eigenvectors are given by:

\begin{equation}
a_2 = - a_1 \frac{\Delta}{\epsilon + 2E} \\
\end{equation}
After normalisation

%\begin{equation}
%a_1 = 1/\sqrt{1+\left(\frac{\Delta}{\epsilon+2E}\right)^2}
%= \left(\epsilon\pm\sqrt{(\epsilon^2+\Delta^2)}\right)/ %\sqrt{{\left(\epsilon\pm\sqrt{(\epsilon^2+\Delta^2\right)^2}+\Delta^2}}
%= \frac{1}{\sqrt{2}}\sqrt{1\pm \frac{\epsilon}{|2E|}}
%\end{equation}

\begin{equation}
a_1 = 1/\sqrt{1+\left(\frac{\Delta}{\epsilon+2E}\right)^2}
= \frac{1}{\sqrt{2}}\sqrt{1\pm \frac{\epsilon}{|2E|}}
\end{equation}

\begin{equation}
a_2 = \pm\sqrt{1-a_1^2} = \pm\frac{1}{\sqrt{2}}\sqrt{1\mp \frac{\epsilon}{|2E|}}
\end{equation}

We finally simplify the notation by fixing the signs of the amplitudes,
\begin{equation}
a_1 = \frac{1}{\sqrt{2}}\sqrt{1 + \frac{\epsilon}{|2E|}} \;;\;
a_2 = \frac{1}{\sqrt{2}}\sqrt{1 - \frac{\epsilon}{|2E|}} \;;\;
\end{equation}
and explicitly writing down all the energy eigenstates,
\begin{eqnarray}
\ket{E_1} = a_1 \ket{0} + a_2 \ket{1} \\
\ket{E_2} = a_2 \ket{0} - a_1 \ket{1}
\end{eqnarray}
where
\begin{equation}
E_1 = - \frac{1}{2}\sqrt{\epsilon_1^2+\Delta_1^2} \;;\;
E_2 = + \frac{1}{2}\sqrt{\epsilon_1^2+\Delta_1^2}
\end{equation}
The sign of $\ket{E_2}$ has been chosen to give the familiar expression for
the superposition at the degeneracy point $\epsilon_1=0$ where
$|a_1|=|a_2|=1/\sqrt{2}$,
\begin{eqnarray}
\ket{E_1} = \frac{1}{\sqrt{2}} (\ket{0} +  \ket{1}) \\
\ket{E_2} = \frac{1}{\sqrt{2}} (\ket{0} -  \ket{1})
\end{eqnarray}

\subsection{The state evolution on the Bloch sphere}
\label{TLSdynamics}

To study the time-evolution of a general state, a convenient way is to expand
in the basis of energy eigenstates,
\begin{eqnarray}\label{evolution}
\ket{\psi(t)} = c_1\ket{E_1} e^{-iE_1t}\ + c_2\ket{E_2} e^{-iE_2t}\
\end{eqnarray}
If we know the coefficients at $t=0$, then we know the time evolution. On the
Bloch sphere this time evolution is represented by rotation of the Bloch
vector with constant angular speed $(E_1-E_2)/\hbar$ around the direction
defined by the energy eigenbasis. Indeed, by introducing parameterization,
$c_1= \cos\theta^\prime,\;c_2=\sin\theta^\prime e^{i\phi\prime}$, we see that
according to Eq. (\ref{evolution}) the polar angle remains constant,
$\theta^\prime=$ const, while the azimuthal angle grows, $\phi^\prime(t) \phi^\prime(0)+ (E_1-E_2)t/\hbar$. The primed angles here refer to a new
coordinate system on the Bloch sphere related to the energy eigenbasis, which
is obtained by rotation from the earlier introduced computational basis, Eq.
(\ref{angles}).

\begin{figure}[h]
\centerline{\epsfxsize=0.35\textwidth\epsfbox{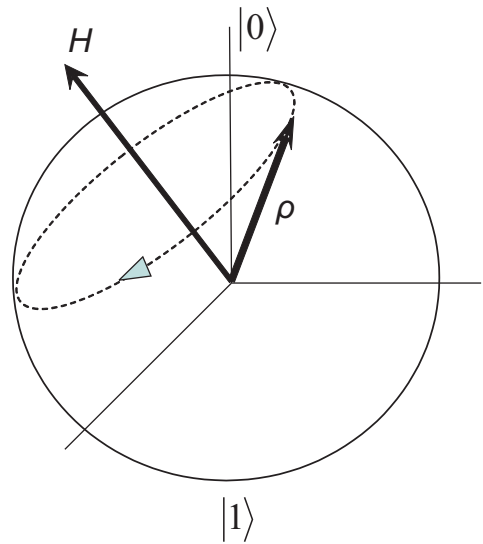}}
\caption{The Bloch sphere: the Bloch vector $\rho$ represents the states of
the two-level system (same as in Fig. 3). The poles of the Bloch sphere correspond to the energy eigenstates; the vector $H$ represents the two-level Hamiltonian.}
\label{figBloch2}
\end{figure}

The dynamics on the Bloch sphere is conveniently described in terms of the
density matrix for a pure quantum state \cite{LandauQM},
\begin{equation}
\label{rhomat}
\hat\rho = |\psi\rangle\langle\psi|.
\end{equation}
This is a $2\times2$ Hermitian matrix whose diagonal elements $\rho_1$ and
$\rho_2$ define occupation probabilities of the basis states, hence satisfying
the normalization condition $\rho_1+\rho_2=1$, while the off-diagonal
elements give information about the phase. The density matrix can be mapped
on a real 3-vector by means of the standard expansion in terms of $\sigma$-matrices,
\begin{equation}
\label{rhomatrix}
\hat\rho = {1\over 2}(1 +  \rho_x\sigma_x + \rho_y\sigma_y + \rho_z\sigma_z).
\end{equation}
Direct calculation of the density matrix Eq.(\ref{rhomat})
using Eq.(\ref{angles}) and comparing with Eq.(\ref{rhomatrix}) shows that the vector
${\mbox{\boldmath$\rho$}} = (\rho_x,\rho_y,\rho_z)$ coincides with the Bloch
vector,
\begin{equation}
\label{rhovector}
{\mbox{\boldmath$\rho$}} = (\sin\theta\cos\phi,\sin\theta\sin\phi,\cos\theta)
\end{equation}
introduced in Fig. \ref{Blochsphere1} and also shown in Fig. \ref{figBloch2}.
In the same $\sigma$-matrix basis, the general two-level Hamiltonian takes the form
\begin{equation}
\label{Hmatrix}
\hat H = (H_x\sigma_x + H_y\sigma_y + H_z\sigma_z).
\end{equation}
giving a 3-vector representation for the Hamiltonian,
\begin{equation}
\label{Hvector}
{\mbox{\boldmath$H$}} = (H_x,H_y,H_z).
\end{equation}
shown in Fig. \ref{figBloch2}.

The time evolution of the density matrix is given by the Liouville equation,
\begin{equation}\label{Liouville}
i\hbar \partial_t\hat \rho = [\hat H ,\hat\rho].
\end{equation}
The vector form of the Liouville equation is readily derived by inserting Eqs.(\ref{Hmatrix}),(\ref{Hvector}) and using the commutation relations among the Pauli matrices,
\begin{equation}\label{Lvector}
\partial_t{\mbox{\boldmath$\rho$}} = {1\over \hbar} [\,{\mbox{\boldmath{$H$}}} \,\times\,
{\mbox{\boldmath$\rho$}}\,].
\end{equation}
This equation coincides with the Bloch equation for a magnetic moment
evolving in a magnetic field, the role of the magnetic moment being played by
the Bloch vector ${\bf\rho}$ which rotates around the effective "magnetic field" ${\bf H}$ associated with the Hamiltonian of the qubit (plus any driving fields) (Fig. \ref{figBloch2}).

\subsection{dc-pulses, sudden switching and free precession}
\label{SectIV_C}

To control the dynamics of the qubit system, one method is to apply dc
(square) pulses which suddenly change the Hamiltonian and, consequently, the
time-evolution operator. Sudden pulse switching means that the time-dependent
Hamiltonian is changed so fast on the time scale of the evolution of the
state vector that the state vector can be treated as time-independent - frozen
- during the switching time interval. This implies that the system is excited
by a Fourier spectrum with an upper cut-off given by the inverse of the
switching time.

In the specific scheme of sudden switching of different terms in the
Hamiltonian using dc-pulses, the initial state is frozen during the switching
event, and begins to evolve in time under the influence of the new
\begin{eqnarray}
\ket{0} = \ket{\psi(0)} = c_1\ket{E_1} + c_2\ket{E_2}
\end{eqnarray}
To find the coefficients we project onto the charge basis, {k}=0,1
\begin{eqnarray}
\langle 0\ket{0} = c_1\langle0\ket{E_1} + c_2\langle0\ket{E_2} \\
\langle 1\ket{0} = c_1\langle 1\ket{E_1} + c_2\langle 1\ket{E_2}
\end{eqnarray}
and use the explict results for the energy eigenstates to calculate the matrix elements, obtaining
\begin{eqnarray}
1 = c_1 a_1 + c_2 a_2 \\
0 = c_1 a_2 - c_2 a_1
\end{eqnarray}
As a result,
\begin{equation}
\ket{0} = a_1 \ket{E_1}+ a_2 \ket{E_2}
\end{equation}
This stationary state then develops in time governed by the constant Hamiltonian as
\begin{eqnarray}
\ket{\psi(t)} = a_1 e^{-iE_1t}\ket{E_1} + a_2 e^{-iE_2t}\ket{E_2}
\end{eqnarray}
Inserting the energy eigenstates we finally obtain the time evolution in the charge basis,
\begin{eqnarray}
\ket{\psi(t)} =  \nonumber\\
 \ket{0} \left[a_1^2 e^{-iE_1t} + a_2^2 e^{iE_1t} \right]
+ \; \ket{1} \left[a_1 a_2 (e^{-iE_1t} - e^{iE_1t}) \right]  \nonumber\\
\end{eqnarray}
The probability amplitudes of finding the system in one of the two charge states is then
\begin{eqnarray}
\langle 0\ket{\psi(t)} = \left[a_1^2 e^{-iE_1t} + a_2^2 e^{iE_1t} \right]\\
\langle 1\ket{\psi(t)} = \left[a_1 a_2 (e^{-iE_1t} - e^{iE_1t}) \right]
\end{eqnarray}
If the system is driven to the degeneracy point $\epsilon_1=0$, where $|a_1|=|a_2|=1/\sqrt{2}$, then
\begin{eqnarray}
\langle 0\ket{\psi(t)} = \cos{E_1t}\\
\langle 1\ket{\psi(t)} = \sin{E_1t}
\end{eqnarray}
In particular, the probability of finding the system in state $|1\rangle$ (level $2$) oscillates like
\begin{eqnarray}
p_2(t) = |\langle 1\ket{\psi(t)}| = \sin^2{E_1t}  \nonumber\\
= \frac{1}{2}\left[1-\cos{(E_2-E_1)t}\right]
\end{eqnarray}
with the frequency of the interlevel distance. On the Bloch sphere, this
describes free precession around the $X$-axis.

\begin{figure}
\centerline{\epsfxsize=0.45\textwidth\epsfbox{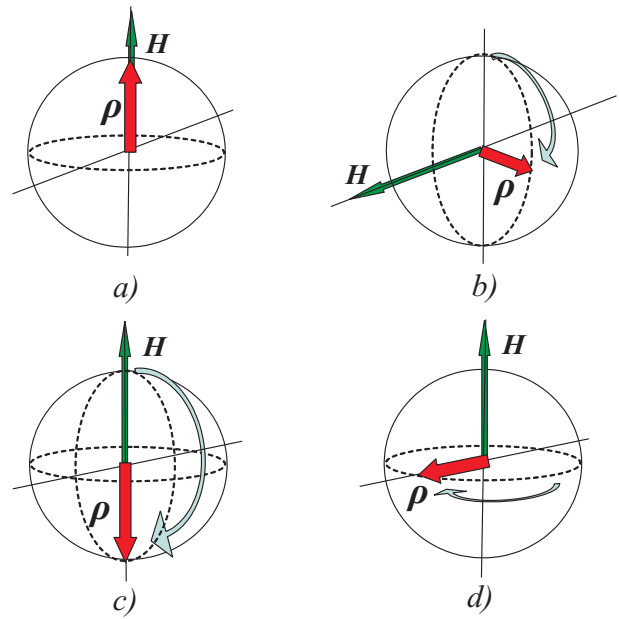}}
\caption{
Qubit operations with dc-pulses: the vector $H$ represents the qubit Hamiltonian, and the vector $\rho$ represents the qubit state. a) The qubit is initialized to the ground state; b) the Hamiltonian vector $H$ is suddenly rotated towards $x$-axis, and the qubit state vector $\rho$ starts to precess around $H$; c) when qubit vector reaches the south pole of the Bloch sphere, the Hamiltonian vector $H$ is switched back to the initial position; the vector $\rho$ remains at the south pole, indicating complete inversion of the level population ($\pi$-pulse); d) if the Hamiltonian vector $H$ is switched back when the qubit vector reaches  the equator  of the Bloch sphere ($\pi/2$-pulse), then the $\rho$ vector remains precessing at the equator, representing equal-weighted superposition of the qubit states (cat states) $|\psi\rangle = |0\rangle + e^{i\phi}|1\rangle $; this operation is the basis for the Hadamard gate.
}
\label{figDCpulse}
\end{figure}

Let us consider, for example, the diagonal qubit Hamiltonian, $\hat H
= (\epsilon/2)\sigma_z$, and apply a pulse $\delta\epsilon$ during a time $\tau$. This operation will shift phases of the qubit eigenstates by,
$\pm\delta\epsilon\tau/2\hbar$. If the applied pulse is such that $\epsilon$
is switched off, and instead, the $\sigma_x$ component, $\Delta$, is switched
on, Fig \ref{figDCpulse}, then the state vector will rotate around the $x$-axis, and after the time $\Delta\tau/2\hbar =\pi$ ($\pi$-pulse) the ground state,
$|+\rangle$, will flip and become, $|+\rangle \rightarrow |-\rangle$. This
manipulation corresponds to the quantum NOT operation. Furthermore, if the
pulse duration is twice smaller ($\pi/2$-pulse), then the ground state vector
will approach the equator of the Bloch sphere and precess along it after the
end of the operation. Such state is an equal-weighted superposition of the
basis states (cat state).

\subsection{Adiabatic switching}

Adiabatic switching represents the opposite limit to sudden switching, namely
that the state develops so fast on the time scale of the Hamiltonian that this can be regarded as "frozen" , i.e. the time-dependence of the
Hamiltonian becomes parametric. This implies that energy is conserved and no
transitions are induced - the system stays in the same energy level (although
the state changes).

\subsection{Harmonic perturbation and Rabi oscillation}
\label{sectionRabi}

A particularly interesting and practically important case concerns harmonic
perturbation with small amplitude $\lambda$ and resonant frequency
$\hbar\omega = E_2-E_1$. Let us consider the situation when the harmonic
perturbation is added to the $z$-component of the Hamiltonian corresponding
to a modulation of the qubit bias with a microwave field.

In the eigenbasis of the non-perturbed qubit, $|E_1\rangle$,
$|E_2\rangle$, the Hamiltonian will take the form,
\begin{equation}
\hat H = E_1\sigma_z + \cos\omega t \left( \lambda_z\,\sigma_z +
\lambda_x\,\sigma_x\right),
\end{equation}
\begin{equation}
\lambda_z= \lambda{\epsilon\over E_2}, \;\;\;\lambda_x =\lambda{\Delta\over
E_2}.
\end{equation}
The first perturbative term determines small periodic oscillations of the
qubit energy splitting, while the second term will induce interlevel
transitions. Despite the amplitude of the perturbation being small,
$\lambda/E_2 \ll 1$, the system will be driven far away from the initial
state because of the resonance. Indeed, let us consider the wave function of
the driven qubit on the form
\begin{equation}
|\psi\rangle = a(t)e^{-iE_1 t/\hbar}|E_1\rangle  +
b(t)e^{-iE_2t/\hbar}|E_2\rangle .
\end{equation}
Substituting this ansatz into the Schr\"odinger equation,
\begin{equation}
i\hbar |\dot\psi\rangle = \hat H(t)\,|\psi\rangle,
\end{equation}
we get the following equations for the coefficients,
\begin{eqnarray}
i\hbar \dot a  = \lambda_x\cos \omega t\,e^{i(E_1-E_2)t/\hbar}\, b ,
\nonumber \\
 i\hbar \dot b = \lambda_x\cos\omega t \,e^{i(E_2-E_1)t/\hbar}\, a.
\label{Rabi_eqs}
\end{eqnarray}
(Here we have neglected a small diagonal  perturbation, $\lambda_z$.)

\begin{figure}[h]
\centerline{\epsfxsize=0.40\textwidth\epsfbox{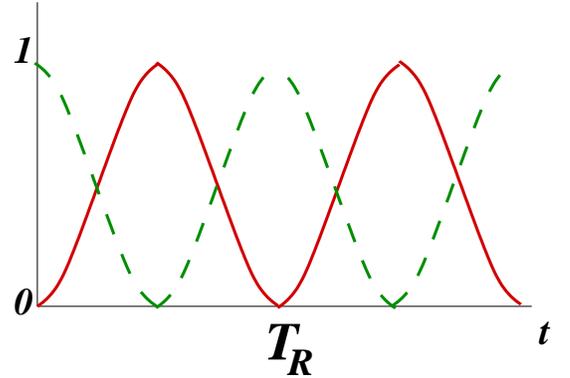}}
\caption{Rabi oscillation of populations of lower level (full line) and
upper level (dashed line) at exact resonance (zero detuning). $T_R= 2\pi/\Omega_R$ is the period of Rabi oscillations.}
\label{figRabi}
\end{figure}

Let us now focus on the slow evolution of the coefficients on the time scale of qubit precession, and average Eqs.(\ref{Rabi_eqs}) over the period of the precession. This approximation is known in the theory of two-level systems as the "rotating wave approximation (RWA)". Then, taking into account the resonance condition, we get the simple equations,
\begin{equation}
i\hbar \dot a  =  {\lambda_x\over 2} \, b ,\;\;\; i\hbar \dot b  {\lambda_x\over 2}\, a,
\end{equation}
whose solutions read,
\begin{eqnarray}
 a^{(1)}(t)  &=&  b^{(1)}(t)  =  e^{- i\lambda_x t/ 2\hbar} ,\nonumber\\
a^{(2)}(t)  &=  - & b^{(2)}(t)  =  e^{i\lambda_x t/ 2\hbar}.
\end{eqnarray}
Thus, the dynamics of a driven qubit is characterized by a linear
combination of the two wave functions,
\begin{eqnarray}\
|\psi^{(1)}\rangle &=& {1\over \sqrt 2}e^{- i\lambda_x t/ 2\hbar}
\left(\,e^{-iE_1t/\hbar}|E_1\rangle  + e^{-iE_2t/\hbar}|E_2\rangle \,
\right),\nonumber\\
 |\psi^{(2)}\rangle &=& {1\over \sqrt
2}e^{i\lambda_x t/ 2\hbar} \left(\,e^{-iE_1t/\hbar}|E_1\rangle -
e^{-iE_2t/\hbar}|E_2\rangle \,\right).\nonumber\\
\end{eqnarray}
Let us assume that the qubit was initially in the ground state,
$|E_1\rangle$, and that the perturbation was switched on instantly. Then the
wave function of the driven qubit will take the form,
\begin{equation}
|\psi\rangle = \cos {\lambda_x t\over 2\hbar}\, e^{-iE_1t/\hbar}|E_1\rangle +
i\sin{\lambda_x t\over 2\hbar} \, e^{-iE_2t/\hbar} |E_2\rangle.
\end{equation}
Correspondingly, the probabilities of the level occupations will oscillate in time,
\begin{eqnarray}
P_1 = \cos^2{\lambda_x t\over 2\hbar}= {1\over 2}\left( 1 +
\cos{\lambda_x t\over \hbar}\right), \nonumber \\
 P_2 = \sin^2{\lambda_x
t\over 2\hbar}= {1\over 2}\left( 1 - \cos{\lambda_x t\over \hbar}\right),
\end{eqnarray}
with small frequency $\Omega_R = \lambda_x/\hbar \ll \omega$, Rabi
oscillations, illustrated in Fig. \ref{figRabi}.

\subsection{Decoherence of qubit systems}

Descriptions of the qubit dynamics in terms of the density matrix and
Liouville equation are more general than descriptions in terms of the wave function, allowing the effects of dissipation to be included. The density matrix
defined in Section \ref{TLSdynamics} for a pure quantum state possesses the
projector operator property, $\hat\rho^2=\hat\rho$. This assumption can be
lifted, and then the density matrix describes a statistical mixture of pure
states, say energy eigenstates,
\begin{equation}\label{rhoE}
\hat\rho=\sum_{i}\rho_{i}|E_i\rangle\langle E_i|.
\end{equation}
The density matrix acquires off-diagonal elements when the basis rotates away
from the energy eigenbasis. Such a mixed state cannot be represented by the
vector on the Bloch sphere; however, its evolution is still described by
the Louville equation (\ref{Liouville}),
\begin{equation}\label{Liouville2}
i\hbar \partial_t\hat \rho = [\hat H ,\hat\rho].
\end{equation}
The density matrix in Eq. (\ref{rhoE}) is the stationary solution of the
Liouville equation. The evolution of an arbitrary density matrix, off-diagonal in the energy eigenbasis, is given by the equations,
\begin{equation}
\rho_1, \rho_2 ={\mbox const}, \;\;\rho_{12}\propto e^{i(E_1-E_2)t/\hbar}.
\end{equation}

Dissipation is included in the density matrix description by extending the
qubit Hamiltonian and including interaction with an environment. The
environment for macroscopic superconducting qubits basically consists of
various dissipative elements in external circuits which provide bias,
control, and measurement of the qubit. The "off-chip" parts of these
circuits are usually kept at room temperature and produce significant
noise. Examples are the fluctuations in the current source producing
magnetic field to bias flux qubits and, similarly, fluctuations of the
voltage source to bias gate of the charge qubits. Electromagnetic
radiation from the qubit during operation is another dissipative
mechanisms. There are also intrinsic microscopic mechanisms of
decoherence, such as fluctuating trapped charges in the substrate of the
charge qubits, and fluctuating trapped magnetic flux in the flux qubits,
believed to produce dangerous 1/f noise. Another intrinsic mechanism is
possibly the losses in the tunnel junction dielectric layer. Various kinds
of environment are commonly modelled with an infinite set of linear
oscillators in thermal equilibrium (thermal bath), linearly coupled to the
qubit (Caldeira-Leggett model \cite{CaldeiraLeggett1981,Leggett1987}). The
extended qubit-plus-environment Hamiltonian has the form in the qubit
energy eigenbasis\cite{Weiss},
\begin{eqnarray}\label{CL}
\hat H = -{1\over 2}E\sigma_z + \sum_i(\lambda_{iz}\sigma_z +
\lambda_{i\perp}\sigma_\perp) X_i \nonumber\\ \sum_i\left({\hat P^2_i\over
2m} + {m\omega_i^2 X_i^2\over 2}\right),
\end{eqnarray}
where $E=E_1-E_2$. The physical effects of the two coupling terms in Eq.
(\ref{CL}) are quite different. The "transverse" coupling term proportional to
$\lambda_\perp$ induces interlevel transitions and eventually leads to the
relaxation. The "longitudinal" coupling term proportional to $\lambda_z$
commutes with the qubit Hamiltonian and thus does not induce interlevel
transitions. However, it randomly changes the level spacing, which eventually
leads to the loss of phase coherence, dephasing. The effect of both
processes, relaxation and dephasing, are referred to as decoherence.

Coupling to the environment leads, in the simplest case, to the following
modification of the Liouville equation\cite{Slichter,Blum},
\begin{equation}
\partial_t \rho_z = -{1\over T_1}(\rho_{z} - \rho^{(0)}_z),\;\;
\end{equation}
\begin{equation}
\partial_t\rho_{12} = {i\over \hbar}E\,\rho_{12} -{1\over T_2}\,\rho_{12}\, .
\end{equation}
This equation is known as the Bloch-Redfield equation. The first equation
describes relaxation of the level population to the equilibrium form,
$\rho^{(0)}_z =-(1/2) \tanh(E/ 2kT)$, $T_1$ being the relaxation time. The
second equation describes disappearance of the off-diagonal matrix element
during characteristic time $T_2$, dephasing.

The relaxation time is determined by the spectral density of the
environmental fluctuations at the qubit frequency,
\begin{equation}
{1\over T_1} = {\lambda_\perp^2\over 2} S_\phi(\omega=E).
\end{equation}
The particular form of the spectral density depends on the properties of the
environment, which are frequently expressed via the impedance (response function) of the environment. The most common environment consists of a pure resistance, in this case, $S_\phi(\omega)\propto \omega$, at low frequencies.

The dephasing time consists of two parts,
\begin{equation}
{1\over T_2} ={1\over 2T_1}+{1\over T_\phi}.
\end{equation}
The first part is generated by the relaxation process, while the second part
results from the pure dephasing due to the longitudinal coupling to the
environment. This pure dephasing part is proportional to the spectral density
of the fluctuation at zero frequency.
\begin{equation}
{1\over T_\phi} = {\lambda_z^2\over 2} S_\phi(\omega=0).
\end{equation}

There is already a vast recent literature on decoherence and noise in superconducting circuits, qubits and detectors, and how to engineer the qubits and environment to minimize decoherence and relaxation \cite{MakhlinRMP2001,Makhlin2000,Falci2002,Paladino2002,Paladino2003,MakhlinShnirman2003,Makhlin2004,Shnirman2004,Falci2004b,Wilhelm2001,Wilhelm2003d,Goorden2003,vdWal2003,Lehnert2003a,Lehnert2003b,Roschier2004,Burkard2004,Burkard2004b,Bertet2004a,Bertet2005,AverinFazio2003,Zazunov2005,Governale2001,StorczWilhelm2003c,RabSverdAverin2004,RabenstAverin2003,Ioffe2004}.
Many of these issues will be at the focus of this article.

\section{Classical superconducting circuits}\label{VClassicalcircuits}

In this section we describe a number of elementary superconducting circuits with tunnel Josephson junctions, which are used as building blocks in qubit applications.
These basic circuits are: single current biased Josephson junction; single
Josephson junction (JJ) included in a superconducting loop (rf SQUID); two
Josephson junctions included in a superconducting loop (dc SQUID); and an ultra-small superconducting island connected to a massive superconducting electrode
via tunnel Josephson junction (Single Cooper pair Box, SCB).

\subsection{Current biased Josephson junction}
\label{sectionJJ}

\begin{figure}
\centerline{\epsfxsize=0.45\textwidth\epsfbox{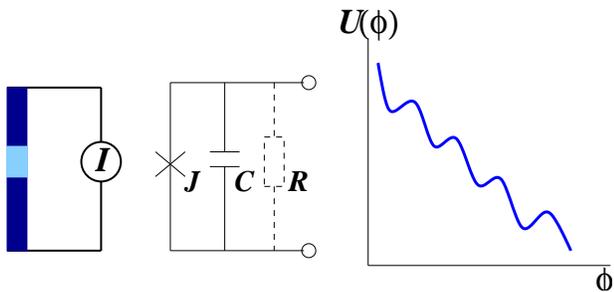}}
\caption{Current-biased Josephson junction (JJ) (left), equivalent circuit
(center), and effective (washboard-like) potential (right).
The superconducting leads are indicated with dark color, and the tunnel junction with light color. }
\label{figJJ}
\end{figure}

The simplest superconducting circuit, shown in Fig. \ref{figJJ}, consists of a tunnel junction with superconducting electrodes, a tunnel Josephson junction, connected to a current source. An equivalent electrical circuit, which represents the junction consists of the three lumped elements connected in parallel: the junction capacitance $C$, the junction resistance $R$, which generally differs from the normal junction resistance $R_N$ and  strongly depends on temperature and applied voltage, and the Josephson element associated with the tunneling through the junction.

The current-voltage relations for the
junction capacitance and resistance have standard forms, $I_C = C\,(dV/
dt)$, and $I_R = V / R$. To write down a similar relation for the Josephson
element, it is necessary to introduce the superconducting phase difference $\phi(t)$ across the junction, often simply referred to as the superconducting phase, which is related to the voltage drop across the junction,
\begin{equation}\label{phi}
\phi(t) = {2e\over\hbar}\,\int V\, dt + \phi,
\end{equation}
where $\phi$ is the time-independent part of the phase difference.
The phase difference can be also related to a magnetic flux,
\begin{equation}\label{phi_e}
\phi = {2e\over \hbar}\, \Phi = 2\pi {\Phi\over\Phi_0},
\end{equation}
where $\Phi_0= h / 2e$ is the magnetic flux quantum. The current through the
Josephson element has the form \cite{Josephson},
\begin{equation}\label{IJ}
I_J = I_c \sin \phi,
\end{equation}
where $I_c$ is the critical Josephson current, i.e. the maximum
non-dissipative current that may flow through the junction. The
microscopic theory of superconductivity \cite{deGennes1966,Tinkham,Barone}
gives the following equation for the Josephson current,
\begin{equation}\label{Ic}
I_c  = {\pi\Delta\over 2eR_N}\, \tanh {\Delta\over 2T},
\end{equation}
where $\Delta$ is the superconducting order parameter, and $T$ is the
temperature. Using these relations and expressing voltage through the
superconducting phase, we can write down Kirchhoff's rule for the circuit,
\begin{equation}\label{KirchhoffI}
{\hbar\over 2e} C \ddot\phi + {\hbar\over 2eR} \dot\phi + I_c
\sin\phi = I_e,
\end{equation}
where $I_e$ is the bias current. This equation describes the dynamics of
the phase, and it has the form of a damped non-linear oscillator. The
role of the non-linear inductance is here played by the Josephson
element.

The dissipation determines the qubit lifetime, and therefore circuits
suitable for qubit applications must have extremely small
dissipation. Let us assume zero level of the dissipation, dropping the resistive term in Eq. (\ref{KirchhoffI}). Then the
circuit dynamic equations, using the mechanical analogy, can be
presented in the Lagrangian form, and, equivalently, in the
Hamiltonian form. The circuit Lagrangian consists of the difference
between the kinetic and potential energies, the electrostatic energy
of the junction capacitors playing the role of kinetic energy, while
the energy of the Josephson current plays the role of potential
energy.

The kinetic energy corresponding to the first term in the Kirchhoff
equation (\ref{KirchhoffI}) reads,
\begin{equation}\label{K}
K(\dot\phi) = \left({\hbar \over 2e}\right)^2 {C \dot\phi^2\over 2}.
\end{equation}
This energy is equal to the electrostatic energy of the junction capacitor,
 $CV^2/2$. It is convenient to introduce the
charging energy of the junction capacitor charged with one electron pair (Cooper pair),
\begin{equation}\label{Ec}
E_C = {(2e)^2\over 2C},
\end{equation}
in which case Eq. (\ref{K}) takes the form
\begin{equation}\label{K2}
K(\dot\phi) = {\hbar^2 \dot\phi^2\over 4E_C}.
\end{equation}
The potential energy corresponds to the last two terms in Eq.
(\ref{KirchhoffI}), and consists of the energy of the Josephson
current, and the magnetic energy of the bias current,
\begin{equation}\label{U}
U(\phi) = E_J(1\, -\,\cos\phi) - {\hbar \over 2e}I_e\phi,
\end{equation}
where $E_J =\hbar/2e\,I_c$ is the Josephson energy. This potential energy has
a form of a washboard (see Fig. \ref{figJJ}). In the absence of
bias current this potential corresponds to a pendulum with the frequency of
small-amplitude oscillations given by
\begin{equation}\label{J}
\omega_J = \sqrt{2eI_c\over \hbar C}.
\end{equation}
This frequency is known as the plasma frequency of the Josephson junction.
When current bias is applied, the pendulum potential becomes tilted, its
minima becoming more shallow, and finally disappearing when the bias current becomes equal to the critical current, $I_e = I_C$. At this point, the plasma oscillations become unstable, which physically corresponds to switching to the dissipative regime and the voltage state.

Now we are ready to write down the Lagrangian for the circuit, which is the
difference between the kinetic and potential energies. Combining Eqs.
(\ref{K2}) and (\ref{U}), we get,
\begin{equation}\label{LagrI}
L (\phi, \dot\phi) =  {\hbar^2 \dot\phi^2\over 4E_C} - E_J(1\, -\,\cos\phi) +
{\hbar \over 2e}I_e\phi.
\end{equation}
It is straightforward to check that the Kirchhoff equation
(\ref{KirchhoffI}) coincides with the dynamic equation following from
the Lagrangian (\ref{LagrI}), using
\begin{equation}
{d\over dt}{\partial L\over \partial \dot\phi} - {\partial L\over
\partial\phi} = 0.
\end{equation}

It is important to emphasize, that the resistance of the junction can only be
neglected for low temperatures, and also only for slow time evolution of the phase; both the temperature and the characteristic frequency must be small compared to the magnitude of the energy gap in the superconductor: $T, \hbar\omega \ll \Delta$. The physical reason behind this constraint concerns the amount of generated
quasiparticle excitations in the system: if the constraint is fulfilled, the
amount of equilibrium and non-equilibrium excitations will be exponentially small. Otherwise, the gap in the spectrum will not play any significant role,
dissipation becomes large, and the advantage of the superconducting state
compared to the normal conducting state will be lost.

\subsection{rf-SQUID}
\label{sectionSQUID}

\begin{figure}
\centerline{\epsfxsize=0.35\textwidth\epsfbox{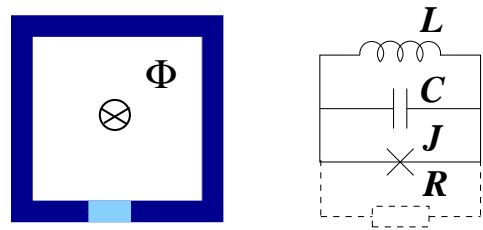}}
\caption{Superconducting quantum interference device - SQUID (left)
consists of a superconducting loop (dark) interrupted by a tunnel junction
(light); magnetic flux $\Phi$ is sent through the loop. Right: equivalent
circuit. }
\label{figSQUID}
\end{figure}

The rf-SQUID is the next important superconducting circuit. It consists of a
tunnel Josephson junction inserted in a superconducting loop, as illustrated in Fig. \ref{figSQUID}. This circuit realizes magnetic flux bias for the Josephson
junction  \cite{Barone}.
To describe this circuit, we introduce the current associated with
the inductance $L$ of the leads,
\begin{equation}\label{IL}
I_L = {\hbar\over 2e L}(\phi - \phi_e),\;\;\; \phi_e = {2e\over
\hbar}\, \Phi_{e}
\end{equation}
where $\Phi_e$ is the external magnetic flux threading the SQUID
loop. The Kirchhoff rule for this circuit takes the form
\begin{equation}\label{Kirchhoff}
{\hbar\over 2e} C \ddot\phi + {\hbar\over 2eR} \dot\phi + I_c
\sin\phi + {\hbar\over 2e L}(\phi - \phi_e) = 0.
\end{equation}
While neglecting the Josephson tunneling ($I_c =0$), this equation describes
a damped linear oscillator of a conventional LC-circuit. The resonant frequency
is then,
\begin{equation}\label{LC}
\omega_{LC} = {1\over \sqrt{LC}},
\end{equation}
and the (weak) damping is $\gamma = 1/RC$.

\begin{figure}
\centerline{\epsfxsize=0.30\textwidth\epsfbox{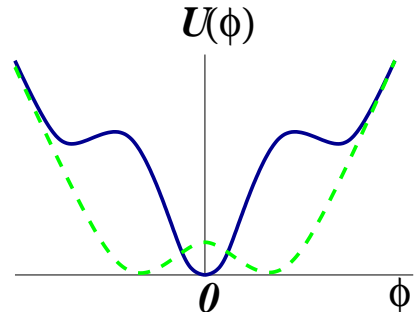}}
\caption{SQUID potential: the full (dark) curve corresponds to integer bias flux (in units of flux quanta), while the dashed (light) curve corresponds to half-integer bias flux.}
\label{figSQUIDpotential}
\end{figure}

In the absence of dissipation, it is straightforward to write down
the Lagrangian of the rf-SQUID,
\begin{equation}\label{Lagr}
L (\phi, \dot\phi) = {\hbar^2\dot\phi^2\over 4E_C} - E_J(1\,
-\,\cos\phi) - E_L {(\phi-\phi_e)^2\over 2}.
\end{equation}
The last term in this equation corresponds to the energy of the
persistent current circulating in the loop,
\begin{equation}
E_L = {\Phi_0^2\over 4\pi^2L}.
\end{equation}
The potential energy $U(\phi)$ corresponding to the last two terms in Eq. (\ref{Lagr}) is schetched in Fig. \ref{figSQUIDpotential}.

For bias flux equal to integer number of flux quanta, or $\phi_e 2\pi
n$, the potential energy of the SQUID has one absolute minimum at $\phi =\phi_e$. For half integer flux quanta the potential energy has two
degenerate minima, which correspond to the two persistent current states
circulating in the SQUID loop in the opposite directions. This configuration
of the potential energy provides the basis for constructing a persistent-current flux qubit (PCQ).

\subsection{dc SQUID}
\label{sectionDCSQUID}

\begin{figure}
\centerline{\epsfxsize=0.30\textwidth\epsfbox{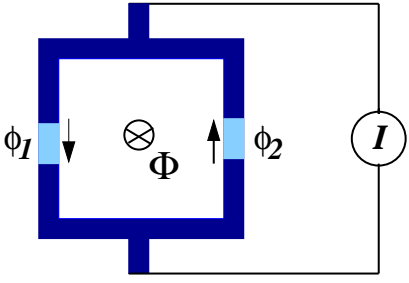}}
\caption{dc SQUID consists of two tunnel junctions included in a
superconducting loop; arrows indicate the direction of the positive
Josephson current. }
\label{figDCSQUID}
\end{figure}

We now consider consider the circuit shown in Fig. \ref{figDCSQUID}
consisting of two Josephson junctions coupled in parallel to a current
source. The new physical feature here, compared to a single current-biased
junction, is the dependence of the effective Josephson energy of the double
junction on the magnetic flux threading the SQUID loop.

Let us evaluate the effective Josephson energy. Now the circuit has two
dynamical variables, superconducting phases, $\phi_{1,2}$ across the two
Josephson junctions. Defining phases as shown in the figure, and
applying considerations from Sections \ref{sectionJJ} and \ref{sectionSQUID}
we find for the static Josephson currents,
\begin{equation}\label{static}
I_{c1}\sin\phi_1 - I_{c2}\sin\phi_2 = I_e,
\end{equation}
where $I_e$ is the biasing current. Let us further assume small inductance of
the SQUID loop and neglect the magnetic energy of circulating currents. Then
the total voltage drop over the two junctions is zero, $V_1 + V_2=0$, and
therefore, $\phi_1 + \phi_2 = \phi_e$, where $\phi_e$ is the biasing phase
related to biasing magnetic flux. Introducing new variables,
\begin{equation}
\phi_\pm = {\phi_1 \pm \phi_2\over 2},
\end{equation}
and taking into account that $2\phi_+ = \phi_e$, we rewrite equation
(\ref{static}) on the form,
\begin{equation}\label{Isquid}
I_c (\phi_e)\sin(\phi_- + \alpha) = I_e,
\end{equation}
where
\begin{equation}
I_c (\phi_e) = \sqrt { I_{c1}^2 + I_{c2}^2 + I_{c1}I_{c2}\cos\phi_e},
\end{equation}
and
\begin{equation}
\tan\alpha = {I_{c1}-I_{c2}\over I_{c1} + I_{c2}} \tan {\phi_e\over 2}.
\end{equation}
For a symmetric SQUID with $I_{c1}=I_{c2}$, giving $\alpha = 0$,  Eq. (\ref{Isquid}) reduces to the form
\begin{equation}\label{Isymm}
2I_c\cos(\phi_e/2)\sin\phi_- = I_e.
\end{equation}
The potential energy generated by Eq. (\ref{Isquid}) has the form,
\begin{equation}\label{USQUID}
U_(\phi) = 2E_J\cos\left({\phi_e\over 2}\right)(1 - \cos \phi_-) - {\hbar
\over 2e}I_e\phi_-,
\end{equation}
which indeed is similar to the potential energy of a single current biased
junction, Eq. (\ref{U}) and Fig. \ref{figSQUIDpotential}, but with flux-controlled critical current. This property of the
SQUID is used in qubit applications for controlling the Josephson coupling,
and also for measuring the qubit flux.

The kinetic energy of the SQUID can readily be written down noticing that it is
associated with the charging energy of the two junction capacitances
connected in parallel,
\begin{equation}
K(\dot\phi) = \left({\hbar \over 2e}\right)^2 (C_1 + C_2){\dot\phi^2 \over
2}.
\end{equation}
Thus the Lagrangian for the SQUID has a form similar to Eq. (\ref{LagrI})
where $E_C =(2e)^2/2(C_1+C_2)$,
\begin{equation}
L(\phi, \dot\phi) =  {\hbar^2 \dot\phi^2\over 4E_C} -
2E_J\cos\left({\phi_e\over 2}\right)(1\, -\,\cos\phi) + {\hbar \over
2e}I_e\phi.
\end{equation}

\subsection{Single Cooper Pair Box (SCB)}
\label{sectionCPB}

\begin{figure}
\centerline{\epsfxsize=0.25\textwidth\epsfbox{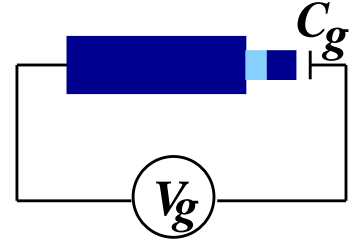}}
\caption{Single Cooper pair box (SCB): a small superconducting island connected to a bulk superconductor via a tunnel junction; the island potential is
controlled by the gate voltage $V_g$.  }
\label{figCPB1}
\end{figure}

There is a particularly important Josephson junction circuit consisting of
a small superconducting island connected via a Josephson tunnel junction to a
large superconducting reservoir (see Fig. \ref{figCPB1}).

The island is capacitively coupled to another massive electrode, which may
act as an electrostatic gate. The voltage source $V_g$ controls the gate
potential. In the normal state, such a circuit is named a Single Electron
Box (SEB) \cite{Kastner1992} for the following reason: if the junction
resistance exceeds the quantum resistance $R_q \approx 26 \;{\mbox
k}\Omega$, and the temperature is small compared to the charging energy of
the island,
 the system is in a Coulomb blockade regime \cite{Giaever1968,Kulik1975} where
 the electrons can only be transferred to the island one by
one, the number of electrons on the island being controlled by the gate
voltage. In the superconducting state, the same circuit is called a Single
Cooper pair Box (SCB) \cite{Lafarge1993a,Lafarge1993b}; for a review see
the book \cite{SingleChargeTunneling}. An experimental SCB device is shown
in Fig. \ref{SCB_Delsing}. In this section we consider a classical
Lagrangian for the circuit. Since the structure now has two
capacitances, one from the tunnel junction, $C$, and another one from the gate,
$C_g$, the electrostatic term in the Hamiltonian must be reconsidered.

Let us first evaluate the electrostatic energy of the SCB. It has the form,
\begin{equation}\label{ChargingSCB}
{CV^2\over 2} + {C_g(V_g - V)^2\over 2},
\end{equation}
where $V$ is the voltage over the tunnel junction. Then the Lagrangian can be
written (omitting the constant term),
\begin{equation}\label{Lcpb}
L(\phi,\dot\phi) = {C_\Sigma\over 2}\left({\hbar\over 2e}\dot\phi - {C_g\over
C_\Sigma}V_g\right)^2 - E_J(1-\cos\phi),
\end{equation}
where $C_\Sigma = C+C_g$.

\begin{figure}
\centerline{\epsfxsize=0.30\textwidth\epsfbox{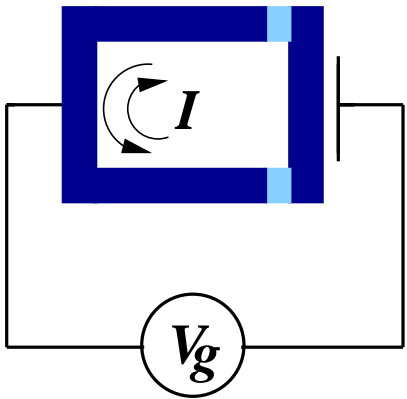}}
\caption{Single Cooper pair transistor (SCT): SCB with loop-shape bulk
electrodes; charge fluctuations on the island produces current fluctuation
in the loop. }
\label{figCPT}
\end{figure}

The interferometer effect of two Josephson junctions connected in
parallel (Fig. \ref{figCPT}) can be used to control the Josephson energy of the single Cooper pair box. This setup can be viewed as a flux-biased dc SQUID
where Josephson junctions have very small capacitances and are placed very
close to each other so that the island confined between them has large
charging energy. The gate electrode is connected to the island to control the
charge. The effect of the gate electrode is only essential for the kinetic
term, and repeating previous analysis we arrive at the following equation for
the kinetic energy (assuming for simplicity identical junctions),
\begin{eqnarray}\label{Kcpt}
K(\dot\phi_-) = {C_\Sigma\over 2}\left({\hbar\over 2e}\dot\phi_- - {C_g\over
C_\Sigma}V_g\right)^2, \nonumber \\
C_\Sigma = 2C + C_g.
\end{eqnarray}
Combining this kinetic energy with the potential energy derived in previous
subsection, we arrive at the Lagrangian of the SCB (cf. Eq. (\ref{Lcpb})) where both the charging energy and the Josephson energy can be controlled,
\begin{eqnarray}\label{Lcpb2}
 L_{} = {C_\Sigma\over 2}\left({\hbar\over 2e}\dot\phi_- -
{C_g\over C_\Sigma}V_g\right)^2 \nonumber \\
+ \; 2E_J\cos\left({\phi_e\over 2}\right)\cos\phi_-.
\end{eqnarray}

\section{Quantum superconducting circuits }
\label{VIQuantumcircuits}

One may look upon the Kirchhoff rules, as well as the circuit Lagrangians,
as the equations describing the dynamics of electromagnetic field in the
presence of the electric current. Generally, this electromagnetic field is
a quantum object, and therefore there must be a quantum generalization of
the equations in the previous section. At first glance, the idea of
quantization of an equation describing a macroscopic circuit containing a
huge amount of electrons may seem absurd. To convince ourselves that the
idea is reasonable, it is useful to recall an early argument in favor of
the quantization of electron dynamics in atoms, and to apply it to the
simplest circuit, an rf-SQUID: When the current oscillations are excited
in the SQUID, it works as an antenna radiating electromagnetic waves.
Since EM waves are quantized, the same should apply to the antenna
dynamics.

To quantize the circuit equation, we follow the conventional way of canonical
quantization: first we introduce the Hamiltonian and then change the classical
momentum to the momentum operator. The Hamiltonian is related to the
Lagrangian as
\begin{equation}\label{HL}
H(p, \phi) = p \dot\phi - L,
\end{equation}
where $p$ is the canonical momentum conjugated to coordinate $\phi$,
\begin{equation}\label{PL}
p = {\partial L \over\partial\dot\phi}.
\end{equation}
For the simplest case of a single junction, Eq. (\ref{K}), the momentum
reads,
\begin{equation}\label{P}
p =  \left({\hbar \over 2e}\right)^2 C \dot\phi.
\end{equation}
 The so defined momentum has a simple interpretation: it is proportional to the charge $q = C V$ on the junction capacitor, $p = (\hbar/2e) q$, or the number $n$ of electronic pairs on the junction capacitor,
\begin{equation}\label{pn}
p = \hbar n.
\end{equation}

The Hamiltonian for the current-biased junction has the form (omitting a
constant),
\begin{equation}\label{Hbiased}
H =  E_C\, n^2 - E_J\,\cos\phi - {\hbar\over 2e}I_e\phi .
\end{equation}
Similarly, the Hamiltonian for the SQUID circuit has the form
\begin{equation}\label{Hsquid}
H(n,\phi) = E_C n^2 - E_J\,\cos\phi + E_L{(\phi-\phi_e)^2\over 2}.
\end{equation}
The dc SQUID considered in the previous Section \ref{sectionDCSQUID} has two
degrees of freedom. The Hamiltonian can be written by generalizing Eqs.
(\ref{Hbiased}), (\ref{Hsquid})  for the phases $\phi_\pm$. In the symmetric
case we have,
\begin{eqnarray}\label{Hsquid2}
H &=&  E_C\,  n_+^2 + E_C\, n_-^2 - 2E_J\,\cos\phi_+\cos\phi_-
\nonumber \\
 &+& E_L{(2\phi_+-\phi_e)^2\over 2} + {\hbar\over 2e}I_e\phi_-.
\end{eqnarray}
For the SCB, Eq. (\ref{Lcpb}), the conjugated momentum has
the form,
\begin{equation}\label{Pcpb}
p = {\hbar C_\Sigma\over 2e} \left( {\hbar\over 2e} \dot\phi - {C_g\over
C_\Sigma}V_g\right),
\end{equation}
and the Hamiltonian reads,
\begin{equation}\label{HQcpb1}
H = E_C (n - {n_g})^2 - E_J\cos\phi,
\end{equation}
where now $E_C=(2e)^2/ 2C_\Sigma$, and $ n =  p/\hbar$ has the
meaning of the number of electron pairs (Cooper pairs) on the island electrode.
$n_g = - C_gV_g/2e$ is the charge on the gate capacitor (in units of Cooper pairs (2e)), which can be tuned by the gate potential, and which
therefore plays the role of external controlling parameter.

The quantum Hamiltonian results from Eq. (\ref{PL}) by substituting the classical
momentum $p$ for the differential operator,
\begin{equation}\label{Phat}
\hat p = -i \hbar \,{\partial\over\partial\phi}.
\end{equation}
Similarly, one can define the charge operator,
\begin{equation}\label{Qhat}
\hat q = - \,2e i\,{\partial\over\partial\phi}, \;\;\;
\end{equation}
and the operator of the pair number,
\begin{equation}\label{Nhat}
 \hat n = - \, i\,{\partial\over\partial\phi}.
\end{equation}
The commutation relation between the phase operator and the pair number
operator has a particularly simple form,
\begin{equation}
[\phi, \hat n ] = \, i.
\end{equation}

The meaning of the quantization procedure is the following: the phase and charge
dynamical variables can not be exactly determined by means of physical
measurements; they are fundamentally random variables with the probability of
realization of certain values given by the modulus square of the wave
function of a particular state,
\begin{equation}
\langle \phi\rangle = \int \psi^\ast (\phi)\,\phi \,\psi (\phi)\,d\phi,\;\;\;
\langle q\rangle = \int \psi^\ast (\phi)\,\hat q \,\psi (\phi)\, d\phi.
\end{equation}
The time evolution of the wave function is given by the Schr\"odinger
equation,
\begin{equation}
i\hbar {\partial\psi(\phi,t)\over\partial t} = \hat H \psi(\phi,t)
\end{equation}
where $\bar H$ is the circuit quantum Hamiltonian. The explicit form of the
quantum Hamiltonian for the circuits considered above, is the following:\\

\noindent
{\em rf-SQUID:}
\begin{equation}\label{HQsquid}
\hat H =  E_C\, \hat n^2 - E_J\,\cos\phi + E_L {(\phi-\phi_e)^2\over 2};
\end{equation}
{\em Current biased JJ:}
\begin{equation}\label{HQJJ}
\hat H =  E_C\, \hat n^2 - E_J\,\cos\phi + {\hbar\over 2e}I_e\phi;
\end{equation}
{\em dc-SQUID:}
\begin{eqnarray}
\hat H &=&  E_C\, \hat n_+^2 + E_C\, \hat n_-^2 -
2E_J\,\cos\phi_+\cos\phi_- \nonumber \\
 &+& E_L{(2\phi_+-\phi_e)^2\over 2} + {\hbar\over 2e}I_e\phi_-.
\end{eqnarray}
and finally the {\em single Cooper pair box (SCB):}
\begin{equation}\label{HQcpb}
\hat H_{} = E_C (\hat n - {n_g})^2 - E_J\cos\phi,
\end{equation}

For junctions connecting macroscopically large electrodes, the charge on
the junction capacitor is a continuous variable. This implies that no
specific boundary conditions on the wave function are imposed. The situation
is different for the SCB: in this case one of the electrodes, the
island, is supposed to be small enough to show pronounced charging effects. If tunneling is forbidden, electrons are trapped on the island, and their
number is always integer (the charge quantization condition). However, there is a difference between the energies of even and odd numbers of electrons on the
island: while an electron pair belongs to the superconducting condensate and
has the additional energy $E_C$, a single electron forms an excitation and thus
its energy consists of the charging energy, $E_C/2$ plus the excitation
energy, $\Delta$ (parity effect)\cite{Lafarge1993a}. To prevent the appearance
of individual electrons on the island and to provide the SCB
regime, the condition $\Delta \gg E_C/2$ must be fulfilled. Thus when the
tunneling is switched on, only Josephson tunneling is allowed since it
transfers Cooper pairs and the number of electrons on the island must change
pairwise, $n_- = \; $integer. In order to provide such a constraint, periodic boundary conditions on the SCB wave function are imposed,
\begin{equation}\label{BC}
\psi(\phi) = \psi(\phi+2\pi).
\end{equation}
This implies that arbitrary state of the SCB is a superposition of the charge
states with integer amount of the Cooper pairs,
\begin{equation}
\psi(\phi) = \sum_n a_n e ^{in\phi}.
\end{equation}

The uncertainty of the dynamical variables are not important as long as the relative mean deviations of dynamical variables are small,
i.e. the amplitudes of the quantum fluctuations are small. In this case, the
particle behaves as a classical particle. It is known from quantum mechanics,
that this corresponds to large mass of the particle, in our case, to large
junction capacitance. Thus we conclude that the quantum effects in the
circuit dynamics are essential when the junction capacitances are
sufficiently small. The qualitative criterion is that the charging energy must be larger than, or comparable to, the Josephson energy of the junction, $E_C \sim E_J$.

Typical Josephson energies of the tunnel junctions in qubit circuits are of
the order of a few degrees Kelvin or less. Bearing in mind that the insulating layers of the junctions have the thickness of few atomic distances, and modeling the junction as a planar capacitor, the estimated junction area should be smaller than a few square micrometers to observe the circuit quantum dynamics.

One of most important consequences of the quantum dynamics is quantization of
the energy of the circuit. Let us consider, for example, the LC circuit. In
the classical case, the amplitude of the plasma oscillations has continuous values. In the quantum case the amplitude of oscillation can only have certain discrete values defined
through the energy spectrum of the oscillator. The linear oscillator is well
studied in the quantum mechanics, and its energy spectrum is very well known,
\begin{equation}
E_n = \hbar\omega_{LC} (n + 1/2),\;\;\; n = 0,1,2,...
\end{equation}

One may ask, why is quantum dynamics never observed in ordinary electrical
circuits? After all, in high-frequency applications, frequencies up to THz
are available, which corresponds to a distance between the quantized
oscillator levels of order 10K, which can be observed at sufficiently low
temperature. For an illuminative discussion on this issue see the paper by
Martinis, Devoret and Clarke \cite{Martinis1987}. According to Ref.
\onlinecite{CaldeiraLeggett1981} it is the dissipation that kills quantum
fluctuations: as known from classical mechanics, the dissipation
(normal resistance) broadens the resonance, and good resonators must have small
resonance width, $\gamma = 1/RC$ compared to the resonance frequency,
$\gamma\ll\omega_{LC}$. It is intuitively clear that the quantization effect
will be destroyed when the level broadening exceeds the level spacing. For
quantum behavior of the circuit, narrow resonances, $\gamma\ll\omega_{LC}$,
are therefore essential. However, even in this case it is hard to observe the
quantum dynamics in linear circuits such as LC-resonators \cite{Martinis1987}
because the expectation values of the linear oscillator follow the
classical time evolution. Thus the presence of non-linear circuit elements is
essential.

The linear oscillator provides the simplest example of a quantum energy
level spectrum: it only consists of discrete levels with equal distance
between the levels. The level spectrum of the Josephson junction
associated with a pendulum potential is more complicated. Firstly, because
of the non-linearity (non-parabolic potential wells), the energy spectrum
is non-equidistant (anharmonic), the high-energy levels being closer to
each other than the low-energy levels. Moreover, for energies larger than
the amplitude of the potential (top of the barrier), $E_J$, the spectrum is
continuous. Secondly, one has to take into account the possibility of a
particle tunneling between neighboring potential wells: this will produce
broadening of the energy levels into energy bands. The level broadening is
determined by the overlap of the wave function tails under the potential
barriers, and it must be small for levels lying very close to the bottom
of the potential wells. Such a situation may only exist if the level
spacing, given by the plasma frequency of the tunnel junction is much
smaller than the Josephson energy, $\hbar\omega_J \ll E_J$, i.e. when $E_C
\ll E_J$. This almost classical regime with Josephson tunneling
dominating over charging effects, is called the phase regime, because
phase fluctuations are small and the superconducting phase is well
defined. In the opposite case, $E_C \gg E_J$, the lowest energy level lies
well above the potential barrier, and this situation corresponds to wide
energy bands separated by small energy gaps. In this case, the wave
function far from the gap edges can be well approximated with a plane
wave,
\begin{equation}
\psi_q(\phi) = \exp{iq\phi\over 2e}.
\end{equation}
This wave function corresponds to an eigenstate of the charge operator with
well defined value of the charge, $q$. Such a regime with small charge
fluctuations is called the charge regime.

\section{Basic qubits}
\label{VIIBasqubits}

The quantum superconducting circuits considered above contain a large
number of energy levels, while for qubit operation only two levels are
required. Moreover, these two qubit levels must be well decoupled from the
other levels in the sense that transitions between qubit levels and the
environment must be much less probable than the transitions between the
qubit levels itself. Typically that means that the qubit should involve a
low-lying pair of levels, well separated from the spectrum of higher
levels, and not being close to resonance with any other transitions.

\subsection{Single Josephson Junction (JJ) qubit}\label{sectionJJq}

The simplest qubit realization is a current biased JJ with large Josephson
energy compared to the charging energy. In the classical regime, the
particle representing the phase either rests at the bottom of one of the
wells of the "washboard" potential (Fig. \ref{figJJ}), or oscillates
within the well. Due to the periodic motion, the average voltage across
the junction is zero, $\overline{\dot\phi}=0$. Strongly excited states,
where the particle may escape from the well, correspond to the dissipative
regime with non-zero average voltage across the junction,
$\overline{\dot\phi} \ne 0$.

\begin{figure}
\centerline{\epsfxsize=0.30\textwidth\epsfbox{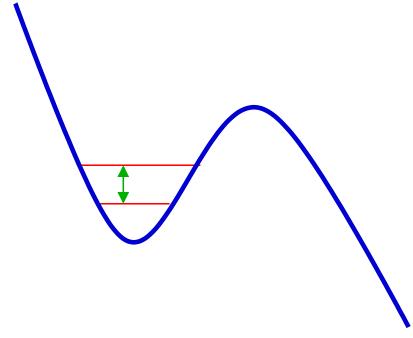}}
\caption{Quantized energy levels in the potential of a current biased Josephson
Junction }
\label{figJJQ}
\end{figure}

In the quantum regime described by the Hamiltonian (\ref{Hbiased}),
\begin{equation}\label{Hbiased1}
\hat H =  E_C\, \hat n^2 - E_J\,\cos\phi - {\hbar\over 2e}I_e\phi ,
\end{equation}
particle confinement, rigorously speaking, is impossible because of
macroscopic quantum tunneling (MQT) through the potential barrier, see Fig.
\ref{figJJQ}. However, the probability of MQT is small and the tunneling
may be neglected if the particle energy is close to the bottom of the local
potential well, i.e. when $E \ll E_J$. To find the conditions for such a
regime, it is convenient to approximate the potential with a parabolic
function, $U(\phi)\approx (1/2)E_J\cos\phi_0\,(\phi - \phi_0)^2$, where
$\phi_0$ corresponds to the potential minimum, $E_J \sin\phi_0 = (\hbar/
2e)I_e$. Then the lowest energy levels, $E_k = \hbar\omega_p(k+1/2)$ are
determined by the plasma frequency, $\omega_p = 2^{1/4}\omega_J (1 -
I_e/I_c)^{1/4}$. It then follows that the levels are close to the bottom
of the potential if $E_C \ll E_J$, i.e. when the Josephson junction is in the
phase regime, and moreover, if the bias current is not too close to the critical
value, $I_e<I_c$.

It is essential for qubit operation that the spectrum in the well is not
equidistant. Then the two lowest energy levels, $k = 0,1$ can be employed for
the qubit operation. Truncating the full Hilbert space of the junction to the
subspace spanned by these two states, $|0\rangle$ and $|1\rangle$, we may
write the qubit Hamiltonian on the form,
\begin{equation}
H_q = -{1\over 2} \epsilon\sigma_z,
\end{equation}
where $\epsilon= E_1 - E_0$.

The interlevel distance is controlled by the bias current. When bias current
approaches the critical current, level broadening due to MQT starts to
play a role, $E_k \,\rightarrow \,E_k + i\Gamma_k/2$. The MQT rate for the
lowest level is given by \cite{Weiss}
\begin{equation}\label{Gmqt}
\Gamma_{MQT} = {52\omega_p\over 2\pi}\sqrt{U_{max}\over \hbar\omega_p}\exp
\left(-{7.2U_{max}\over \hbar\omega_p}\right),
\end{equation}
where $U_{max} = 2\sqrt 2 (\Phi_0/2\pi)(1 - I_e/I_c)^{3/2}$ is the height of
the potential barrier at given bias current.

\subsection{Charge qubits}
\subsubsection{Single Cooper pair Box - SCB}

\begin{figure}
\centerline{\epsfxsize=0.30\textwidth\epsfbox{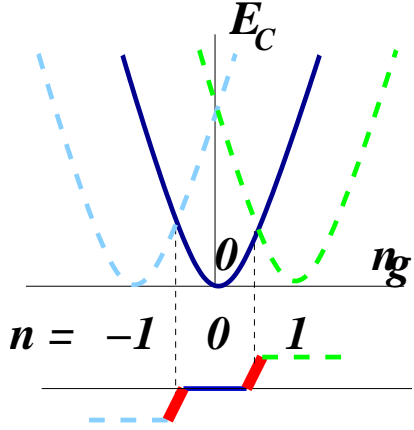}}
\caption{Single Cooper pair box (SCB): charging energy (upper panel), and charge on the island (lower panel) vs gate potential. Washed-out onsets of the charge
steps indicate quantum fluctuations of the charge on the island.}
\label{figCPB}
\end{figure}

An elementary charge qubit can be made with the SCB operating in the charge
regime, $E_C \gg E_J$. Neglecting the Josephson coupling implies the complete
isolation of the island of the SCB, with a specific number of Cooper pairs trapped on the island. Correspondingly, the eigenfunctions,
\begin{equation}
E_C(\hat n - n_g)^2 |n\rangle = E_n |n \rangle,
\end{equation}
correspond to the charge states $n = 0,1,2...$, with the energy
spectrum $E_n = E_C(n - n_g)^2$, as shown in Fig. \ref{figCPB}. The ground
state energy oscillates with the gate voltage, and the number of
Cooper pairs in the ground state increases. There are, however,
specific values of the gate voltage, e.g. $n_g =  1/2$ where the
charge states $|0\rangle$ and $|1\rangle$ become degenerate.
Switching on a small Josephson coupling will then lift the degeneracy,
forming a tight two-level system.

The qubit Hamiltonian is derived by projecting the full Hamiltonian
(\ref{HQcpb}) on the two charge states, $|0\rangle$, $|1\rangle$, leading to
\begin{equation}\label{HSCB}
\hat H_{SCB}= -{1\over 2} (\epsilon \;\sigma_z + \Delta \;\sigma_x),
\end{equation}
where $\epsilon= E_C(1- 2n_g)$, and $\Delta = E_J$. The qubit level
energies are then given by the equation
\begin{equation}
E_{1,2} = \mp {1\over2}\sqrt {E_C^2(1 - 2n_g)^2 + E_J^2},
\end{equation}
the interlevel distance being controlled by the gate voltage. At
the degeneracy point, $n_g=1/2$, the diagonal part of the qubit
Hamiltonian vanishes, the levels being separated by the Josephson
energy, $E_J$, and the qubit eigenstates corresponding to the cat
states, $|E_1\rangle, |E_2\rangle = |0\rangle \mp |1\rangle$. For theses states,
the average charge on the island is zero, while it changes to $\mp
2e$ far from the degeneracy point, where the qubit eigenstates
approach pure charge states.

\begin{figure}
\centerline{\epsfxsize=0.30\textwidth\epsfbox{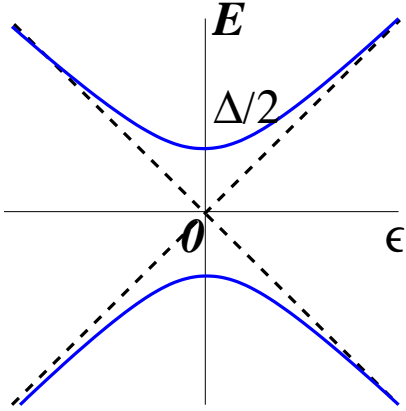}}
\caption{Energy spectrum of the SCB (solid lines): it results from
hybridization of the charge states (dashed lines). }
\label{figSpectrum}
\end{figure}

The SCB was first experimentally realized by Lafarge et al.
\cite{Lafarge1993b}, observing the Coulomb staircase with steps of $2e$
and the superposition of the charge states, see also \cite{Bouchiat1998}.
Realization of the first charge qubit by manipulation of the SCB and
observation of Rabi oscillations was done by Nakamura et al.
\cite{Nakamura1999,Nakamura2002a,Nakamura2002b}, and further investigated
theoretically by Choi et al. \cite{Choi2001}.

\subsubsection{Single Cooper pair Transistor - SCT}
\label{sectionSCT}

In the SCB, charge fluctuations on the island generate fluctuating current
between the island and large electrode. In the two-junction setup discused at
the end of Section \ref{sectionCPB}, an interesting question concerns how the
current is distributed between the two junctions. The answer to this question
is apparently equivalent to evaluating the persistent current circulating
in the SQUID loop. This current was neglected so far because the associated induced flux $\tilde\phi = 2\phi_+ - \, \phi_e$ was assumed to be
frozen, $\tilde\phi=0$, in the limit of infinitely small SQUID inductance,
$L=0$. Let us now lift this assumption and allow fluctuation of the induced
flux; the Hamiltonian will then take the form (\ref{Hsquid2}), in which a gate
potential is included in the charging term of the SCB, and the term
containing external current is dropped,
\begin{eqnarray}\label{HQSCT}
\hat H_{SCT} = E_C(\hat n_-  - n_g)^2 + E_C \hat n^2_+  \nonumber\\ -
2E_J\cos\phi_+\cos\phi_- + E_L{(2\phi_+-\phi_e)^2\over 2}.
\end{eqnarray}
($\hat n_+ = -i\partial/\partial \phi_+$).  For small but non-zero
inductance, the amplitude of the induced phase is small, $\tilde\phi
= 2\phi_+ -\phi_e \ll 1$, and the cosine term containing $\phi_+$ can
be expanded, yielding the equation
\begin{equation}\label{HcptLin}
\hat H_{SCT} = \hat H_{SCB}(\phi_-) + \hat H_{osc}(\tilde\phi) + \hat
H_{int} \; .
\end{equation}
$\hat H_{SCB}(\phi_-)$ is the SCB Hamiltonian (\ref{HQcpb}) with the
flux dependent Josephson energy, $E_J(\phi_e) = 2E_J\cos(\phi_e/2)$. $\hat
H_{osc}(\tilde\phi)$ describes the linear oscillator associated with the
variable $\tilde\phi$,
\begin{equation}
\hat H_{osc}(\tilde\phi) = 4E_C\hat{\tilde n^2} + E_L{\tilde\phi^2\over 2},
\end{equation}
and the interaction term reads,
\begin{equation}\label{Hint}
\hat H_{int} = E_J\sin\left({\phi_e\over 2}\right)\cos(\phi_-) \, \tilde\phi \;.
\end{equation}
Thus, the circuit consists of the non-linear oscillator of the SCB linearly
coupled to the linear oscillator of the SQUID loop. This coupling gives the
possibility to measure the charge state of the SCB by measuring the
persistent currents and the induced flux.

Truncating Eq. (\ref{HcptLin}) we finally arrive at the Hamiltonian which is
formally equivalent to the spin-oscillator Hamiltonian,
\begin{equation}\label{HQOSC-2}
\hat H_{SCT} = -{1\over 2}\left(\epsilon\sigma_z + \Delta(\phi_e)\sigma_x
\right)+ \lambda\tilde\phi\sigma_x + H_{osc} \;.
\end{equation}
In this equation, $\Delta(\phi_e) = 2E_J\cos(\phi_e/2)$, and $\lambda = E_J\sin(\phi_e/2)$.

\subsection{Flux qubit}
\label{sectionFluxqubit}

\subsubsection{Quantum rf-SQUID}

\begin{figure}
\centerline{\epsfxsize=0.30\textwidth\epsfbox{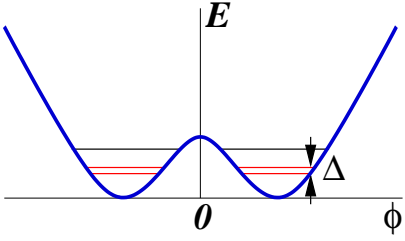}}
\caption{ Double-well potential of the rf-SQUID with degenerate
quantum levels in the wells (black). Macroscopic quantum tunneling (MQT) through the potential barrier introduces a level splitting $\Delta$, and the lowest
level pair forms a qubit.}
\label{figDwell}
\end{figure}

An elementary flux qubit can be constructed from an rf-SQUID operating in the phase regime, $E_J \gg E_C$. Let us consider the Hamiltonian
(\ref{Hsquid}) at $\phi_e = \pi$, i.e. at half integer bias magnetic flux. The potential, $U(\phi)$, shown in Fig. \ref{figDwell}
has two identical wells with equal energy levels when MQT between the wells
is neglected (phase regime, $\omega_J \ll E_J$). These levels are connected with
current fluctuations within each well around averaged values corresponding
to clockwise and counterclockwise persistent currents circulating in the loop
(the flux states). Let us consider the lowest, doubly degenerate, energy level.
When the tunneling is switched on, the levels split, and a tight two-level
system is formed with the level spacing determined by the MQT rate, which is
much smaller than the level spacing in the well.
In the case that the tunneling barrier is much smaller than the Josephson energy, the potential in Eq. (\ref{U}) can be approximated,
\begin{eqnarray}\label{Ufluxqubit}
U(\phi) &=&  E_J (1-\cos\phi) + E_L
{(\phi-\phi_e)^2\over 2}\nonumber\\
 & \approx & E_L \left(-\varepsilon {\tilde\phi^2\over 2} - f\tilde\phi + { 1+\varepsilon\over
24}\tilde\phi^4\right),
\end{eqnarray}
 where $\tilde\phi = \phi - \pi$, $f=\phi_e-\pi$, and where
\begin{equation}
\varepsilon = {E_J\over E_L} -1 \ll 1.
\end{equation}
 determines the height of the tunnel barrier.

\begin{figure}
\centerline{\epsfxsize=0.30\textwidth\epsfbox{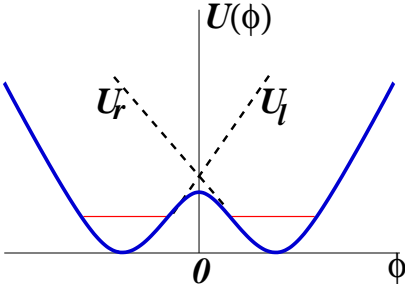}}
\caption{Flux qubit: truncation of the junction Hamiltonian; dashed lines
indicate potentials of the left and right wells with ground energy levels.}
\label{figDWell2}
\end{figure}

 The qubit Hamiltonian is derived by projecting the whole Hilbert
space of the full Hamiltonian (\ref{Hsquid}) on the subspace spanned by these
two levels. The starting point of the truncation procedure is to approximate
the double well potential with $U_l$ and $U_r$, as shown in Fig.
\ref{figDWell2}, to confine the particle to the left or to the right well,
respectively. The corresponding ground state wave functions $|l\rangle$ and
$|r\rangle$ satisfy the stationary Schr\"odinger equation,
\begin{equation}
\hat H_{l}|l\rangle = E_{l} |l\rangle,\;\;\;
 \hat H_{r}|r\rangle = E_{r} |r\rangle.
\end{equation}
The averaged induced flux for these states, $\phi_l$ and  $\phi_r$ have
opposite signs, manifesting opposite directions of the circulating persistent
currents. Let us allow the bias flux to deviate slightly from the half integer
value, $\phi_e = \pi + f$, so that the ground state energies are not equal
but still close to each other, $E_l \approx E_r$. The tunneling will
hybridize the levels, and we can approximate the true eigenfunction,
$|E\rangle$,
\begin{equation}\label{Hfull}
\hat H|E\rangle = E |E\rangle,
\end{equation}
 with a superposition,
\begin{equation}
|E\rangle  = a |l\rangle + b |r\rangle.
\end{equation}
The qubit Hamiltonian is given by the matrix elements of the full
Hamiltonian, Eq. (\ref{Hfull}), with respect to the states $|l\rangle$ and
$|r\rangle$,
\begin{eqnarray}
H_{ll} = E_l + \langle l|U-U_l|l\rangle, \nonumber \\
 H_{rr} = E_r + \langle r|U-U_r|r\rangle, \nonumber\\
 H_{rl} = E_l\langle r|l\rangle + \langle r|U-U_l|l\rangle.
\end{eqnarray}
In the diagonal matrix elements, the second terms are small because the wave
functions are exponentially small in the region where the deviation of the
approximated potential from the true one is appreciable. The off diagonal
matrix element is exponentially small because of small overlap of the ground
state wave functions in the left and right wells, and also here the main contribution comes from the first term. Since the wave functions can be chosen real, the truncated Hamiltonian is symmetric, $H_{lr}=H_{rl}$. Then introducing
$\epsilon = E_r-E_l$, and $\Delta/2 = H_{rl}$, we arrive at the Hamiltonian
of the flux qubit,
\begin{equation}
\hat H = -{1\over 2}(\epsilon\sigma_z + \Delta\sigma_x).
\end{equation}

The Hamiltonian of the flux qubit is formally equivalent to that of the charge qubit, Eq. (\ref{HSCB}), but the physical meaning of the terms is rather different. The flux qubit Hamiltonian is written in the flux basis, i.e. the basis of the states with certain averaged induced flux, $\phi_l$ and $\phi_r$ (rather than the charge basis of the charge qubit).

The energy spectrum of the flux qubit is obviously the same as that of the charge qubit,
\begin{equation}
E_{1,2} = \mp {1\over 2}\sqrt { \epsilon^2 + \Delta^2},
\end{equation}
as shown in Fig. \ref{figSpectrum}. However, for the flux qubit the dashed lines indicate persistent current states in the absence of macroscopic tunneling, and the current degeneracy point ($\epsilon = 0$) corresponds to a half-integer bias flux. The energy levels are controlled by the bias magnetic flux (instead of the gate voltage for the charge qubit).

At the flux degeneracy point, $f=0$
($\phi_e=\pi$), the level spacing is determined by the small amplitude of
tunneling through macroscopic potential barrier, and the wave functions
correspond to the cat states, which are equally weighted superpositions of
the flux states. Far from the degeneracy point, the qubit states are almost
pure flux states.

The possibility to achieve quantum coherence of macroscopic current states in
an rf-SQUID with a small capacitance Josephson junction was first pointed out
in 1984 by Leggett \cite{LeggettGarg1985}. However, successful experimental
observation of the effect was achieved only in 2000 by Friedman et al.
\cite{Friedman2000}.

\subsubsection {3-junction SQUID - persistent current qubit (PCQ)}
\begin{figure}
\centerline{\epsfxsize=0.20\textwidth\epsfbox{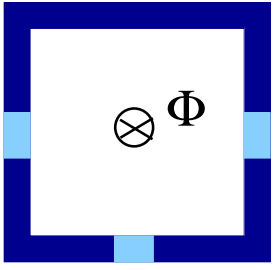}}
\caption{Persistent current flux qubit (PCQ) with 3 junctions. The side junctions are identical, while the central junction has smaller area. }
\label{Delft_PCQ}
\end{figure}

\begin{figure}
\centerline{\epsfxsize=0.45\textwidth\epsfbox{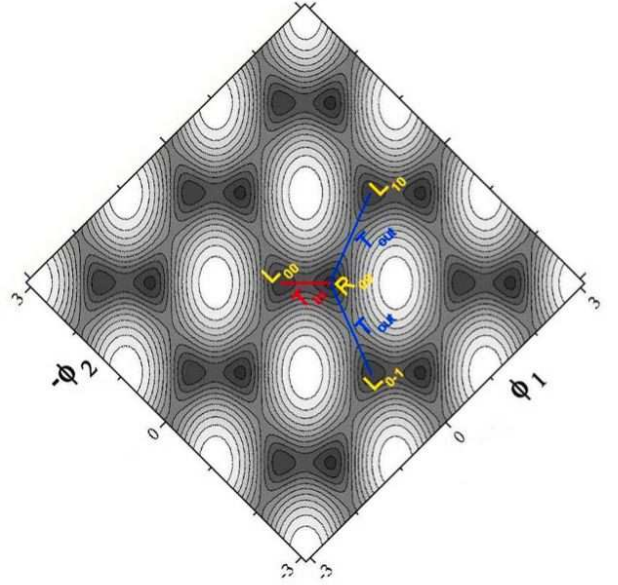}}
\caption{Potential energy landscape $U(\phi_1, \phi_2)$ = $U(\phi_+, \phi_-)$ of the Delft qubit as a function of the two independent phase variables $\phi_1$ and $\phi_2$, or equivalently, $\phi_+$ (horizontal axis) and $\phi_-$ (vertical axis).
Black represents the bottom of the potential wells, and white the top of the potential barriers. The qubit double-well potential is determined by the potential landscape centered around the origin in the horizontal direction, and typically has the shape shown in Fig. 16.
{\em Courtesy of C.H. van der Wal.}
}
\label{3Jpotential_vdWal}
\end{figure}

The main drawback of the flux qubit with a single Josephson junction
(rf-SQUID) described above concerns the large inductance of the qubit loop,
the energy of which must be comparable to the Josephson energy to form the
required double-well potential profile. This implies large size of the qubit
loop, which makes the qubit vulnerable to dephasing by magnetic fluctuations
of the environment. One way to overcome this difficulty was pointed out by Mooij
et al. \cite{Mooij1999}, replacing the large loop inductance by the Josephson
inductance of an additional tunnel junction, as shown in Fig. \ref{Delft_PCQ},
The design employs three tunnel junctions connected in series in a
superconducting loop. The inductive energy of the loop is chosen to be much smaller than the Josephson energy of the junctions. The two junctions are supposed to be identical while the third junction is supposed to have smaller area, and therefore smaller Josephson and larger charging energy. The
Hamiltonian has the form,
\begin{eqnarray}
\hat H = E_C [\hat n_1^2 + \hat n_2^2 + \hat n_3^2/(1/2+\varepsilon)] -
\nonumber\\ E_J [\cos\phi_1 + \cos\phi_2 + (1/2+\varepsilon)\cos\phi_3].
\end{eqnarray}
To explain the idea, let us consider the potential energy. The three phases
are not independent and satisfy the relation $\phi_1 + \phi_2 + \phi_3 = \phi_e$.
Let us suppose that the qubit is biased at half integer flux quantum, $\phi_e = \pi$. Then introducing new
variables, $\phi_\pm = (\phi_1 \pm \phi_2)/2$, we have
\begin{equation}
U(\phi_+, \phi_-) = - E_J [ 2\cos\phi_- \cos\phi_+ \, - (1/2 +
\varepsilon)\cos 2\phi_+].
\end{equation}
The potential landscape is shown in Fig. \ref{3Jpotential_vdWal}, and the qubit potential consists of the double well structure near the points $(\phi_+,\phi_-) = (0,0)$. An approximate form of the potential energy is given by
\begin{equation}
U(\phi_+, 0) \approx E_J \left( -2\varepsilon \phi_+^2 + {\phi_+^4\over
4}\right).
\end{equation}

Each well in this structure corresponds to clock- and counterclockwise
currents circulating in the loop. The amplitude of the structure is given by
the parameter $\epsilon E_J$, and for $\epsilon \ll 1$ the tunneling between
these wells dominates. Thus this qubit is qualitatively similar to
the single-junction qubit described above, but the quantitative parameters are
different and can be significantly optimized.

\subsection{Potential qubits}

The superconducting qubits that have been discussed in previous
sections exploit the fundamental quantum uncertainty between electric
charge and magnetic flux. This uncertainty appears already in the
dynamics of a single Josephson junction (JJ), which is the basis for
elementary JJ qubits. There are however other possibilities. One
of them is to delocalize quantum information in a JJ network by
choosing global quantum states of the network as a computational
basis. Recently, some rather complicated JJ networks have been
discussed, which have the unusual property of degenerate ground state,
which might be employed for efficient qubit protection against
decoherence \cite{IoffeNature2002,FeigelmanPRL2004}.

An alternative possibility of further miniaturization of superconducting qubits could be to replace the standard Josephson junction by a quantum point contact (QPC), using the microscopic conducting modes in the JJ QPC, the bound Andreev states, as a computational basis, allowing control of intrinsic decoherence inside the junction \cite{Zazunov2005,Zazunov2003}.

To explain the physics of this type of qubit, let us consider an rf SQUID
(see Fig. \ref{figSQUID}) with a junction that has such a small cross section that the quantization of electronic modes in the (transverse) direction
perpendicular to the current flow becomes pronounced. In such a junction,
quantum point contact (QPC), the Josephson current is carried by a number of
independent conducting electronic modes, each of which can be considered an
elementary microscopic Josephson junction characterized by its own
transparency. The number of modes is proportional to the ratio of the
junction cross section and the area of the atomic cell (determined by the Fermi wavelength) of the junction material. In atomic-sized QPCs with only a few conducting modes, the Josephson current can be appreciable if the conducting modes are transparent (open modes). If the junction is fully transparent (reflectivity $R=0$) then
current is a well defined quantity. This will correspond to a persistent
current with certain direction circulating in the qubit loop. On the other hand, for a finite reflectivity ($R\neq0$), the electronic back scattering
will induce hybridization of the persistent current states giving rise to
strong quantum fluctuation of the current.

Such a quantum regime is distinctly different from the macroscopic
quantum coherent regime of the flux qubit described in Section
\ref{VIIBasqubits}, where the quantum hybridization of the
persistent current states is provided by charge fluctuations on the
junction capacitor. Clearly, charging effects will not play any
essential role in quantum point contacts, and the leading role
belongs to the microscopic mechanism of electron back scattering.
This mechanism is only pronounced in quantum point contacts: in
classical (large area)  contacts, such as junctions of macroscopic
qubits with areas of several square micrometers, the current is
carried by a large number ($>10^4$) of  statistically independent
conducting modes, and fluctuation of the net Josephson current is
negligibly small.

\begin{figure}
\centerline{\epsfxsize=0.35\textwidth\epsfbox{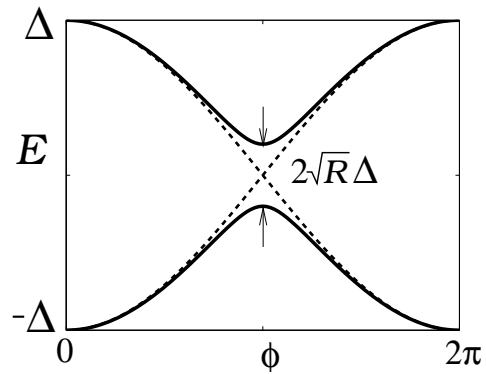}}
\caption{Energy spectrum of microscopic bound Andreev levels; the level
splitting is determined by the conact reflectivity.}
\label{figAL}
\end{figure}
In QPCs, the Josephson effect is associated with microscopic Andreev levels,
localized in the junction area, which transport Cooper pairs from one
junction electrode to the other\cite{Furusaki90,Shumeiko93}. As shown in Fig. \ref{figAL}, the Andreev
levels, two levels per conducting mode, lie within the superconducting gap
and have the phase-dependent energy spectrum,
\begin{equation}\label{Ea}
E_a=\pm \Delta\sqrt{\cos^2(\phi/ 2)+R\sin^2(\phi/ 2)},
\end{equation}
(here $\Delta$ is the superconducting order parameter in the junction
electrodes). For very small reflectivity, $R\ll 1$, and phase close to $\pi$
(half integer flux bias) the Andreev two-level system is well isolated from
the continuum states. The expectation value for the Josephson current carried
by the level is determined by the Andreev level spectrum,
\begin{equation}\label{Ia}
I_a = {2e\over \hbar} {dE_a\over d\phi_e},
\end{equation}
and it has different sign for the upper and lower level. Since the state of
the Andreev two-level system is determined by the phase difference and
related to the Josephson current, the state can be manipulated by driving
magnetic flux through the SQUID loop, and read out by measuring circulating
persistent current\cite{Moriond2001,SQUID2002}.

This microscopic physics underlines recent proposal for Andreev level qubit
\cite{Zazunov2005,Zazunov2003}. The qubit is similar to the macroscopic flux qubits with respect to how it is manipulated and measured, but the great difference is that the quantum information is stored in the microscopic quantum states. This difference is reflected in the more complex form of the qubit Hamiltonian, which consists of the two-level Hamiltonian of the Andreev levels strongly coupled to the quantum oscillator describing phase fluctuations,
\begin{equation}
\hat H = \Delta \, e^{-i \sigma_x \sqrt{R} \, \phi /2} \left( \cos
\frac{\phi}{2} \, \sigma_z + \sqrt{R} \, \sin \frac{\phi}{2} \,
\sigma_y \right) + \hat H_{osc}[\phi],\label{Ha}
\end{equation}
$\hat H_{osc}[\phi] = E_C\hat n + (E_L/2)(\phi-\phi_e)^2$. Comparing this
equation with e.g. the SCT Hamiltonian (\ref{HcptLin}), we find that the
truncated Hamiltonian of the SCB is replaced here by the Andreev level
Hamiltonian.

\section{Qubit readout and measurement of quantum information}
\label{SectVIII}

In this section we present a number of proposed, and realized, schemes for
measuring quantum states of various superconducting qubits.

\subsection {Readout: why, when and how?}

As already mentioned in Section \ref{Readout1}, the ultimate objective
of a qubit readout device is to distinguish the {\em eigenstates} of a qubit
in a single measurement "without destroying the qubit", a so called
"single-shot" quantum non-demolition (QND) projective measurement. This
objective is essential for several reasons: state preparation for computation, readout for error correction during the calculation, and readout of results at the end of the calculation. Strictly speaking, the QND property is only needed if the qubit must be left in an eigenstate after the readout. In a broader sense, readout of a specific qubit must of course not destroy any other qubits in the system.

It must be carefully noted that one cannot "read out the {\em state} of a qubit" in a single measurement - this is prohibited by quantum mechanics.
It takes repeated measurements on a large number of replicas of the quantum state to
characterize the state of the qubit (Eq. (\ref{angles})) - "quantum tomography" \cite{LiuWeiNori2004}.

The measurement connects the qubit with the open system of the detector,
which collapses the combined system of qubit and measurement device to one of
its common eigenstates. If the coupling between the qubit and the detector is
weak, the eigenstates are approximately those of the qubit. In general
however, one must consider the eigenstates of the total qubit-detector system
and manipulate gate voltages and fluxes such that the readout measurement is
performed in a convenient energy eigenbasis (see e.g. \cite{MakhlinRMP2001,Wilhelm2003a}).

Even under ideal conditions, a single-shot measurement can only determine the
population of an eigenstate if the system is prepared in an eigenstate: then
the answer will always be either "0" or "1". If an ideal single-shot
measurement is used to read out a qubit superposition state, e.g. during Rabi
oscillation, then again the answer can only be  "0" or "1". To determine the
qubit population (i.e. the $|a_1|^2$ and $|a_2|^2$  probabilities) requires
repetition of the measurement to obtain the expectation value. During the
intermediate stages of quantum computation one must therefore not perform a
measurement on a qubit unless one knows, because of the design and timing of
the algorithm, that this qubit is in an energy eigenstate. Then the value is
predetermined and  the qubit left in the eigenstate (Stern-Gerlach-style).

On the other hand, to extract the desired final result it may be necessary to
create an ensemble of calculations to be able to perform a complete measurement
to determine the expectation values of variables of interest, performing quantum state tomography \cite{LiuWeiNori2004}.

\subsection{Direct qubit measurement}

\begin{figure}[t]
\centerline{\epsfysize=0.30\textwidth \epsfbox{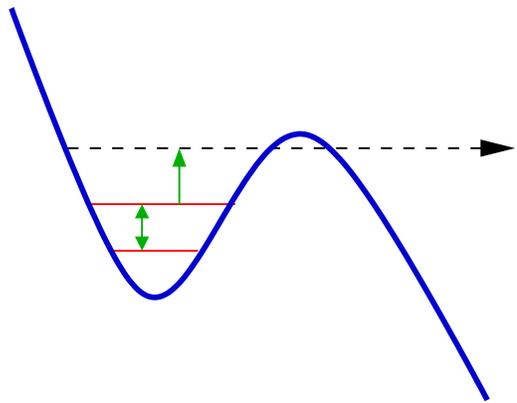}}
\caption{Measurement of the phase qubit. Long-living levels form the qubit, while the dashed line indicates a leaky level with large energy.}
\label{fig8JJQ2}
\end{figure}

Direct destructive measurement of the qubit can be illustrated with the
example of a single JJ (phase) qubit, Section \ref{sectionJJq}. After the
manipulation has been performed (e.g. Rabi oscillation), the qubit is left
in a superposition of the upper and lower energy states. To determine the
probability of the upper state, one slowly increases the bias current until it reaches such a value that the upper energy level equals Or gets close to) the top of the potential barrier, see Fig. \ref{fig8JJQ2}. Then the junction, being at the upper energy level, will switch from the Josephson branch to the
dissipative branch, and this can be detected by measuring the finite average
voltage appearing across the junction (voltage state). If the qubit is in the lower energy state the qubit will remain on the Josephson branch and a finite
voltage will not be detected (zero-voltage state). An alternative method to activate switching \cite{Martinis2002} is to apply an rf signal with resonant frequency (instead of tilting the junction potential) in order to excite the upper energy level and to induce the switching event, see Fig. \ref{fig8JJQ2}
(also illustrating a standard readout method in atomic physics).

It is obvious that, in this example, the qubit upper energy state is always
destroyed by the measurement. Single-shot measurement is possible provided
the MQT rate for the lower energy level is sufficiently small to prevent the
junction switching during the measurement time. It is also essential to keep
a sufficiently small rate of interlevel transitions induced by fluctuations
of the bias current and by the current ramping.

A similar kind of direct destructive measurement was performed by Nakamura et
al. \cite{Nakamura1999} to detect the state of the charge qubit. The qubit
operation was performed at the charge degeneracy point, $u_g=1$, where the
level splitting is minimal. An applied gate voltage then shifted the SCB working point (Fig. \ref{figSpectrum}), inducing a large level splitting of the pure charge states $\ket{0}$ and $\ket{1}$ (the measurement preparation stage).
In this process the upper $\ket{1}$ charge state went above the
threshold for Cooper pair decay, creating two quasi-particles which
immediately tunnel out via the probe junction into the leads. These
quasi-particles were measured as a contribution to the classical charge
current by repeating the experiment many times. Obviously, this type of
measurement is also destructive.

\subsection{Measurement of charge qubit with SET}

\begin{figure}[t]
\centerline{\epsfysize=0.30\textwidth \epsfbox{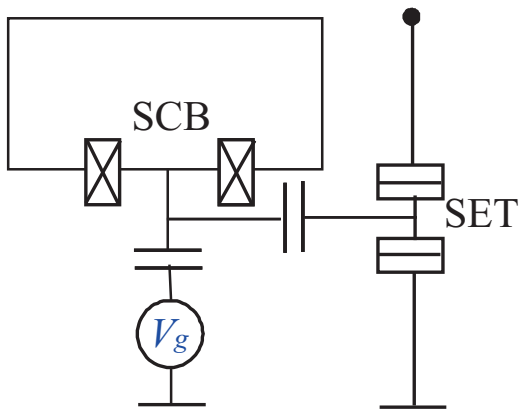}}
\caption{Single Electron Transistor (SET) capacitively coupled to an SCB.}
\label{SCB+rfSET}
\end{figure}

Non-destructive measurement of the charge qubit has been implemented by
connecting the qubit capacitively to a SET electrometer \cite{Aassime2001}. The
idea of this method is to use a qubit island as an additional SET gate (Fig. \ref{SCB+rfSET}), controlling the dc current through the SET depending on the state of the qubit. When the measurement is to be performed, a driving
voltage is applied to the SET, and the dc current is measured. Another
version of the measurement procedure is to apply rf bias to the SET (rf-SET
\cite{Aassime2001,DevSchoel2000,Schoelkopf1998,Aassime2001b}) in Fig. \ref{SCB+rfSET}, and to
measure the dissipative or inductive response. In both cases the transmissivity
will show two distinct values correlated with the two states of the qubit.
Yet another version has recently been developed by the NEC group
\cite{Astafiev2004} to perform single-shot readout: the Cooper pair on the
SCB island then tunnels out onto a trap island (instead of the leads) used as a gate to control the current through the SET.

The physics of the SET-based readout has been extensively studied
theoretically (see \cite{MakhlinRMP2001,Johansson2002,Kaeck2003} and
references therein). A similar idea of controlling the transmission of a
quantum point contact (QPC) (instead of an SET) capacitively coupled to a charge
qubit has also been extensively discussed in literature
\cite{Averin2000,KorotkovAverin2001,Averin2002,Goan2001a,Goan2001b,Shelankov2003,Rammer2004}.

\begin{figure}[t]
\centerline{\epsfysize=0.14\textwidth \epsfbox{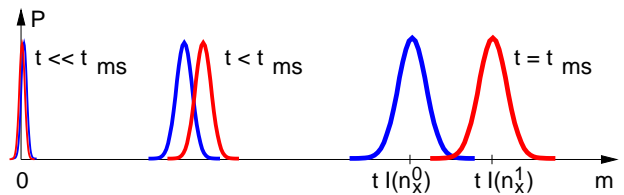}}
\caption{
Probability distributions P of counted electrons as functions of
time after the turning on the measurement beam of electrons.
{\em Courtesy of G. Johansson, Chalmers.}
}
\label{Measurementtime}
\end{figure}

The induced charge on the SET gate depends on the state of the qubit,
affecting the SET working point and determining the conductivity and the
average current. The development of the probability distributions of counted
electrons with time is shown in Fig. \ref{Measurementtime}.

As the number of
counted electrons grows, the distributions separate and become
distinguishable, the distance between the peaks developing as $\sim N$ and the
width $\sim \sqrt N$. Detailed investigations \cite{Rammer2004} show that the
two electron-number probability distributions correlate with the probability
of finding the qubit in either of two energy levels. The long-time
development depends on the intensity and frequency distribution of the
back-action noise from the electron current. With very weak detector back
action, the qubit can relax to $\ket{0}$ during the natural relaxation time
$T_1$. With very strong back-action noise at the qubit frequency, the qubit
may become saturated in a 50/50 mixed state.

\subsection{Measurement via coupled oscillator}

Another method of qubit read out that has attracted much attention concerns the
measurement of the properties of a linear or non-linear oscillator coupled to
a qubit. This method is employed for the measurement of induced magnetic
flux and persistent current in the loop of flux qubits and  charge-phase
qubits, as well as for charge measurement on charge qubits. With this method, the qubit affects the characteristics of the coupled oscillator, e.g. changes the shape of the oscillator potential, after which the oscillator can be probed to detect the changes. There are two versions of the method: resonant spectroscopy of a linear tank circuit/cavity, and threshold detection using biased JJ or SQUID magnetometer.

The first method uses the fact that the resonance frequency of a linear
oscillator weakly coupled to the qubit undergoes a shift depending on the
qubit state. The effect is most easily explained by considering the SCT
Hamiltonian, Eq. (\ref{HQOSC}),
\begin{eqnarray}\label{HQOSC2}
\hat H_{SCT} = -{1\over 2}\left(\epsilon\sigma_z + \Delta(\phi_e)\sigma_x
\right)+ \lambda(\phi_e)\tilde\phi\sigma_x \nonumber\\ + 4E_C\hat{\tilde n^2}
+ {1\over2}E_L\tilde\phi^2.
\end{eqnarray}
Let us proceed to the qubit energy basis, in which casee the qubit Hamiltonian takes the form $-(E/2)\sigma_z,\;\; E=\sqrt{\epsilon^2+\Delta^2}$.
The interaction
term in the qubit eigenbasis will consist of two parts, the longitudinal
part, $\lambda_z\tilde\phi\sigma_z$, $\lambda_z= (\Delta/E)\lambda$, and the
transverse part, $\lambda_x\tilde\phi\sigma_x$, $\lambda_x = (\epsilon/E)\lambda$. In the limit of weak coupling the transverse part of
interaction is the most essential. In the absence of interaction
($\phi_e=0$), the energy spectrum of the qubit + oscillator system is
\begin{equation}
%E_{n\sigma}= - \sigma E + \hbar\omega(n + {1\over 2}), \;\;\;\sigma
%=\pm,
E_{n\mp}=  \mp {E\over 2} + \hbar\omega(n + {1\over 2}),
\end{equation}
where $\hbar\omega=\sqrt {8E_CE_L}$ is the plasma frequency of the
oscillator. The effect of weak coupling is enhanced in the vicinity of the
resonance, when the oscillator plasma frequency is close to the qubit level
spacing, $\hbar\omega\approx E$. Let us assume, however, that the coupling
energy is smaller than the deviation from the resonance, $\lambda_x \ll
|\hbar\omega-E|$. Then the spectrum of the interacting system in the lowest
perturbative order will acquire a shift,
\begin{equation}\label{shiftE}
\delta E_{n\pm}= \pm \; (n+1) \; {\lambda_x^2\hbar\omega\over
E_L(\hbar\omega-E)}.
\end{equation}
This shift is proportional to the first power of the oscillator quantum number
$n$, which implies that the oscillator frequency acquires a shift (the
frequency of the qubit is also shifted \cite{Hekking2001,MarquardtBruder2001,Girvin2003,Blais2004,Rau2004}). Since the
sign of the oscillator frequency shift is different for the different qubit
states, it is possible to distinguish the state of the qubit by probing this
frequency shift.

In the case of the SCT, the LC oscillator is a generic part of the circuit. It
is equally possible to use an additional LC oscillator inductively coupled to
a qubit. This type of device has been described by Zorin \cite{Zorin2002} for SCT readout, and recently implemented for flux qubits by Il'ichev et al. \cite{Ilichev2003,Izmalkov2004}.

\begin{figure}[t]
\centerline{\epsfysize=0.35\textwidth \epsfbox{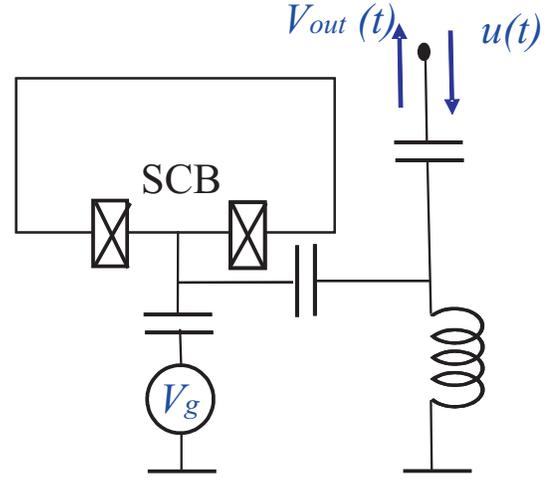}}
\caption{SCT qubit coupled to a readout oscillator. The qubit is operated by
input pulses $u(t)$. The readout oscillator is controlled and driven by ac
microwave pulses $V_g(t)$. The output signal will be ac voltage pulses
$V_{out}(t)$, the amplitude or phase of which may discriminate between the
qubit "0" and "1" states.}
\label{SCT_oscreadout}
\end{figure}

Figure \ref{SCT_oscreadout} illustrates another case, namely a charge qubit
capacitively coupled to an oscillator, again providing energy resolution for
discriminating the two qubit levels \cite{Roschier2005}. Analysis of this circuit is similar to
the one discussed below in the context of qubit coupling via oscillators,
Section \ref{SectIX}. The resulting Hamiltonian is similar to Eq.
(\ref{SCB_Osc}), namely,
\begin{equation}\label{HSCT}
\hat H = \hat H_{SCB} + \lambda\sigma_y \phi + \hat H_{osc}.
\end{equation}
In comparison with the case of the SCT, Eq. (\ref{HSCT}) has a different
form of the coupling term, which does not change during rotation to the
qubit eigenbasis. Therefore the coupling constant $\lambda$ directly
enters Eq. (\ref{shiftE}). Recently, this type of read out has been
implemented for a charge qubit by capacitively coupling the SCB of the
qubit to a superconducting strip resonator \cite{Wallraff2004,Schuster2004,Wallraff2005}.

\subsection{Threshold detection}

\begin{figure}[t]
\centerline{\epsfysize=0.25\textwidth \epsfbox{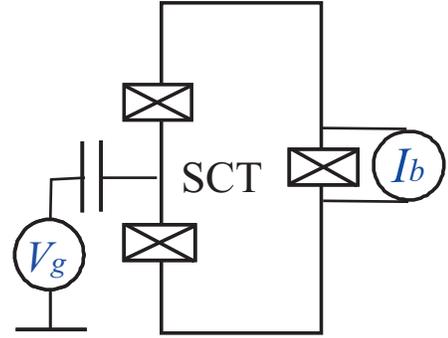}}
\caption{SCT qubit coupled to a JJ readout quantum oscillator. The JJ
oscillator is controlled by dc/ac current pulses $I_b(t)$ adding to the
circulating currents in the loop due to the SCT qubit. The output will be
dc/ac voltage pulses $V_{out}(t)$ discriminating between the qubit "0" and
"1" states.}
\label{SCT_JJreadout}
\end{figure}

To illustrate the threshold-detection method, let us consider an SCT qubit
with a third Josephson junction inserted in the qubit loop, as shown in Fig.
\ref{SCT_JJreadout}.

When the measurement of the qubit state is to be
performed, a bias current is sent through the additional junction. This
current is then added to the qubit-state dependent persistent current
circulating in the qubit loop. If the qubit and readout currents flow in the
same direction, the critical current of the readout JJ is exceeded, which
induces the junction switching to the resistive branch, sending out a voltage
pulse. This effect is used to distinguish the qubit states. The method has
been extensively used experimentally by Vion et
al.\cite{Vion2002,Vion2003,Collin2004}.

\begin{figure}[t]
\centerline{\epsfysize=0.30\textwidth \epsfbox{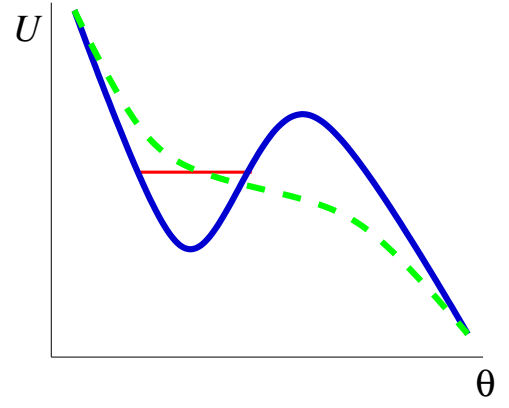}}
\caption{Josephson potential energy of the measurement junction during the
measurement: for the "0" qubit eigenstate  there is a well (full line)
confining a level, while for the "1" qubit state there is no well
(dashed line). }
\label{fig8TwoPotentials}
\end{figure}

To describe the circuit, we add the Lagrangian of a biased JJ, Eq.
(\ref{LagrI}),
\begin{equation}
L= {\hbar^2 \dot\phi^2\over 4E_C^m} + E_J^m\,\cos\phi + {\hbar\over
2e}I_e\phi,
\end{equation}
to the SCT Lagrangian (generalized Eq. (\ref{Lcpb2}), cf. Eqs.
(\ref{Hsquid2}), (\ref{HQSCT})),
\begin{eqnarray}
L_{SCT} = {\hbar^2\over 4E_{C}}\left(\dot\phi_- - {2eC_g\over \hbar
C_\Sigma}V_g\right)^2 + {\hbar^2\over
4E_{C}}\dot\phi_+^2\nonumber\\
- 2E_J\cos\phi_+\cos\phi_- - {1\over 2}E_L\tilde\phi^2
\end{eqnarray}
(here we have neglected a small contribution of the gate capacitance to the qubit charging energy). The phase quantization condition will now read:
$2\phi_+ + \phi = \phi_e + \tilde\phi$. The measurement junction will be assumed in the phase regime, $E_J^m \gg E_C^m$, and moreover, the inductive energy
will be the largest energy in the circuit, $E_L \gg E_J^m$. The latter
implies that the induced phase is negligibly small and can be dropped from
the phase quantization condition. We also assume that $\phi_e = 0$, thus
$2\phi_+ + \phi = 0$. Then, after having omitted the  variable $\phi_+$, the
kinetic energy term of the qubit can be combined with the much larger
kinetic energy of the measurement junction leading to insignificant
renormalization of the measurement junction capacitance, $C^m + C/4
\rightarrow C^m$. As a result, the total Hamiltonian of the circuit will
take the form,
\begin{eqnarray}\label{HQSCT3}
\hat H = E_C(\hat n_-  - n_g)^2 -  2E_J\cos\left({\phi\over
2}\right)\cos\phi_- \nonumber\\
 + E_C^m\,\hat n^2 - E_J^m\cos\phi - {\hbar\over 2e}I_e\phi.
\end{eqnarray}
Since the measurement junction is supposed to be almost classical,
its phase is fairly close to the minimum of the junction potential.
During qubit operation, the bias current is zero;  hence the phase of
the measurement junction is zero. When the measurement is made, the
current is ramped to a large value close to the critical current of the measurement junction, $I_e = (2e/\hbar)E_J^m -\delta
I$, tilting the junction potential and shifting the
minimum towards $\pi/2$. Introducing a new variable $\phi = \pi +
\theta$, we expand the potential with respect to small $\theta \ll
1$ and, truncating the qubit part, we obtain
\begin{eqnarray}\label{HQSCT4}
\hat H = -{\epsilon\over 2}\,\sigma_z - {\Delta\over 2}\left(1 -
{\theta\over 2}\right) \sigma_x  \nonumber \\ + E_C^m \,\hat n^2 -
E_J^m {\theta^3\over 6} + {\hbar\over 2e}\delta I \; \theta,
\end{eqnarray}
where $\Delta = 2\sqrt E_J$. The ramping is supposed to be adiabatic
so that the phase remains at the minimum point. Let us analyze the
behavior of the potential minimum by omitting a small kinetic term
and diagonalizing the Hamiltonian (\ref{HQSCT4}). The corresponding
eigenenergies depend on $\theta$,
\begin{equation}
E_\pm (\theta) = \mp {E\over 2} - E_J^m {\theta^3\over 6} +
\left({\hbar\over 2e}\delta I \pm {\Delta^2\over 4E}\right)\theta ,
\end{equation}
as shown in Fig. \ref {fig8TwoPotentials}.
Then within the interval of the bias currents, $|\delta I| \leq
-(2e/\hbar)(\Delta^2/ 4E)$, the potential energy corresponding to the
ground state has a local minimum, while for the excited state it does
not. This implies that when the junction is in the ground state,
no voltage will be generated. However, if the junction is in the excited state, it will switch to the resistive branch, generating a voltage pulse that can be detected.

With the discussed setup the direction of the persistent current is measured. It is also possible to arrange the measurement of the flux by using a dc
SQUID as a threshold detector. Such a setup is suitable for the
measurement of flux qubits. Let us consider, for example, the three
junction flux qubit from Section \ref{sectionFluxqubit} inductively coupled
to a dc SQUID. Then, under certain assumptions, the Hamiltonian of the
system can be reduced to the following form:
\begin{eqnarray}
\hat H = -{1\over 2}(\epsilon\sigma_z + \Delta  \sigma_x ) \nonumber
\\ + E_C^s \,\hat n2 - (E_J^s + \lambda \sigma_z)\cos\phi - {\hbar\over
2e}\delta I \; \phi,
\end{eqnarray}
where $E_J^s$ is an effective (bias flux dependent) Josephson energy
of the SQUID, Eq. (\ref{USQUID}), and $\lambda$ is an effective
coupling constant proportional to the mutual inductance of the qubit
and the SQUID loops.

\section{Physical coupling schemes for two qubits}
\label{SectIX}

\subsection{General principles}

A generic scheme for coupling qubits is based on the physical interaction of
linear and non-linear oscillators constituting a superconducting circuit.
The Hamiltonians for the SCB, rf-SQUID, and plain JJ contain
quadratic terms representing the kinetic and potential energies plus the
non-linear Josephson energy term, which is quadratic when expanded to lowest
order,
\begin{equation}
\hat H = E_C(\hat n - n_g)^2 + E_J(1\, -\,\cos\phi)
\end{equation}
\begin{equation}
\hat H =  E_C\, \hat n^2 + E_J(1\, -\,\cos\phi) + E_L {(\phi-\phi_e)^2\over2}
\end{equation}
\begin{equation}
\hat H =  E_C\, \hat n^2 + E_J(1\, -\,\cos\phi) + {\hbar\over 2e}I_e\phi .
\end{equation}
In a multi-qubit system the induced gate charge in the SCB, or the flux
through the SQUID loop, or the phase in the Josephson energy, will be a sum
of contributions from several (in principle, all) qubits. The energy of the
system can therefore not be described as the sum of two independent qubits
because of the quadratic dependence, and the cross terms represent
interaction energies of different kinds: capacitive, inductive and
phase/current. Moreover, using JJ circuits as non-linear coupling elements we
have the advantage that the direct physical coupling strength may be
controlled, e.g tuning the inductance via current biased JJs, or tuning the
capacitance by a voltage biased SCB.

\subsection{Inductive coupling of flux qubits }
\begin{figure}[t]
\centerline{\epsfxsize=0.40\textwidth\epsfbox{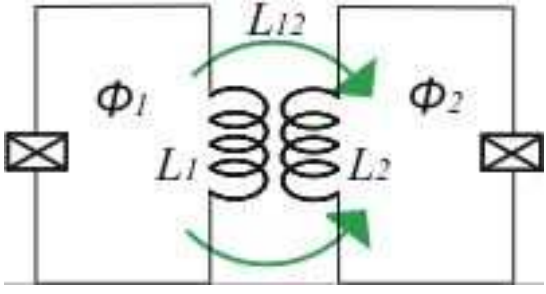}}
\caption{Fixed inductive (flux) coupling of elementary flux qubit.
The loops can be separate, or have a common leg like in the figure.}
\end{figure}

A common way of coupling flux qubits is the inductive coupling: magnetic
flux induced by one qubit threads the loop of another qubit, changing the
effective external flux. This effect is taken into account by introducing the
inductance matrix $L_{ik}$, which connects flux in the $i$-th loop with the
current circulating in the $k$-th loop,
\begin{equation}
\Phi_i = \sum_k L_{ik} I_k.
\end{equation}
The off-diagonal element of this matrix, $L_{12}$, is the mutual inductance
which is responsible for the interaction. By using the inductance matrix, the
magnetic part of the potential energy in Eq. (\ref{HQsquid}) can be
generalized to the case of two coupled qubits,
\begin{equation}
{1\over 2}\left({\hbar\over 2e}\right)^2\sum_{ik} (L^{-1})_{ik} (\phi_i-
\phi_{ei})(\phi_k - \phi_{ek}).
\end{equation}
Then following the truncation procedure explained in
Section \ref{sectionFluxqubit}, we calculate the matrix elements,
\begin{equation}
\langle l|\tilde\phi-f|l\rangle,\;\;\langle r|\tilde\phi-f|r\rangle,\;\;
\langle l|\tilde\phi-f|r\rangle,
\end{equation}
for each qubit. The last matrix element is exponentially small, while the
first two ones are approximately equal to the minimum points of the potential
energy, $\phi_l$ and $\phi_r$, respectively. This implies that the truncated
interaction basically has the $zz$-form,
\begin{eqnarray}\label{Lcoupling}
\hat H_{int} = \lambda\sigma_{z1} \sigma_{z2}, \nonumber\\
\lambda = {1\over 8}\left({\hbar\over 2e}\right)^2(L^{-1})_{12}\,
(\phi_l-\phi_r)_1 (\phi_l- \phi_r)_2.
\end{eqnarray}

\subsection{Capacitive coupling of charge qubits}

\begin{figure}[t]
\centerline{\epsfxsize=0.40\textwidth\epsfbox{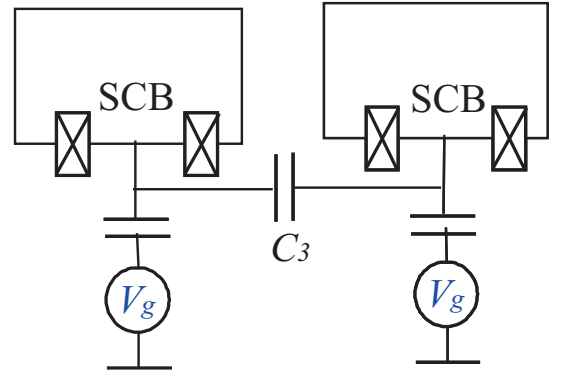}}
\caption{Fixed capacitive coupling of charge qubits}
\label{figCcoupling}
\end{figure}

One of the simplest coupling schemes is the capacitive coupling of charge
qubits. Such a coupling is realized by connecting the islands of two SCBs
via a small capacitor, as illustrated in Fig. \ref{figCcoupling}.
This will introduce an additional
term in the Lagrangian of the two non-interacting SCBs, Eq.
(\ref{ChargingSCB}), namely the charging energy of the capacitor $C_3$,
\begin{equation}\label{C3}
\delta L = {C_3V_3^2\over 2}.
\end{equation}
The voltage drop $V_3$ over the capacitor is expressed via the phase
differences across the qubit junctions,
\begin{equation}\label{V3}
V_3 = {\hbar\over 2e}(\dot\phi_1-\dot\phi_2),
\end{equation}
and thus the kinetic part of the Lagrangian (\ref{Lcpb}) will take the
form
\begin{eqnarray}
K(\dot\phi_1,\dot\phi_2) = {1\over2}\left({\hbar\over 2e}\right)^2
\sum_{i,k} C_{ik}\dot\phi_i \dot\phi_k \nonumber \\
- {\hbar\over 2e}\sum_i^2 C_{gi}V_{gi}\dot\phi_i,
\end{eqnarray}
where the capacitance matrix elements are $C_{ii} = C_{\Sigma i}+ C_3$, and
$C_{12}= C_3$. Then proceeding to the circuit quantum Hamiltonian as
described in Section \ref{VIQuantumcircuits}, we find the interaction term,
\begin{equation}\label{Hnn}
\hat H_{int} = 2e^2(C^{-1})_{12}\hat n_1\hat n_2.
\end{equation}
This interaction term is diagonal in the charge basis, and therefore leads to
the $zz$-interaction after truncation,
\begin{equation}\label{Hzz}
\hat H_{int} = \lambda \sigma_{z1}\sigma_{z2}, \;\;\; \lambda = {e^2\over
2}(C^{-1})_{12}.
\end{equation}
The qubit Hamiltonians are given by Eq. (\ref{HSCB}) with charging energies renormalized by the coupling capacitor.

\subsection{JJ phase coupling of charge qubits}

\begin{figure}[t]
\centerline{\epsfxsize=0.40\textwidth\epsfbox{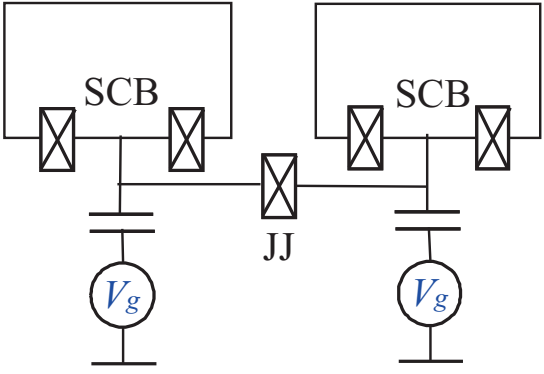}}
\caption{Fixed phase coupling of charge qubits}
\label{SCT_JJfixcoupl}
\end{figure}

Instead of the capacitor, the charge qubits can be connected via a Josephson
junction \cite{Siewert2001a}, as illustrated in Fig. \ref{SCT_JJfixcoupl},

In this case, the Josephson energy of the coupling junction $E_{J3}\cos(\phi_1
-\phi_2)$ must be added to the Lagrangian in addition to the charging energy.
This interaction term is apparently off-diagonal in the charge basis and, after truncation, gives rise to $xx$- and $yy$-couplings,
\begin{equation}
\hat H_{int} = \lambda(\sigma_{x1}\sigma_{x2} + \sigma_{y1}\sigma_{y2}),
\;\;\; \lambda={E_{J3}\over 4},
\end{equation}
or equivalently,
\begin{equation}
\hat H_{int} = 2\lambda(\sigma_{+,1}\sigma_{-,2}+ \sigma_{-,1}\sigma_{+,2}).
\end{equation}

\subsection{Capacitive coupling of single JJs}

\begin{figure}
\centerline{\epsfxsize=0.30\textwidth\epsfbox{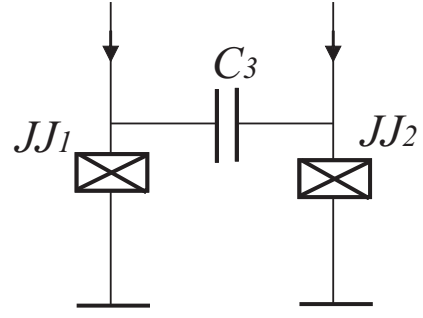}}
\caption{Capacitive coupling of single JJ qubits}
\label{JJ_2q}
\end{figure}

Capacitive coupling of JJ qubits, illustrated in Fig. \ref{JJ_2q} is described in a way similar to the charge qubit, in terms of the Lagrangian Eqs. (\ref{C3}), (\ref{V3}), and the resulting interaction Hamiltonian has the form given in Eq. (\ref{Hnn}).

Generally, in the qubit eigenbasis, $|0\rangle$ and $|1\rangle$, all matrix
elements of the interaction Hamiltonian are non-zero. However, if we adopt a
parabolic approximation for the Josephson potential, then the
diagonal matrix elements turn to zero, $n_{00}=n_{11}=0$, while the off-diagonal matrix elements remain finite, $n_{01}=-n_{10}= -i(E_J/E_C)^{1/4}$.
Then, after truncation, the charge number operator $\hat n$ turns to $\sigma_y$, and the qubit-qubit interaction takes the $yy$-form,
\begin{equation}
\hat H_{int} = \lambda\sigma_{y1}\sigma_{y2},\;\;\;
\lambda=2e^2\sqrt{\hbar^2\omega_{p1}\omega_{p2}\over
E_{C1}E_{C2}}\,(C^{-1})_{12}.
\end{equation}

\subsection{Coupling via oscillators}

Besides the direct coupling schemes described above, several schemes of coupling
qubits via auxiliary oscillators have been considered \cite{MakhlinRMP2001}. Such schemes provide more flexibility, e.g. to control qubit interaction, to couple two remote qubits, and to connect several qubits. Moreover, in many advanced qubits, the qubit variables are generically connected to the outside world via an oscillator (e.g. the Delft and Saclay qubits). To explain the principles of such a coupling, we consider the coupling scheme for charge qubits suggested by Shnirman et al.\cite{Shnirman1997}.

\subsubsection{Coupling of charge (SCB, SCT) qubits}
\label{sectionCouplingOscSCB}
\begin{figure}[t]
\centerline{\epsfxsize=0.45\textwidth\epsfbox{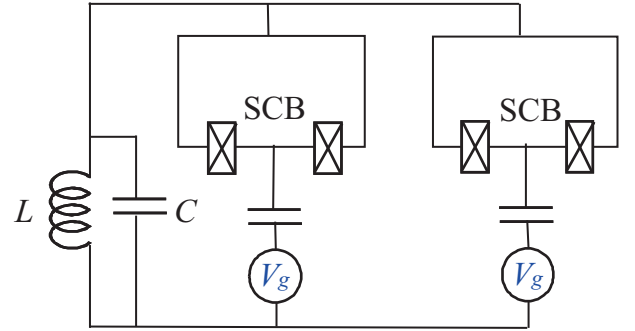}}
\caption{Two charge qubits coupled to a common LC-oscillator.}
\label{figOsccoupling}
\end{figure}

In this circuit the island of each SCB is connected to ground via a common $LC$-oscillator, as illustrated in Fig. \ref{figOsccoupling}. The kinetic
energy (\ref{ChargingSCB}) of a single qubit should now be modified
taking into account the additional phase difference $\phi$ across the oscillator,
\begin{equation}
K(\dot\phi_{-,i};\dot\phi) = {1\over 2} \left({\hbar\over 2e}\right)^2
\left[2C\dot\phi_{-,i}^2 + C_{g}(V_{gi} - \dot\phi -
\dot\phi_{-,i})^2\right].
\end{equation}
The cross term in this equation can be made to vanish by a change of qubit
variable,
\begin{equation}
\phi_{-,i} =  \phi_i - a \phi, \;\;\;a = {C_{g}\over C_\Sigma}.
\end{equation}
The kinetic energy will then split into two independent parts, the kinetic energy of the qubit in Eq. (\ref{Lcpb}), and an additional quadratic term,
\begin{equation}\label{addition}
 {1\over 2} \left({\hbar\over 2e}\right)^2
{CC_g\over C_{\Sigma}}\dot\phi^2 ,
\end{equation}
which should be combined with the kinetic energy of the oscillator, leading to renormalization of  oscillator capacitance.

Expanding the Josephson energy, after the change of variable, gives
\begin{equation}
E_{Ji}\cos(\phi_i - a\phi) \approx E_{Ji}\cos\phi_i -
E_{Ji}a\phi\sin\phi_i \, .
\end{equation}
provided the amplitude of the oscillations of $\phi$ is small. The last term
in this equation describes the linear coupling of the qubit to the
$LC$-oscillator.

Collecting all the terms in the Lagrangian and performing quantization and
truncation procedures, we arrive at the following Hamiltonian of the qubits
coupled to the oscillator (this is similar to Eq. (\ref{HSCT}) for the
SCT),
\begin{equation}
\label{SCB_Osc} \hat H = \sum_{i=1,2} (\hat H_{SCB,i} +
\lambda_i\sigma_{yi}\phi) + \hat H_{osc},
\end{equation}
where $\hat H_{SCB}^{(i)}$ is given by Eq. (\ref{HSCB}), and
\begin{equation}\label{lambda}
\lambda_i= {E_{Ji}C_{g}\over C_\Sigma},
\end{equation}
is the coupling strength, and
\begin{equation}
\hat H_{osc} = E_{Cosc}\hat n^2 + E_L\phi^2/2,
\end{equation}
is the oscillator Hamiltonian where the term in Eq. (\ref{addition}) has
been included.

The physics of the qubit coupling in this scheme is the following: quantum
fluctuation of the charge of one qubit produces a displacement of the oscillator, which perturbs the other qubit. If the plasma frequency of the $LC$ oscillator is much larger than the frequencies of all qubits, then virtual excitation of
the oscillator will produce a direct effective qubit-qubit coupling, the
oscillator staying in the ground state during all qubit operations. To
provide a small amplitude of the zero-point fluctuations, the oscillator plasma
frequency should be small compared to the inductive energy, or $E_{Cosc}\ll
E_L$. Then the fast fluctuations can be averaged out. Noticing that the
displacement does not change the oscillator ground state energy, which then
drops out after the averaging, we finally arrive at the Hamiltonian of the
direct effective qubit coupling,
\begin{equation}
\hat H_{int} = - {\lambda_1\lambda_2\over E_L}\, \sigma_{y1}\sigma_{y2}.
\end{equation}
for the oscillator-coupled charge qubits in Fig. \ref{figOsccoupling}.

\subsubsection{Current coupling of SCT qubits}
\label{sectionCouplingOscSCT}

\begin{figure}[t]
\centerline{\epsfxsize=0.30\textwidth\epsfbox{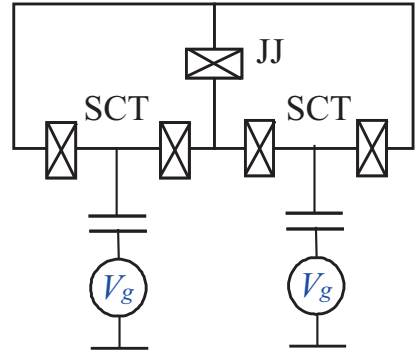}}
\caption{Charge (charge-phase) qubits coupled via a common Josephson junction
providing phase coupling of the two circuits}
\label{SCT_JJcoupl}
\end{figure}

Charge qubits based on SCTs can be coupled by connecting loops of neighboring
qubits by a large Josephson junction in the common link
\cite{YouPRL2002,You2003a,Wang2004,Lantz2004,Wallquist2004,Wallquist2005,WeiLiuNori2004b}, as illustrated in Fig. \ref{SCT_JJcoupl},

The idea is similar to the previous one: to couple qubit variables to a new
variable, the phase of the coupling Josephson junction, then to arrange the
phase regime for the junction with large plasma frequency ($E_{Ccoupl}\ll
E_{Jcoupl}$), and then to average out the additional phase. Technically, the
circuit is described using the SCT Hamiltonian, Eqs. (\ref{HQSCT}),
(\ref{HcptLin}), for each qubit,
\begin{eqnarray}\label{HQSCT1}
\hat H_{SCT} = E_C(\hat n_-  - n_g)^2 + E_C \hat n^2_+  \nonumber\\ -
2E_J\cos\phi_+\cos\phi_- + E_L{(2\phi_+-\phi_e)^2\over 2},
\end{eqnarray}
and adding the Hamiltonian of the coupling junction,
\begin{equation}
\hat H_c = E_{C,c}\hat n_c^2 - E_{J,c}\cos\phi_{c}.
\end{equation}
The phase $\phi_{c}$ across the coupling junction must be added to the flux
quantization condition in each qubit loop; e.g., for the first qubit
$2\phi_{+,1} + \phi_{c}= \phi_{e,1} + \tilde\phi_1$ (for the second qubit the
sign of $\phi_{c}$ will be minus). Assuming small inductive energy, $E_L\ll
E_{J,c}$, we may neglect $\tilde\phi$; then assuming the flux regime for the
coupling Josephson junction we adopt a parabolic approximation for the junction
potential, $E_{J,c}\phi_c^2/2$.

With these approximations, the Hamiltonian of the first qubit plus coupling junction will a take form similar to Eq. (\ref{HQSCT1}) where $E_{J,c}$ will substitute for $E_L$, and $\phi_c$ will substitute for$2\phi_+-\phi_e$.
Finally assuming the amplitude of the $\phi_c$-oscillations to
be small, we proceed as in the previous subsection, i.e. expand the cosine
term obtaining linear coupling between the SCB and the oscillator, truncate
the full Hamiltonian, and average out the oscillator. This will yield the
following interaction term,
\begin{equation}\label{EJcoupling}
\hat H_{int} =  {\lambda_1\lambda_2\over E_{J,c} }
\sigma_{x1}\sigma_{x2},\;\;\;\lambda_{i}= E_J\sin{\phi_{i}\over 2}
\end{equation}

This coupling scheme also applies to flux qubits: in this case, the
coupling will have the same form as in Eq. (\ref{Lcoupling}), but the strength
will be determined by the Josephson energy of the coupling junction, cf. Eq.
(\ref{EJcoupling}), rather than by the mutual inductance.

\subsection{Variable coupling schemes}

Computing with quantum gate networks basically assumes that one-and two-qubit gates can be turned on and off at will. This can be achieved by tuning qubits with fixed, finite coupling in and out of resonance, in NMR-style computing \cite{Rigetti2005}.

Here we shall discuss an alternative way, namely to vary the strength of the physical coupling between nearest-neighbour qubits, as discussed in a number of recent papers \cite{YouPRL2002,You2003a,Lantz2004,Wallquist2004,Wallquist2005,StorczWilhelm2003b,Blais2003,AverinBruder2003,Strauch2003}.

\subsubsection{Variable inductive coupling}

\begin{figure}[t]
\centerline{\epsfxsize=0.45\textwidth\epsfbox{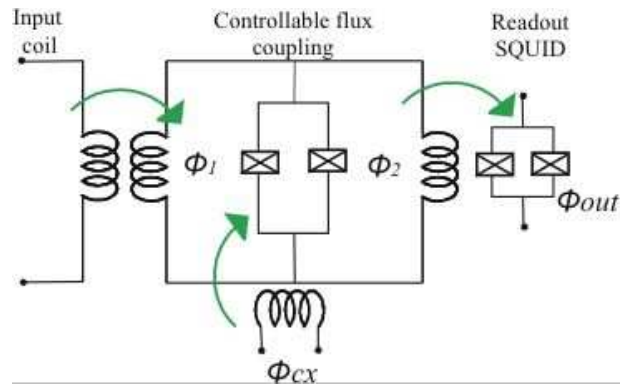}}
\caption{Flux transformer with variable coupling controlled by a SQUID.}
\label{Cosmelli}
\end{figure}

To achieve variable inductive coupling of flux qubits one has to be
able to the control the mutual inductance of the qubit loops. This can be done
by different kinds of controllable switches (SQUIDS, transistors) \cite{StorczWilhelm2003b} in the circuit. In a recent experiment, a variable flux transformer was implemented as a coupling element (see Fig.
\ref{Cosmelli}) by controlling the transforming ratio \cite{Cosmelli2004}.
The flux transformer is a superconducting loop strongly inductively
coupled to the qubit loops, which are distant from each other so that the
direct mutual qubit inductance is negligibly small. Because of the effect of
quantization of magnetic flux in the transformer loop \cite{Tinkham}, the local
variation of magnetic flux $\Phi_1$ induced by one qubit will affect a local
magnetic flux $\Phi_2$ in the vicinity of the other qubit creating effective
qubit-qubit coupling. When a dc SQUID is inserted in the transformer loop, as
shown in Fig. \ref{Cosmelli}, it will shortcircuit the transformer loop, and the
transformer ratio $\Phi_2/\Phi_1$ will change. The effect depends on the
current flowing through the SQUID, and is proportional to the critical
current of the SQUID. The latter is controlled by applying a magnetic flux
$\Phi_{cx}$ to the SQUID loop, as explained in Section \ref{sectionDCSQUID}
and shown in Fig. \ref{Cosmelli}. Quantitatively, the dependence
of the transformer ratio on the controlling flux is given by the equation
\cite{Cosmelli2004},
\begin{equation}
{\Phi_2\over \Phi_1} = \left( 1 + {E_J\over E_L}\cos
{\pi\Phi_{cx}\over \Phi_0}\right)^{-1},
\end{equation}
where $E_J$ is the Josephson energy of the SQUID junction, and $E_L$ is the
inductive energy of the transformer.

\subsubsection{Variable Josephson coupling}

A variable Josephson coupling is obtained when a single Josephson junction is
substituted by a symmetric dc SQUID whose effective Josephson energy
$2E_J\cos(\phi_e/2)$ depends on the magnetic flux threading the SQUID loop
(see the discussion in Section \ref{sectionDCSQUID}). This property is commonly
used to control level spacing in both flux and charge qubits introduced in
Section \ref{sectionCPB}, and it can also be used to switch on and off
qubit-qubit couplings. For example, the coupling of the charge-phase qubits via
Josephson junction in Fig. \ref{SCT_JJcoupl} can be made variable by
substituting the single coupling junction with a dc SQUID \cite{YouPRL2002,You2003a}.

The coupling scheme discussed in Section \ref{sectionCouplingOscSCB} is made
controllable by using a dc SQUID design for the SCB as explained in
Section \ref{sectionCPB}). Indeed, since the coupling strength depends on the
Josephson energy of the qubit junction, Eq. (\ref{lambda}), this solution
provides variable coupling of the qubits. Similarly, the coupling of the SCTs
considered in Section \ref{sectionCouplingOscSCT} can be made controllable by
employing a dc SQUID as a coupling element. A disadvantage of this solution is
that the qubit parameters will vary simultaneously with varying of the
coupling strength. A more general drawback of the dc SQUID-based controllable
coupling is the necessity to apply magnetic field locally, which might be
difficult to achieve without disturbing other elements of the circuit. This is however an experimental question, and what are practical solutions in the long run remains to be seen.

\subsubsection{Variable phase coupling}

\begin{figure}[t]
\centerline{\epsfxsize=0.30\textwidth\epsfbox{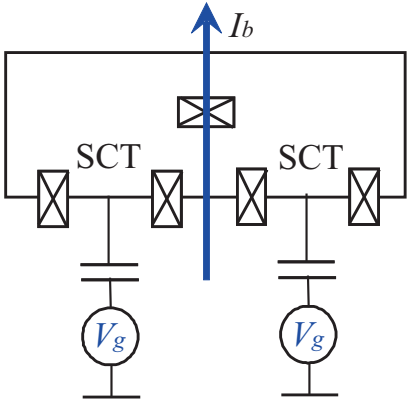}}
\caption{Coupled charge qubits with current-controlled phase coupling: the
arrow indicates the direction of the controlling bias current.}
\label{SCT_JJcoupl2}
\end{figure}

An alternative solution for varying the coupling
is based on the idea of controlling the properties of the
Josephson junction by applying external dc current \cite{Lantz2004,Wallquist2004,Wallquist2005}, as illustrated in
Fig. \ref{SCT_JJcoupl2}.

Let us consider the
coupling scheme of Section \ref{sectionCouplingOscSCT}: the coupling strength
here depends on the plasma frequency of the coupling Josephson junction,
which in turn depends on the form of the local minimum of the junction
potential energy. This form can be changed by tilting the junction potential
by applying external bias current (Fig. \ref{SCT_JJcoupl2}), as discussed in
Section \ref{sectionJJ}. The role of the external phase bias, $\phi_e$, will
now be played by the minimum point $\phi_0$ of the tilted potential
determined by the applied bias current, $E_{J,c}\sin\phi_0 = (\hbar/ 2e)
I_e$. Then the interaction term will read,
\begin{equation}
\hat H_{int} = \lambda \sigma_{x1}\sigma_{x2}, \;\;\; \lambda =
{E_J^2\sin^2(\phi_0/2)\over E_{J,c}\cos\phi_0},
\end{equation}
and local magnetic field biasing is not required.

\subsubsection{Variable capacitive coupling}

\begin{figure}[t]
\centerline{\epsfxsize=0.35\textwidth\epsfbox{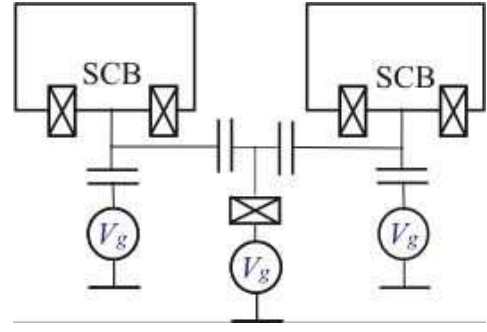}}
\caption{Variable capacitance tuned by a voltage-controlled SCB.}
\label{AverinBruder}
\end{figure}

Variable capacitive coupling of charge qubits based on a quite different physical mechanism of interacting SCB charges has been proposed in Ref.
\cite{AverinBruder2003}. The SCBs are then connected via the circuit
presented in Fig. \ref{AverinBruder}.

The Hamiltonian of this circuit, including the charge qubits, has the form
\begin{equation}
\hat H = \sum _i\hat H_{SCB,i} + E_C(\hat n-q(\hat n_1 + \hat n_2))^2 -
E_J\cos\phi,
\end{equation}
where $E_C$ and $E_J\sim E_c$ are the charging and Josephson energies
of the coupling junction, and $\hat n$ and $\phi$ are the
charge and the phase of the coupling junction. The function $q$ is a linear
function of the qubit charges, $\hat n_1$, and $\hat n_2$, and it
also depends on the gate voltages of the qubits and the coupling
junction. In contrast to the previous scheme, here the coupling
junction is not supposed to be in the phase regime; however, it is
still supposed to be fast, $E_J\gg E_{Ji}$. Then the energy gap in
the spectrum of the coupling junction is much bigger than the qubit
energy, and the junction will stay in the ground state during qubit
operations. Then after truncation, and averaging out the coupling
junction, the Hamiltonian of the circuit will take the form,
\begin{equation}
\hat H = \sum _i\hat H_{SCB,i} + \epsilon_0\left(\sigma_{z1} +
\sigma_{z2}\right),
\end{equation}
where the qubit Hamiltonian is given by Eq. (\ref{HSCB}), and the function
$\epsilon_0$ is the ground state energy of the coupling junction. The
latter can be generally presented as a linear combination of terms proportional to $\sigma_{z1}\sigma_{z2}$ and $\sigma_{z1} + \sigma_{z2}$,
\begin{equation}
\epsilon_0 \left(\sigma_{z1} + \sigma_{z2}\right) = \alpha + \nu \sigma_{z1}
\sigma_{z2} + \beta \,\left(\sigma_{z1} + \sigma_{z2}\right).
\end{equation}
with coefficients depending on the gate potentials. The second term
in this expression gives the $zz$-coupling (in the charge basis), and
the coupling constant $\nu$ may, according to the analysis of
Ref. \cite{AverinBruder2003}, take on both positive and negative
values depending on the coupling junction gate voltage. In
particular it may turn to zero, implying qubit decoupling.

\subsection{Two qubits coupled via a resonator}

In the previous discussion, the coupling oscillator plays a passive
role, being enslaved by the qubit dynamics. However, if the oscillator
is tuned into resonance with a qubit, then the oscillator dynamics
will become essential, leading to qubit-oscillator entanglement.
In this case, the approximation of direct qubit-qubit coupling is not
appropriate; instead, manipulations explicitly involving the
oscillator must be considered.

Let us consider, as an example, operations with two charge qubits
capacitively coupled to the oscillator. Assuming the qubits to be
biased at the degeneracy point and proceeding to the qubit eigenbasis
(phase basis in this case), we write the Hamiltonian on the form (cf.
Eq. (\ref{Hint}),
\begin{equation}
\hat{H} = -\sum_{i=1,2} \left(\frac{\Delta_i}{2}\sigma_{zi}\;-\; \lambda_i
\sigma_x\,\phi_i\right) \; + \; \hat H_{osc}[\phi].
\end{equation}
Let us consider the following manipulation involving the variation of
the oscillator frequency (cf. Ref \cite{Blais2003}): at time $t=0$, the oscillator frequency is off-resonance with both qubits,
\begin{equation}
\hbar\omega(0) \;< \;\Delta_1 \;< \; \Delta_2.
\end{equation}
Then the frequency is rapidly ramped so that the oscillator becomes
resonant with the first qubit,
\begin{equation}
\hbar\omega(t_1) \;= \;\Delta_1 ,
\end{equation}
the frequency remaining constant for a while. Then the frequency is ramped again and brought into resonance with the second qubit,
\begin{equation}
\hbar\omega(t_2) \;= \;\Delta_2.
\end{equation}
Finally, after a certain time it is ramped further so that the oscillator gets
out of resonance with both qubits at the end,
\begin{equation}
\hbar\omega(t>t_3) \;> \;\Delta_2 .
\end{equation}

When passing through the resonance, the oscillator is hybridized with
the corresponding qubit, and after passing the resonance, the
oscillator and qubit have become entangled. For example, let us prepare
our system at $t=0$ in the {\em excited} state $\psi(0) \; =\;
|100\rangle \; = \; |1\rangle |0\rangle |0\rangle $, where the first
number denotes the state of the oscillator (first excited level),
and the last numbers denote the (ground) states of the first and second
qubits, respectively. After the first operation, the oscillator
will be entangled with the first qubit,
\begin{eqnarray}
\psi(t_1<t<t_2) \; = \;\left( \cos\theta_1|10\rangle + \sin\theta_1
e^{i\alpha} |01\rangle \right)\; |0\rangle. \nonumber\\
\end{eqnarray}
After the second manipulation, the state $|100\rangle$ will be
entangled with state  $|001\rangle$,
\begin{eqnarray}
\psi(t>t_3) \; = \; \cos\theta_1 \left( \cos\theta_2 |100\rangle  +
\sin\theta_2 e^{i\beta} |001\rangle \right) \nonumber \\
+ \; \sin\theta_1 e^{i\alpha} |010\rangle. \;\;
\end{eqnarray}
To ensure that there are no more resonances during the described manipulations, it is sufficient to require $\hbar\omega(0) \;> \;\Delta_2 \;- \; \Delta_1$.

If the controlling pulses are chosen so that $\theta_2 = \pi/2$, then
the initial excited state will be eliminated form the final
superposition, and we'll get entangled states of the qubits, while
the oscillator will return to the ground state,
\begin{equation}
\psi(t>t_3) \; = \; |0\rangle \left(\cos\theta_1 e^{i\beta}
|01\rangle ) \; + \; \sin\theta_1 e^{i\alpha} |10\rangle\right).
\end{equation}
The manipulation should not necessarily be step-like, it is
sufficient to pass the resonance rapidly enough to provide the
Landau-Zener transition, i.e. the speed of the frequency ramping
should be comparable to the qubit level splittings.

References to recent work on the entanglement of qubits and oscillators will be given in Section IV C.

\section{Dynamics of multi-qubit systems}

\subsection{General N-qubit formulation}

A general N-qubit Hamiltonian with general qubit-qubit coupling can be written on the form
\begin{eqnarray}
\hat{H} = -\sum_i (\frac{\epsilon_i}{2}\sigma_{zi}
+\frac{\Delta_i}{2}\sigma_{xi}) \;+\; \frac{1}{2} \sum_{i,j;\nu}
\lambda_{\nu,ij}\;\sigma_{\nu i}\sigma_{\nu j} \nonumber \\
\end{eqnarray}
To solve the Schr\"odinger equation
\begin{equation}
\hat{H} \ket{\psi} = E\ket{\psi}
\end{equation}
we expand the N-qubit state in a complete basis, e.g. the $2^N$ basis states of the $(\sigma_z)^N$ operator,
\begin{eqnarray}
\ket{\psi} = \sum a_q\ket{q} \nonumber \\
= a_0\ket{0..00} + a_1\ket{0..01} + a_2\ket{0..10} + ... a_{\tiny{(2^N-1)}}\ket{1..11} \nonumber \\
\end{eqnarray}
and project onto the basis states
\begin{equation}
\hat{H} \sum_p\ket{p}\bra{p}\psi\rangle = E \ket{\psi}
\end{equation}
obtaining the ususal matrix equation
\begin{equation}
\sum_p \bra{q} \hat{H} \ket{p} a_p = E  a_q
\end{equation}
where $a_q = \bra{q}\psi\rangle$, with typical matrix elements given by
\begin{equation}
H_{qp} = \bra{q} \hat{H}  \ket{p} =  \bra{1..01} \hat{H}  \ket{0..11}\\
\end{equation}

\subsection{Two qubits, longitudinal (diagonal) coupling}

The first case is an Ising-type model Hamiltonian, relevant for
capacitively or inductively coupled flux qubits,
\begin{equation}
H = -\sum_{i=1,2} (\frac{\epsilon_i}{2}\sigma_{zi}
+\frac{\Delta_i}{2}\sigma_{xi})\;+\; \lambda_{12} \; \sigma_{z1}\sigma_{z2}
\end{equation}

We expand a general 2-qubit state in the the $\sigma_z$ basis $q=\{kl\}$,

\begin{equation}
\ket{\psi} = a_1\ket{00} + a_2\ket{01} + a_3\ket{10} + a_4\ket{11} \end{equation}

With $q=\{kl\}$ and $p=\{mn\}$, we get

\begin{eqnarray}
H_{qp} = -\bra{kl} \sum_{i=1,2} \frac{\epsilon_i}{2}\sigma_{zi}
+\frac{\Delta_i}{2}\sigma_{xi})\;+\; \lambda_{12} \; \sigma_{z1}\sigma_{z2}
\ket{mn} \nonumber  \\
= -(\frac{\epsilon_1}{2}\bra{k}\sigma_{z1}\ket{m}
+\frac{\Delta_1}{2}\bra{k}\sigma_{x1}\ket{m}) \bra{l}{n}\rangle \nonumber  \\
- \bra{k}{m}\rangle(\frac{\epsilon_2}{2}\bra{l}\sigma_{z2}\ket{n})+
\frac{\Delta_2}{2}\bra{l}\sigma_{x2}\ket{n}) \nonumber  \\
 \;+\; \lambda_{12}\; \bra{k}\sigma_{z1}\ket{m}\bra{l}\sigma_{z2}\ket{n} \nonumber  \\
\end{eqnarray}

The longitudinal zz-coupling only connects basis states with the same indices (diagonal terms), $\ket{kl} \leftrightarrow \ket{kl}$. Evaluation of the matrix elements results in the Hamiltonian matrix (setting $\lambda_{12}=\lambda, \;
\epsilon_1+\epsilon_2 = 2 \epsilon, \; \epsilon_1-\epsilon_2 = 2\Delta\epsilon$)

\begin{equation}
{\hat H}=
\left(\begin{array}{cccc}
\epsilon +\lambda &  -\frac{1}{2}\Delta_2  &  -\frac{1}{2}\Delta_1  &  0 \\
-\frac{1}{2}\Delta_2 & \Delta\epsilon -\lambda & 0 &  -\frac{1}{2}\Delta_1  \\
-\frac{1}{2}\Delta_1 & 0 &  -\Delta\epsilon -\lambda &  -\frac{1}{2}\Delta_2  \\
  0 &  -\frac{1}{2}\Delta_1  &  -\frac{1}{2}\Delta_2  & -\epsilon +\lambda
\end{array}\right)
\label{hamzz}
\end{equation}

The one-body operators can only connect single-particle states, and therefore the rest of the state must be unchanged (unit overlap). Zero overlap matrix elements

\begin{equation}
\bra{k}{m}\rangle = \delta_{k,m}, \;\; \bra{l}{n}\rangle = \delta_{l,n}
\end{equation}
therefore make some matrix elements zero.

The two-body operators describing the qubit-qubit coupling connect two-particle states with up to two different indices (single-particle basis states). In the present case, the qubit coupling Hamiltonian $H_2 = \lambda \; \sigma_{z1}\sigma_{z2}$ has the charge basis as eigenstates and cannot couple different charge states. It therefore only contributes to shifting the energy eigenvalues on the diagonal by $\pm\lambda$. In contrast, qubit coupling Hamiltonians involving $\sigma_{x1}\sigma_{x2}$ or $\sigma_{y1}\sigma_{y2}$ couplings contribute to the off-diagonal matrix elements (in the charge basis), and specifically remove the zero matrix elements above.

In the examples below we will specifically consider the case when $\epsilon_1 -\epsilon_2 = 2\Delta\epsilon = 0$, which can be achieved by tuning the charging energies via fabrication or gate voltage bias.
The eigenvalues of the matrix Schr\"odinger equation with the Hamiltonian given by  (\ref{hamzz}) are then determined by
\begin{eqnarray}
det\;({\hat H}-E) = 0 =\;[(\lambda^2-E^2)+\frac{1}{4}(\Delta_1+\Delta_2)^2]  \nonumber \\
\times \;[(\lambda^2-E^2)+\frac{1}{4}(\Delta_1-\Delta_2)^2]
- \;\epsilon^2\;(\lambda+E)^2 \nonumber \\
\end{eqnarray}
\subsubsection{Biasing far away from the degeneracy point}

Far away from the degeneracy points in the limit $\epsilon \gg \Delta_1, \Delta_2$, the Hamiltonian matrix is diagonal and we are dealing with pure charge states. The secular equation factorises according to
\begin{eqnarray}
det\;({\hat H}-E) = \nonumber \\
= [(\lambda^2-E^2)+\epsilon(\lambda+E)][(\lambda^2-E^2)-\epsilon(\lambda+E)] = 0
\nonumber \\
\end{eqnarray}
giving the eigenvalues
\begin{equation}
E_1 = \epsilon+\lambda \;,\;\; E_{2,3}= -\lambda\;,\;\; E_4 = -\epsilon+\lambda
\end{equation}
The corresponding eigenvectors and 2-qubit states are given by \\

$E_1 = \epsilon+\lambda:$
\begin{equation}
a_2=a_3=a_4 = 0
\end{equation}
\begin{equation}
\ket{\psi_1} = \ket{00}
\end{equation}

$ E_{2,3}= -\lambda:$
\begin{equation}
a_1=a_4 = 0
\end{equation}
\begin{equation}
\ket{\psi_2}=\ket{01}\;,\;\;(a_3=0)\;;\;\; \ket{\psi_3}=\ket{10}\;,\;\;(a_2=0)
\end{equation} \\

$E_4 = -\epsilon+\lambda:$
\begin{equation}
a_1=a_2=a_3 = 0 \\
\end{equation}
\begin{equation}
\ket{\psi_4} = \ket{11}
\end{equation}

\subsubsection{Biasing at the degeneracy point}

Parking both qubits at the degeneracy point, $\epsilon =0$, the secular equation factorises in a different way,
\begin{eqnarray}
det\;({\cal H}-E) = 0 = [(\lambda^2-E^2)+\frac{1}{4}(\Delta_1+\Delta_2)^2]
\nonumber \\
\times [(\lambda^2-E^2)+\frac{1}{4}(\Delta_1-\Delta_2)^2] \nonumber \\
\end{eqnarray}
again giving a set of exact energy eigenvalues
\begin{eqnarray}
E_{1,4}  = \pm \sqrt{\lambda^2+\frac{1}{4}(\Delta_1+\Delta_2)^2}\\
E_{2,3}  = \pm \sqrt{\lambda^2+\frac{1}{4}(\Delta_1-\Delta_2)^2}
\end{eqnarray}
where $E_1 < E_2 < E_3 < E_4$. For $E=E_{1,4}$, the corresponding eigenvectors and 2-qubit states are given by
\begin{equation}
a_1=a_4 \;,\;\;a_2=a_3; \;,\;\;
a_2 = - a_1\frac{\Delta_1+\Delta_2}{\lambda+2E} \\
\end{equation}
\begin{equation}
\ket{E_1}= a_1(\ket{00}+\ket{11}) + a_2(\ket{01}+\ket{10}) \\
\label{E1}
\end{equation}
\begin{equation}
\ket{E_4}= a_2(\ket{00}+\ket{11}) - a_1(\ket{01}+\ket{10})
\label{E4}
\end{equation}
After normalisation
\begin{equation}
a_1 = \frac{1}{2}\sqrt{1+\frac{\lambda}{|2E_1|}}\;;\;
a_2 = \frac{1}{2}\sqrt{1-\frac{\lambda}{|2E_1|}}
\end{equation}

For $E=E_{2,3}$, the corresponding eigenvectors and 2-qubit states are given by
\begin{equation}
b_1=-b_4 \;,\;\;b_2=-b_3; \;,\;\;
b_2 = - b_1\frac{\Delta_1-\Delta_2}{\lambda+2E} \\
\end{equation}
\begin{equation}
\ket{E_2}= b_1(\ket{00}-\ket{11}) + b_2(\ket{01}-\ket{10}) \\
\label{E2}
\end{equation}
\vspace{-0.8cm}
\begin{equation}
\ket{E_3}= b_2(\ket{00}-\ket{11}) - b_1(\ket{01}-\ket{10})
\label{E3}
\end{equation}
After normalisation
\begin{equation}
b_1 = \frac{1}{2} \sqrt{1+\frac{\lambda}{|2E_2|}}\;;\;
b_2 = \frac{1}{2} \sqrt{1-\frac{\lambda}{|2E_2|}}
\end{equation}

\subsubsection{Two-qubit dynamics under dc-pulse excitation}
\label{2q_dynamics}
We now investigate the dynamics of the two-qubit state in the specific case that the dc-pulse is applied equally to both gates, $\epsilon_1(t)=\epsilon_2(t)$, where $\epsilon_1=\epsilon_1(n_{g1})=E_C(1-2n_{g1})$, $\epsilon_2=\epsilon_2(n_{g2})=E_C(1-2n_{g2})$, taking the system from the the pure charge-state region to the co-degeneracy point and back. This is the precise analogy of the single-qubit dc-pulse scheme discussed in
Section \ref{SectIV_C}.

Since the charge state $\ket{00}$ of the starting point does not have time to evolve during the steep rise of the dc-pulse, it remains frozen and forms the initial state $\ket{00}$ at time $t=0$ at the co-degeneracy point, where it can be expanded in the energy eigenbasis,
\begin{eqnarray}
\ket{00} = \ket{\psi(0)} = c_1\ket{E_1} + c_2\ket{E_2} + c_3 \ket{E_3} + c_4 \ket{E_4} \nonumber \\
\end{eqnarray}
To find the coefficients we project onto the charge basis, {kl}=0,1
\begin{eqnarray}
\langle kl\ket{00} = c_1\langle kl\ket{E_1} + c_2\langle kl\ket{E_2} + c_3\langle kl\ket{E_3} + c_4 \langle kl \ket{E_4} \nonumber \\
\end{eqnarray}
and use the explict results for the energy eigenstates to calculate the matrix elements, obtaining

\begin{equation}
\ket{00} = a_1 \ket{E_1} + b_1 \ket{E_2} - b_2 \ket{E_3} - a_2 \ket{E_4}
\end{equation}
For $t > 0$ this stationary state then develops in time governed by the constant Hamiltonian as
\begin{eqnarray}
\ket{\psi(t)} = a_1 e^{-iE_1t}\ket{E_1} + b_1 e^{-iE_2t}\ket{E_2} -  \nonumber \\
- b_2 e^{+iE_2t}\ket{E_3} - a_2 e^{+iE_1t}\ket{E_4}
\end{eqnarray}
Inserting the energy eigenstates from expressions (\ref{E1}),(\ref{E4}),(\ref{E2}),(\ref{E3})  we  obtain the time evolution in the charge basis,
\begin{eqnarray}
\ket{\psi(t)} = \nonumber \\
\ket{00} \left[a_1^2 e^{-iE_1t} +a_2^2 e^{iE_1t} + b_1^2 e^{-iE_2t} +b_2^2 e^{iE_2t} \right] \nonumber \\
+\; \ket{01} \left[a_1 a_2 (e^{-iE_1t}+e^{iE_1t}) + b_1 b_2 (e^{-iE_2t}+e^{iE_2t})
\right] \nonumber \\
+\; \ket{10} \left[a_1 a_2 (e^{-iE_1t}+e^{iE_1t}) - b_1 b_2 (e^{-iE_2t}+e^{iE_2t})
\right] \nonumber \\
+\; \ket{11} \left[a_1^2 e^{-iE_1t} +a_2^2 e^{iE_1t} - b_1^2 e^{-iE_2t} - b_2^2 e^{iE_2t} \right] \nonumber \\
\end{eqnarray}
This state could be the input state for another gate operation, where some other part of the Hamiltonian is suddenly varied. It might also be that this is the state to be measured on. For that purpose one needs in principle the probabilities
\begin{eqnarray}
p_1(t) = |\langle 00\ket{\psi(t)}|^2 \\
p_2(t) = |\langle 01\ket{\psi(t)}|^2 \\
p_3(t) = |\langle 10\ket{\psi(t)}|^2 \\
p_4(t) = |\langle 11\ket{\psi(t)}|^2
\end{eqnarray}
As one example, we will calculate $p_3(t) + p_4(t)$, which represents the probability of finding the qubit $1$ in the upper state $|1\rangle$, independently of the state of qubit $1$ (the states of which are summed over).
\begin{eqnarray}
p_3(t) = |\left[(a_1^2+a_2^2)\cos{E_1t} - (b_1^2+b_2^2)\cos{E_2t}\right] \nonumber \\
- i \left[(a_1^2-a_2^2)\sin{E_1t} - (b_1^2-b_2^2)\sin{E_2t}\right]|^2  \;\;\; \\
p_4(t) = |\left[2 i a_1 a_2 \sin{E_1t} -2 i  b_1 b_2 \sin{E_2t} \right]|^2  \;\;\;
\end{eqnarray}
Adding $p_3(t)$  and $p_4(t)$ we obtain
\begin{eqnarray}
p_3(t)+ p_4(t) = \nonumber \\
= \frac{1}{2} - \frac{1}{4}\cos{E_1t}\cos{E_2t} + \frac{1}{4}\chi \sin{E_1t}\sin{E_2t}  \nonumber \\
= \frac{1}{2} - \frac{1}{4} (1+\chi)\cos{E_+t} + (1-\chi)\cos{E_-t} \;\;\;
\label{beatingosc}
\end{eqnarray}
where $E_+=E_2+E_1, \; E_-=E_2-E_1$, and where $\chi =(1-\frac{\lambda^2}{2|E_1E_2|})$.

\subsection{Two qubits, transverse x-x coupling}

This model is relevant for charge qubits in current-coupled loops. The Hamiltonian is now given by

\begin{equation}
\hat{H} = -\sum_{i=1,2} (\frac{\epsilon_i}{2}\sigma_{zi} +\frac{\Delta_i}{2}\sigma_{xi}\;+\; \lambda \; \sigma_{x1}\sigma_{x2}
\end{equation}

Proceeding as before we have

\begin{eqnarray}
H_{qp} = \bra{kl} \sum_{i=1,2} \frac{\epsilon_i}{2}\sigma_{zi} -\frac{\Delta_i}{2}\sigma_{xi})\;+\; \lambda \; \sigma_{z1}\sigma_{z2} \ket{mn} \nonumber \\
= (\frac{\epsilon_1}{2}\bra{k}\sigma_z^{(1)}\ket{m} -\frac{\Delta_1}{2}\bra{k}\sigma_{x1}\ket{m}) \bra{l}{n}\rangle  \nonumber \\
+ \bra{k}{m}\rangle (\frac{\epsilon_2}{2}\bra{l}\sigma_{z2}\ket{n}) \frac{\Delta_2}{2}\bra{l}\sigma_{x2}\ket{n}) \nonumber \\
+ \; \lambda \; \bra{k}\sigma_{x1}\ket{m}\bra{l}\sigma_{x2}\ket{n} \nonumber \\
\end{eqnarray}
This coupling only connects basis states which differ in both indices ("anti-diagonal" terms), i.e. $\ket{00} \leftrightarrow \ket{11}$ and $\ket{01} \leftrightarrow \ket{10}$, meaning that interaction will only appear on the anti-diagonal. Evaluation of the matrix elements result in the Hamiltonian matrix (again setting $\lambda_{12}=\lambda, \; \epsilon_1+\epsilon_2 = 2 \epsilon, \; \epsilon_1-\epsilon_2 = 2\Delta\epsilon$)
\begin{equation}
{\hat H}=
\left(\begin{array}{cccc}
\epsilon &  -\frac{1}{2}\Delta_2  &  -\frac{1}{2}\Delta_1  & \lambda \\
-\frac{1}{2}\Delta_2  &  \Delta\epsilon_1 & \lambda & -\frac{1}{2}\Delta_1  \\
-\frac{1}{2}\Delta_1  &  \lambda &  -\Delta\epsilon &  -\frac{1}{2}\Delta_2  \\
 \lambda &  -\frac{1}{2}\Delta_1  &  -\frac{1}{2}\Delta_2  & -\epsilon
\end{array}\right),
\end{equation}

\section{Experiments with single qubits and readout devices}

In this section we shall describe a few experiments with single-qubits that
represent the current state-of-the-art and quite likely will be central
components in the developmement of multi-qubit systems during the next five
to ten years. The first experiment presents Rabi oscillations induced and
observed in the elementary phase qubit and readout oscillator formed by a
single JJ-junction
\cite{Martinis2002,Martinis2003,Simmonds2004,Cooper2004,Claudon2004}. The
next example describes a series of recent experiments with a flux qubit
\cite{Chiorescu2003} coupled to different kinds of SQUID oscillator readout
devices \cite{Chiorescu2004,Bertet2004,Lupascu2003}. A further example will
discuss the charge-phase qubit coupled to a JJ-junction oscillator
\cite{Vion2002} and the recent demonstration of extensive NMR-style operation
of this qubit \cite{Collin2004}. The last example will present the case of a
charge qubit (single Cooper pair box, SCB) coupled to a microwave stripline
oscillator \cite{Girvin2003,Blais2004,Wallraff2004,Schuster2004},
representing a solid-state analogue of "cavity QED".

Before describing experiments and results, however, we will discuss in some
detail the measurement procedures that give information about resonance line
profiles, Rabi oscillations, and relaxation and decohrence times. The
illustrations will be chosen from Vion et al. \cite{Vion2002} and related work for the case of
the charge-phase qubit, but the examples are relevant for all types of
qubits, representing fundmental procedures for studying quantum systems.

\subsection {Readout detectors}

Before discussing some of the actual experiments, it is convenient to
describe some of basic readout-detector principles which more or less
the same for the SET, rf-SET, JJ and SQUID devices. A typical pulse scheme
for exciting a qubit and reading out the response is shown in Fig.
\ref{CEA_pulses}:
\begin{figure}
\centerline{\epsfxsize=0.35\textwidth\epsfbox{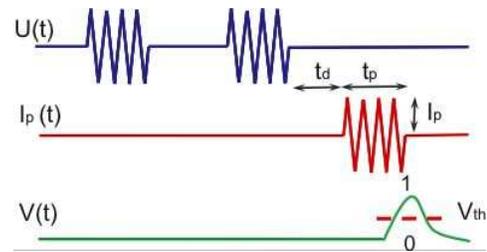}}
\caption{
Control pulse sequences involved in quantum state manipulations and measurement.
Top: microwave voltage pulses $u(t)$ are applied to the control gate for state manipulation. Middle: a readout dc ac pulse (DCP) or ac pulse (ACP) $I_b(t)$ is applied to the threshold detector/discriminator a time $t_d$ after the last microwave pulse.
Bottom: output signal V(t) from the detector. The occurence of a output pulse depends on the occupation probabilities of the energy eigenstates. A discriminator with threshold $V_{th}$ converts V(t) into a boolean 0/1 output for statistical analysis.}
\label{CEA_pulses}
\end{figure}

In Fig. \ref{CEA_pulses} the readout control pulse can be a dc pulse (DCP) or ac pulse (ACP). A DCP readout most often leads to an output voltage pulse, which may be quite destructive for the quantum system. An ACP readout presents a much weaker perturbation by probing the ac-response of an oscillator coupled to the qubit, creating much less back action, at best representing QND readout. \\

\subsubsection {Spectroscopic detection of Rabi oscillation}

In the simplest use of the classical oscillator, it does not discriminate between the two different qubit states, but only between energies of radiation emitted by a lossy resonator coupled to the qubit. In this way it is possible to detect the "low-frequency" Rabi oscillation of a qubit driven by {\em continuous} (i.e. not pulsed) high-frequency radiation tuned in the vicinity of the qubit transition energy. If the oscillator is tunable, the resonance window can be swept past the Rabi line. Alternatively, the Rabi frequency can be tuned and swept past the oscillator window by changing the qubit pumping power \cite{Ilichev2003}.

\subsubsection{Charge qubit energy level occupation from counting electrons: rf-SET}

In this case, the charge qubit is interacting with a beam of electrons
passing through a single-electron transistor (SET) coupled to a charge qubit
(e.g. the rf-SET, \cite{Aassime2001}), as discussed in Section \ref{SectVIII} and illustrated in Fig. \ref{SCB+rfSET}.
In these cases the transmissivity of the electrons will show two distinct values correlated with the two states of the qubit.

\subsubsection{Coupled qubit-classical-oscillator system:
switching detectors with dc-pulse (DCP) output}

In Section \ref{SectVIII} we analyzed the case of an SCT qubit current-coupled
to a JJ-oscillator (Fig. \ref{SCT_JJreadout}) and discussed the Hamiltonian
of the coupled qubit-JJ-oscillator system. The effect of the qubit was to
deform the oscillator potential in different ways depending on the state of
the qubit. The effect can then be probed in a number of ways, by input and
output dc and ac voltage and current pulses,  to determine the occupation of
the qubit energy levels.

Using non-linear oscillators like single JJs or SQUIDS one can achieve threshold and switching behaviour where the JJ/SQUID switches out of the zero-voltage state, resulting in an output dc-voltage pulse.
\begin{figure}
\centerline{\epsfxsize=0.30\textwidth\epsfbox{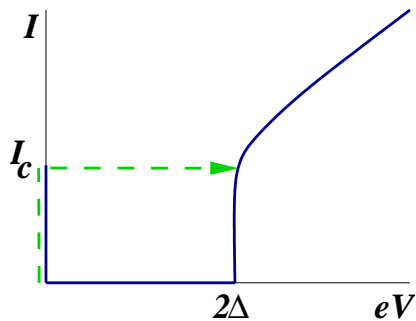}}
\caption{Current-voltage characteristic of tunnel junction (solid line)
consists of the  Josephson branch - vertical line at $V=0$, and the
dissipative branch - curve at $eV \geq 2\Delta$. When the current is
ramped, the junction stays at the Josephson branch and when the current
approaches the critical value $I_c$, the junction switches to the
dissipative branch (dashed line).}
\label{JJ_IVC}
\end{figure}

{\em Switching JJ:} The method is based on the dependence of the critical current of the JJ on the state of the qubit, and consists of applying a short current DCP to the JJ at a value $I_b$ during a time $\Delta t$, so that the JJ will switch out of its zero-voltage state with a probability $P_{sw}(I_b)$. For well-chosen parameters, the detection efficiency can approach unity. The switching probability then directly measures the qubit's energy level population.

\begin{figure}
\centerline{\epsfxsize=0.40\textwidth\epsfbox{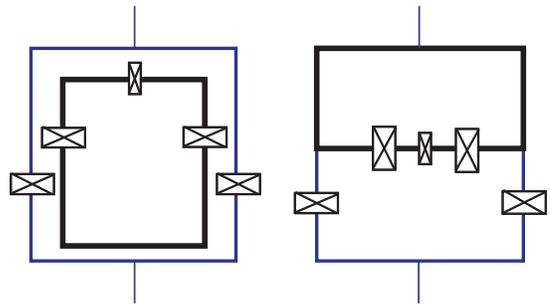}}
\caption{Schematics of readout dc SQUID coupled to flux qubit; left - inductive
coupling, right - direct phase coupling}
\label{Delft_qubit_SQUID}
\end{figure}

{\em Switching SQUID:} In the experiments on flux qubits by the Delft group,
two kinds of physical coupling of the SQUID to the qubit have been implemented, namely inductive coupling (Fig. \ref{Delft_qubit_SQUID} (left))
\cite{vdWal2000,Lupascu2003}
and direct coupling (Fig. \ref{Delft_qubit_SQUID} (right)):
\cite{Chiorescu2003,Chiorescu2004,Bertet2004}
The critical current of the SQUID depends on the flux threading the loop, and therefore is different for different qubit states.
The problem is to detect a two percent variation in the SQUID critical current associated with a transition between the qubit states in a time shorter than the qubit energy relaxation time $T_1$.
The SQUID behaves as an oscillator with a characteristic plasma frequency $\omega_p = [(L+ L _J )C_{sh}]^{-1/2}$. This frequency
depends on the bias current $I_b$ and on the critical current $I_C$ via the
Josephson inductance $L_J =\Phi_0/2\pi I_C \sqrt{1-I^2_b/I^2_c}$ (the shunt capacitor with capacitance $C_{sh}$ and lead inductance $L$ is used to "tune" $\omega_p$). Thus, the plasma frequency takes different values $\omega^{(0)}_p$ or $\omega^{(1)}_p$ depending on the state of qubit, representing two different shapes of the SQUID oscillator potential.

In the dc-pulse-triggered switching SQUID \cite{vdWal2000,Chiorescu2003,Chiorescu2004}, a dc-current readout-pulse is applied after the operation pulse(s) (Fig. \ref{CEA_pulses}), setting a switching threshold for the critical current. The circulating qubit current for one qubit state will then add to the critical current and make the SQUID switch to the voltage state, while the other qubit state will reduce the current and leave the SQUID in the zero-voltage state.

In an application of ac-pulse-triggered switching SQUID \cite{Bertet2004}, readout relies on resonant activation by a microwave pulse at a frequency close to $\omega_p$, adjusting the power so that the SQUID switches to the finite voltage state by resonant activation if the qubit is in state $\ket{0}$, whereas it stays in the zero-voltage state if it is in state $\ket{1}$. The resonant activation scheme is similar to the readout scheme used by Martinis et al. \cite{Martinis2002,Martinis2003,Simmonds2004,Cooper2004}. (see Section {\ref{SectNIST}).

\subsubsection{Coupled qubit-classical-oscillator system: ac-pulse (ACP) non-switching detectors}

This implementation of ACP readout uses the qubit-SQUID combination
\cite{vdWal2000} shown in Fig. \ref{Delft_qubit_SQUID} (left), but with ACP instead of DCP readout, implementing a nondestructive dispersive method for the readout of the flux qubit \cite{Lupascu2003}. The detection is based on the measurement of the Josephson inductance of a dc-SQUID inductively coupled to the qubit. Using this method, Lupascu et al. \cite{Lupascu2003} measured the spectrum of the qubit resonance line and obtained relaxation times around 80 $\mu s$, much longer than observed with DCP.

A related readout scheme was recently implemented by Siddiqi et al. \cite{Siddiqi2003} using two different oscillation states of the non-linear JJ in the zero-voltage state.

\subsection{Operation and measurement procedures}

A number of operation and readout pulses can be applied to a qubit circuit in
order to measure various properties. The number of applied microwave pulses
can vary depending on what quantities are to be measured: resonance line
profile, relaxation time, Rabi oscillation, Ramsey interference or Spin Echo,
as discussed below.

\subsubsection{Resonance line profiles and $T_2$ decoherence times}

To study the resonance line profile, one applies a single long weak microwave
pulse with given frequency, followed by a readout pulse. The procedure is
then repeated for a spectrum of frequencies. The Rabi oscillation amplitude,
the upper state population, and the detector switching probability p(t) will
depend on the detuning and will grow towards resonance. The linewidth gives
directly the total inverse decoherence lifetime $1/T_2 = 1/2T_1 +
1/T_{\phi}$. The decoherence-time contributions from relaxation ($1/T_1$) and
dephasing ($1/T_{\phi}$) can be (approximately) separately measured, as
discussed below.

\begin{figure}
\centerline{\epsfxsize=0.50\textwidth\epsfbox{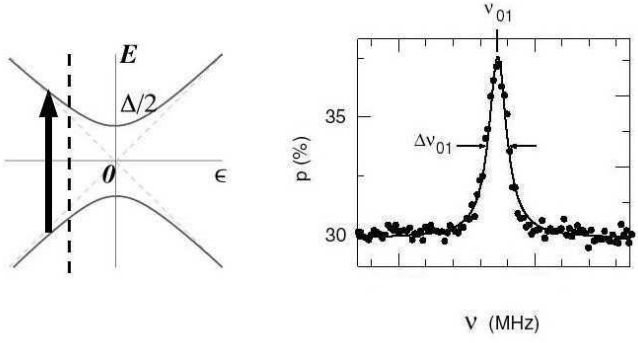}}
\caption{Left: qubit energy level scheme. The vertical dashed line marks the
qubit working point and transition energy. The arrow marks the detuned
microwave excitation. Right: population of the upper level as a function of
the detuning; the inverse of the half width (FWHM) of the resonance line
gives the total decoherence time $T_2$.
 {\em Courtesy of D. Esteve, CEA-Saclay.}
}.
\label{Readout_resonance}
\end{figure}

\subsubsection{$T_1$ relaxation times}

To determine the $T_1$ relaxation time one measures the decacy of the
population of the upper $\ket{1}$ state after a long microwave pulse
saturating the transition, varying the delay time $t_d$ of the detector
readout pulse. The measured $T_1$ =1.8 microseconds is so far the best value
for the Quantronium charge-phase qubit.
\begin{figure}
\centerline{\epsfxsize=0.45\textwidth\epsfbox{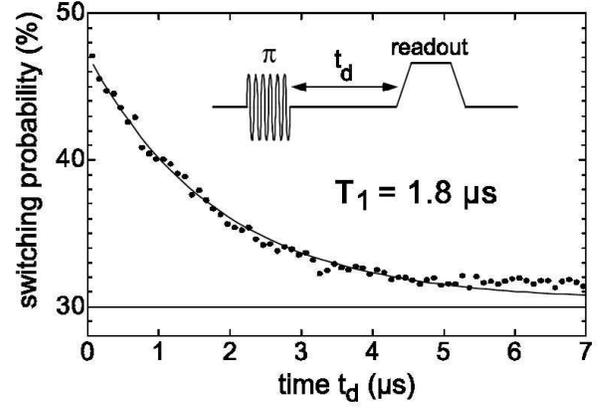}}
\caption{
Decay of the switching probability of the charge-qubit readout junction as a function of the delay time $t_d$ between the excitation and readout pulses. {\em Courtesy of D. Esteve, CEA-Saclay.}
}
\label{CEA_T1}
\end{figure}

\subsubsection{Rabi oscillations and $T_{2,Rabi}$ decoherence time}

To study Rabi oscillations (frequency $\Omega \sim u$, the amplitude of
driving field) one turns on a resonant microwave pulse for a given time
$t_{\mu w}$ and measures the upper $\ket{1}$ state population (probability)
$p_1(t)$ after a given (short) delay time $t_d$. If the systems is perfectly
coherent, the state vector will develop as $\cos \Omega t \;\ket{0} + \sin
\Omega t \;\ket{1}$, and the population of the upper state will then
oscillate as $\sin^2 \Omega t $ between 0 and 1. In the presence of
decoherence, the amplitude of the oscillation of $p_1(t)$ will decay on a
time scale $T_{Rabi}$ towards the average value $p_1(t=\infty) = 0.5$. This
corresponds to incoherent saturation of the 0 to 1 transition.
\begin{figure}
\centerline{\epsfxsize=0.45\textwidth\epsfbox{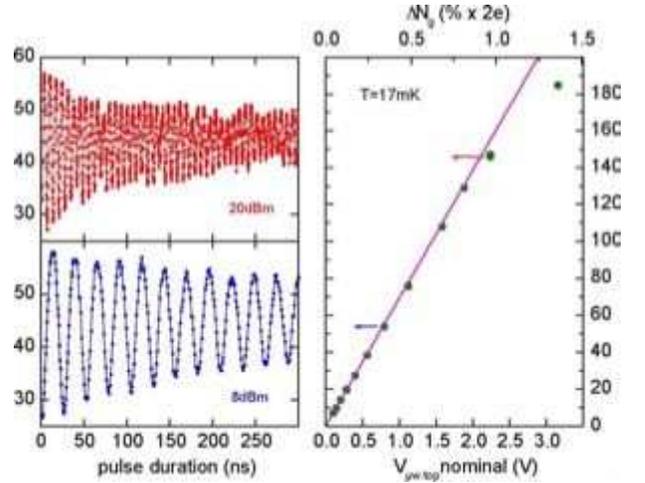}}
\caption{Left: Rabi oscillations of the switching probability
measured just after a resonant microwave pulse of duration.
Right: Measured Rabi frequency (dots) varies linearly with microwave amplitude (voltage) as expected.
{\em Courtesy of D. Esteve, CEA-Saclay.}
}
\label{Rabi}
\end{figure}

\subsubsection{Ramsey interference, dephasing and $T_{2,Ramsey}$ decoherence time}

The Ramsey interference experiment measures the decoherence time of the
non-driven, freely precessing, qubit. In this experiment a $\pi/2$ microwave
pulse around the x-axis induces Rabi oscillation that tips the spin from the
north pole down to the equator. The spin vector rotates in the x-y plane, and
after a given time $\Delta t$, another $\pi/2$  microwave pulse is applied,
immediately followed by a readout pulse.

\begin{figure}
\centerline{\epsfxsize=0.45\textwidth\epsfbox{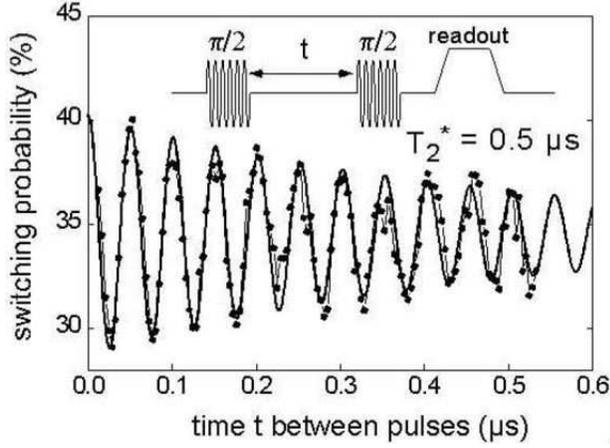}}
\caption{
Ramsey fringes of the switching probability after two phase-coherent microwave $\pi/2$ pulses separated by the time delay $t$. The continuous line represents a fit by exponentially damped cosine function with time constant $T_2^* = T_{\phi} = 0.5 \mu s$. The oscillation period coincides with the inverse of the detuning frequency $\delta$ (here $\delta=\nu -\nu_{01} = 20.6$ MHz).
{\em Courtesy of D. Esteve, CEA-Saclay.}
}
\label{Ramsey}
\end{figure}
Since the $\pi/2$ pulses are detuned by $\delta$ from the qubit $\ket{0} \rightarrow \ket{1}$
transition frequency, the qubit will precess with frequency $\delta$ relative to the
rotating frame of the driving field. Since the second microwave pulse will be applied
in the plane of the rotating frame, it will have a projection $\cos \delta t$ on the
qubit vector and will drive the qubit towards the north or south poles, resulting in a
specific time-independent final superposition state
$\cos \delta t \;\ket{0} + \sin \delta t \;\ket{1}$ of the qubit at the end of the
last $\pi/2$ pulse. The readout pulse then catches the qubit in this superposition state
and forces it to decay if the qubit is in the upper $\ket{1}$ state. The probability will
oscillate with the detuning frequency, and a single-shot experiment will then detect the
upper state with this probability. Repeating the experiment many times for different pi/2
pulse separation $\Delta t$ will then give $\ket{0}$ or $\ket{1}$ with probabilities
$\cos^2 \delta t$ and $\sin^2 \delta t$. Taking the average, and then varying the
pulse separation, will trace out the Ramsey interference oscillatory signal. Dephasing
will make the signa decay on the timescale $T_{\phi}$. \\

\subsubsection{Spin-echo}

The spin-echo and Ramsey pulse sequences differ in that a $\pi$-pulse around the x-axis is added in between the two $\pi/2$-pulses in the spin-echo experiment, as shown in Fig.\ref{Echo}. As in the Ramsey experiment, the first $\pi/2$-pulse makes the Bloch vector start rotating in the equatorial x-y plane with frequency $E/\hbar = \nu_{01}$. The effect of the $\pi$-pulse is now to flip the entire x-y plane with the rotating Bloch vector around the x-axis, reflecting the Bloch vector in the x-z plane. The Bloch vector then continues to rotate in the x-y plane in the same direction. Finally a second $\pi/2$-pulse is applied to project the state on the z-axis.
\begin{figure}
\centerline{\epsfxsize=0.50\textwidth\epsfbox{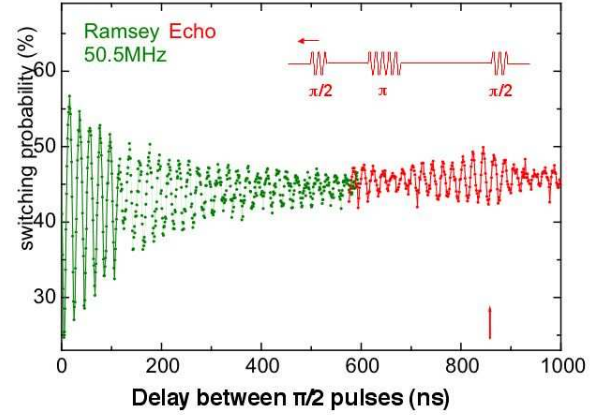}}
\caption{
Spin-echo experiment. The left part shows the basic Ramsey oscillation. The right part shows the echo signal appearing in the time window around twice the time delay between the first $\pi/2$-pulse and the $\pi$-pulse.
{\em Courtesy of D. Esteve, CEA-Saclay.}
}
\label{Echo}
\end{figure}

If two Bloch vectors with slightly different frequency start rotating at the same time in the x-y plane, they will move with different angular speeds. The effect of the $\pi$-pulse at time $\Delta t$ will be to permute the Bloch vectors, and then let the motion continue in the same direction. This is similar to reversing the motion and letting the Bloch vectors back-trace. The net result is that the two Bloch vectors re-align after time $2\Delta t$.

In NMR experiments, the different Bloch vectors correspond to different spins in the ensemble. In the case of a single qubit, the implication is that in a series of repeated experiments, the result will be insensitive to small variations $\delta E$ of the qubit energy {\em between} measurements, as long as the energy (rotation frequency) is constant {\em during} one and the same measurement. If fluctuations occur during one measurement, then this cannot be corrected for. The spin-echo procedure can therefore remove the measurement-related line-broadening associated with slow fluctuations of the qubit precession, and allow observation of the intrinsic coherence time of the qubit.

\subsection{NIST Current-biased Josephson Junction Qubit}
\label{SectNIST}

Several experimental groups have realized the Josephson Junction (JJ) qubit \cite{Yu2002,Martinis2002,Martinis2003,Simmonds2004,Cooper2004,Claudon2004}.
Here we describe the experiment performed at
NIST\cite{Simmonds2004,Cooper2004}. In this experiment the junction
parameters and bias current were chosen such that a small number of well defined
levels were formed in the potential well (Fig. \ref{fig8JJQ2}), with the
interlevel frequencies, $\omega_{01}/2\pi = 6.9$GHz and $\omega_{12}/2\pi 6.28$GHz, the quality factor was Q=380. The experiment was performed at very
low temperature, T= 25 mK.

The qubit was driven from the ground state, $|0\rangle$, to the upper state, $|1\rangle$, by the resonance rf pulse with frequency $\omega_{01}$, and then the occupation of the upper qubit level was measured. The measurement was performed by exciting qubit further from the upper level to the auxiliary level with higher escape rate by applying the second resonance pulse with frequency $\omega_{12}$. During the whole operation, across the junction only oscillating voltage develops with zero average value over the period. When the tunneling event occurred, the junction switched to the dissipative regime, and finite dc voltage appeared across the junction, which was detected.
Alternatively, post-measurement classical states "0" and "1" differ in flux by $\Phi_0$, which is readily measured by a readout SQUID.

The relaxation rate was measured in the standard way by applying a Rabi pumping pulse followed by a measuring pulse with a certain delay. Non-exponential relaxation was observed, first rapid, $\sim 20$ nsec, and then more slow, $ \sim 300$ nsec.
By reducing the length of the pumping pulse down to 25 nsec,
i.e. below the relaxation time, Rabi oscillations were observed.
In this experiment the amplitude of the pumping pulse rather than duration was varied, which affected the Rabi frequency and allowed the observation of oscillation of the level population for fixed duration of the pumping pulse.

The Grenoble group \cite{Claudon2004} has observed coherent oscillations in a multi-level quantum system, formed by a current-biased dc SQUID. These oscillations have been induced by applying resonant microwave pulses of flux. Quantum measurement is performed by a nanosecond flux pulse that projects the final state onto one of two different voltage states of the dc SQUID, which can be read out. The number of quantum states involved in the coherent oscillations increases with increasing microwave power. The dependence of the oscillation frequency on microwave power deviates strongly from the linear regime expected for a two-level system and can be very well explained by a theoretical model taking into account the anharmonicity of the multi-level system.

\subsection{Flux qubits}

\subsubsection{Delft 3-junction persistent current qubit with dc-pulse (DCP) readout}

The original design of the 3-junction qubit (Fig. \ref{3J_vdWal} was published by Mooij et al. in 1999 \cite{Mooij1999}, and the first spectroscopic measurements by van der Wal et al. in 2000 \cite{vdWal2000} (and simultaneously by Friedman et al. \cite{Friedman2000} for a single-junction rf-SQUID). Recently the Delft group has also investigated designs  where the 3-junction qubit is sharing a loop with the measurement SQUID to increase the coupling strength \cite{Chiorescu2003}, as shown in Fig. \ref{Chiorescu_circuit}.
\begin{figure}
\centerline{\epsfxsize=0.45\textwidth\epsfbox{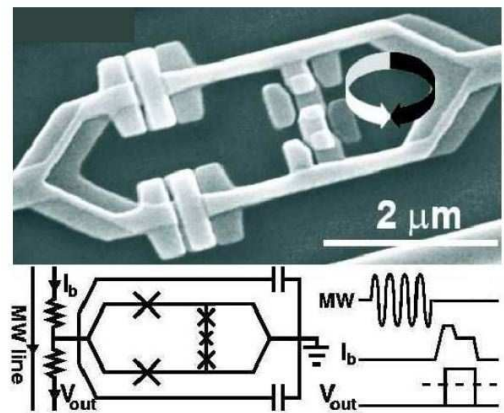}}
\caption{
Upper panel: Scanning electron micrograph of a small-loop flux-qubit with three Josephson junctions of critical current $\sim 0.5$ mA, and an attached large-loop SQUID with two big Josephson junctions of critical current $\sim 2.2$ mA. Arrows indicate the two directions of the persistent current in the qubit.
Lower panels: Schematic of the on-chip circuit; crosses represent the Josephson junctions. The SQUID is shunted by two capacitors to reduce the SQUID plasma frequency and biased through a small resistor to avoid parasitic resonances in the leads. Symmetry of the circuit is introduced to suppress excitation of the SQUID from the qubit-control pulses. The MW line provides microwave current bursts inducing oscillating magnetic fields in the qubit loop. The current line provides the measuring pulse $I_b$ and the voltage line allows the readout of the switching pulse $V{out}$.
{\em Adapted from Chiorescu et al. \cite{Chiorescu2003}}
}
\label{Chiorescu_circuit}
\end{figure}

To observe and study Rabi oscillations, the qubit was biased at the degeneracy point and the qubit $\ket{0} \rightarrow \ket{1}$ transition excited by a pulse of 5.71 GHz microwave radiation of variable length, followed by a bias-current ($I_b$) readout pulse applied to the SQUID (Fig. \ref{Chiorescu_circuit}, lower left panel). The first (high) part of the readout pulse (about 10 ns) has two functions: It displaces the qubit away from the degeneracy point so that the qubit eigenstates carry finite current, and it tilts the SQUID potential so that the SQUID can escape to the voltage state if the qubit is in its upper state. The purpose of the long lower plateau of the the readout pulse is to prevent the SQUID from returning ("retrapping") to the zero-voltage state.
With these operation and readout techniques, Rabi (driven) oscillations, Ramsey (free) oscillations and spin-echos of the qubit were observed\cite{Chiorescu2003}, giving a Ramsey free oscillation dephasing time of  20 $ns$  and spin echo dephasing time of 30 $ns$.

As mentioned above, relaxation times around 80 $\mu s$ have been measured with ACP readout \cite{Lupascu2003}, demonstrating that the properties of readout devices are critically important for observing intrinsic qubit decoherence times. Recent further improvements have resulted in Ramsey decoherence times $T_{2,Ramsey}$ of 200 $ns$, Rabi (driven) decoherence times $T_{2,Rabi}$ of 5 $\mu s$, and relaxation times $T_1$ of more than 100 $\mu s$.

\subsection {Charge-phase qubit}

\subsubsection {General considerations}

As described in Section VII, the charge-phase qubit circuit consists of a
single-Cooper-pair transistor (SCT) in a superconducting loop,
\begin{figure}[t]
\centerline{\epsfysize=0.35\textwidth \epsfbox{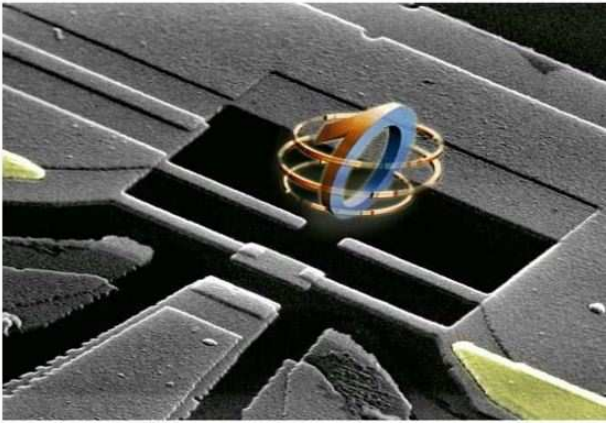}}
\caption{
AFM picture of the charge-phase qubit ("quantronium") corresponding to the circuit scheme in Fig. \ref{SCT_JJreadout}. The working point is controlled by two external "knobs", a voltage gate controlling the induced charge ($n_g$) on the SCT island, and a current gate controlling the total phase across the SCT via the external flux ($\phi_e$) threading the loop. The large readout JJ is seen in the left part of the figure.
{\em Courtesy of D. Esteve, CEA-Saclay.}
}
\label{Saclay_charge-phase}
\end{figure}
The Hamiltonian for the SCT part of this circuit is given by
\begin{equation}\label{HQOSC}
\hat H_{SCT} = -{1\over 2}\left[\;\epsilon(n_g)\;\sigma_z + \Delta(\phi_e)\;\sigma_x\;\right]
\end{equation}
where the charging and tunneling parameters are themselves functions of
external control parameters, gate charge $ n_g$ and loop flux $\phi_e$,
$\epsilon(n_g) = E_C(1-2n_g)$ and $\Delta(\phi_e) = 2E_J\cos(\phi_e/2)$. The
qubit energy levels
\begin{equation}
E_{1,2} = \mp {1\over2}\sqrt{\epsilon(n_g)^2 + \Delta(\phi_e)^2}
\end{equation}
then form a 2-dimensional landscape as functions of the gate charge $n_g(V)$
and phase $\phi_e(\Phi)$, which are functions of the gate voltage $V = C_g
2en_g$ and gate magnetic flux $\Phi = (\hbar/2e) \phi_e$.
\begin{figure}[t]
\centerline{\epsfysize=0.40\textwidth \epsfbox{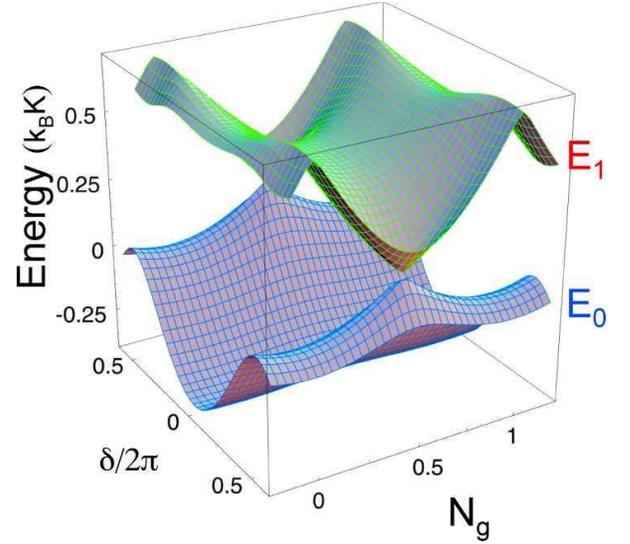}}
\caption{
Charge-phase qubit energy surface. ($N_g=n_g$; $\delta/2\pi=\phi_e$.)
{\em Courtesy of D. Esteve, CEA-Saclay.}
}
\label{CEA_qubit_energysurfaces}
\end{figure}
The energy level surfaces are therefore functions of two parameters, gate
voltage and flux, giving us two independent knobs for controlling the working
point of the (charge-phase) qubit. Expanding the energy in Taylor series
around some bias working point $(V_b,\Phi_b)$, one obtains
\begin{eqnarray}
\delta E(V,\Phi) =  E (V,\Phi)- E (V_b,\Phi_b) \nonumber \\
= \frac{\delta E}{\delta V}\delta V + \frac{\delta E}{\delta \Phi}\delta \Phi
+ {1\over2}\frac{{\delta}^2 E}{{\delta V}^2} {\delta V}^2
+ {1\over2}\frac{{\delta}^2 E}{{\delta \Phi}^2} {\delta \Phi}^2 \;\;\;
\end{eqnarray}
The derivatives are response functions for charge, current, capacitance and inductance (omitting the cross term) ($i=1,2$),
\begin{eqnarray}
\delta E_i(V,\Phi) = Q_i \;\delta V + I_i \;\delta \Phi + C_i \;{\delta V}^2 + L_i \; {\delta \Phi}^2 \;\;\;
\end{eqnarray}
or, equivalently,
\begin{eqnarray}
\delta E_i (n_g,\phi_e)
= \; \tilde Q_i \;\delta n_g + \tilde I_i \;\delta \phi_e + \tilde C_i \;{\delta n_g}^2 + \tilde L_i \;{\delta \phi_e}^2 \nonumber \\
\end{eqnarray}

On the energy level surfaces (Fig. \ref{CEA_qubit_energysurfaces}), the special point $(n_g,\phi_e) = (0.5,0)$ is an extreme point with zero first derivative. This means that the energies of the $\ket{0}$ and $\ket{1}$ states will be invariant to first order to small variations of charge and phase, which will minimize the qubit sensitivity to fluctuations of the working point caused by noise.

The point $(n_g,\phi_e) = (0.5,0)$ is often referred to as the "degeneracy" point because the charging energy is zero, $\epsilon(n_g) = E_C(1-2n_g) = 0$.
The level splitting at this point is determined by the Josephson tunneling interaction $E_J$, and is a function of the external bias $\phi_e$) $\Delta(\phi_e) = 2E_J\cos(\phi_e/2)$.
\begin{figure}[t]
\centerline{\epsfysize=0.40\textwidth \epsfbox{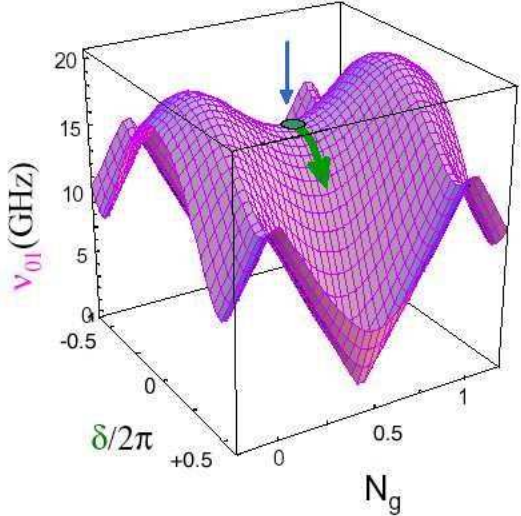}}
\caption{
The charge-phase qubit frequency surface. ($N_g=n_g$; $\delta/2\pi=\phi_e$.)
{\em Courtesy of D. Esteve, CEA-Saclay.}
}
\label{Saclay_ny01}
\end{figure}

Figure \ref{Saclay_ny01} shows the frequency surface $\nu_{01}={\Delta E}/h$, $\Delta E = E_2 -E_1$.
During operation, the qubit is preferably parked at the degeneracy point $(n_g,\phi_e) = (0.5,0)$ where $Q = I = 0$, in order to minimize the decohering influence of noise. In order to induce a qubit response, for gate operation or readout, one must therefore either (i) move the bias point away from point $(0.5,0)$ to have finite first-order response with $Q \ne 0$ or $I \ne 0$, or (ii) stay at $(0.5,0)$ and apply a perturbing ac-field that makes the second-order response significant.

\subsubsection {The CEA-Saclay "quantronium" charge-phase qubit}

The "quantronium" charge-phase circuit is given by Fig. \ref{Saclay_charge-phase} adding a large readout Josephson junction in the loop (cf. Fig. \ref{SCT_JJreadout}).
The minimum linewidth (Fig. \ref{Readout_resonance}) corresponds to a Q-value of 20000 and a decoherence time $T_2$ = 0.5 $\mu$s. With the measured relaxation time $T_1$=1.8 $\mu$s (Fig. \ref{CEA_T1}), the dephasing time can be estimated to $T_\phi$=0.8 $\mu$s.

A set of results for Rabi oscillations of the Saclay quantronium qubit is shown in Fig. \ref{Rabi}.
The decoherence time represents decoherence under driving conditions. Another, more fundamental, measure of decoherence is the free precession dephasing time $T_2$ when the qubit is left to itself. This is measured in in the Ramsey two-pulse experiment, as already described above in Fig. \ref{Ramsey} showing the CEA-Saclay data \cite{Vion2002,Vion2003}, giving $T_2$ = 0.5 $\mu$s. The spin-echo results in Ref. \cite{Vion2003} gave a lifetime as long as $T_2$ = 1$\mu$s.

In a recent systematic experimental and theoretical investigation of a specific charge-phase device, investigating the effects of relaxation and dephasing on Rabi oscillation, Ramsey fringes and spin-echos \cite{Ithier2005}, one obtains the following picture of different coherence times for the quantronium charge-phase qubit:
\begin{figure}
\centerline{\epsfxsize=0.45\textwidth\epsfbox{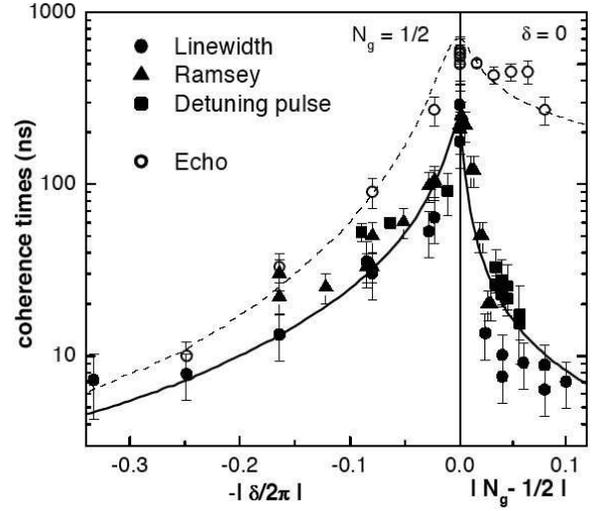}}
\caption{
Free-evolution decoherence times for the quantronium charge-phase SCT qubit. \cite{Ithier2005} ($N_g=n_g$; $\delta/2\pi=\phi_e$.) Full and dashed curves represent results of theoretical modeling.
{\em Courtesy of D. Esteve, CEA-Saclay.}
}
\label{CEA_coherence_times}
\end{figure}

Fluctuations of charge $\delta n_g$ and flux $\delta\phi_e$ will shift energy levels and make the qubit transition energy $\sqrt{\epsilon(n_g+\delta n_g)^2 + \Delta(\phi_e+\delta\phi_e)^2}$ fluctuate.
However, the charge degeneracy point is a saddle point, which means that at that working point the qubit transition frequency is insensitive to low-frequency noise to first order, allowing long coherence times.
As can be seen in Fig. \ref{CEA_coherence_times},
the coherence is sharply peaked ound the "magic" degeneracy point. The spin-echo experiment indicates the presence of slow charge fluctuations from perturbing impurity two-level systems (TLS) (1/f noise) and that long coherence time can be recovered by spin-echo techniques until the decoherence becomes too rapid at significant distances from the magic point along the charge axis. In contrast, in the experiments with this device the decoherence due to phase fluctuations could not be significantly compensated for by spin-echo techniques, indicating that the phase noise (current and flux noise) in this device is high-frequency noise.

\section{Experiments with qubits coupled to quantum oscillators}
\label{SectXIII}

\subsection {General discussion}

The present development of quantum information processing with Josephson Junctions (JJ-QIP) goes in the direction of coupling qubits with quantum oscillators, for operation, readout and memory. In Section \ref{SectVIII} we discussed the SCT, i.e. a single Cooper-pair transistor in a superconducting loop \cite{Zorin2002}, providing one typical form of the Hamiltonian for a qubit-oscillator coupled system. We also showed with perturbation theory how the Hamiltonian gave rise to qubit-dependent deformed oscillator potential and shifted oscillator frequency. In \ref{SectVIII}, in addition we discussed the SCT charge-coupled to an LC-oscillator, which is also representative for a flux qubit coupled to a SQUID oscillator, and which describes a charge qubit in a microwave cavity.

To connect to the language of quantum optics and cavity QED and the current work on solid-state applications, we now explicitly introduce quantization of the oscillator. Quantizing the oscillator, $\tilde\phi \sim (a^+ + a)$, a representative form of the qubit-oscillator Hamiltonian reads:
\begin{eqnarray}
\hat H = \hat H_q + \hat H_{int}+ \hat H_{osc} \\
\hat H_q = -{1\over 2}E\;\sigma_z \\
\hat H_{int} = g \;\sigma_x \;(a^+ + a) \\
\hat H_{osc} = \hbar \omega \;(a^+a + \frac{1}{2})
\end{eqnarray}
Introducing the step operators $\sigma_{\pm} = \sigma_x \pm i\; \sigma_y$, the interaction term can be written as
\begin{equation}
\hat H_{int} = g \; (\sigma_+ \;a + \sigma_- \;a^+) + g \; (\sigma_+ \;a^+ + \sigma_- \;a)
\end{equation}
The first term
\begin{equation}
\hat H_{int} = g  \; (\sigma_+ \;a + \sigma_- \;a^+)
\end{equation}
represents the resonant (co-rotating) part of the interaction, while the second term represents the non-resonant counter-rotating part. In the rotating-wave approximation (RWA) one only keeps the first term, which gives the Jaynes-Cummings model \cite{JaynesCummings1963,Stenholm1973,ShoreKnight1993,GerryKnight2004}.
Diagonalizing the Jaynes-Cummings Hamiltonian to second order by a unitary transformation gives
\begin{equation}
H = -{1\over 2}(E + {g^2\over\delta})\;\sigma_z +
(\hbar\omega + {g^2\over\delta}\;\sigma_z)\;a^+a
\label{JC2ndorder}
\end{equation}
where $\delta= E - \hbar\omega$ is the so called detuning.
The result illustrates what we have already discussed in detail, namely that (i) the qubit transition energy E is shifted (renormalized) by the coupling to the oscillator, and (ii) the oscillator energy $\hbar\omega$ is shifted by the qubit in different directions depending on the state of the qubit, which allows discriminating the two qubit states.

Figure \ref{Qubit-osc_levelstructure} shows the basic level structure of the qubit-oscillator system in the cases of (a) weak and (b) strong coupling.
\begin{figure}[t]
\centerline{\epsfysize=0.23\textwidth \epsfbox{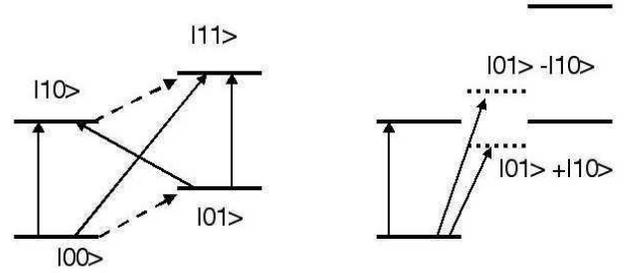}}
\caption{Qubit-oscillator level structure. The notation for the states is: $\ket{qubit; oscillator}= \ket{0/1; n=0,1,..}$; (a) $E \approx 2\hbar\omega$: very large "detuning", weak coupling. (b) $E \approx \hbar\omega$: resonance, strong coupling and hybridization, level ("vacuum Rabi") splitting.}
\label{Qubit-osc_levelstructure}
\end{figure}

What is weak or strong coupling is determined by the strength of the level hybridization, which in the end depends on qubit-oscillator detuning and level degeneracies. The perturbative Hamiltonian in Eq. (\ref{JC2ndorder}) is valid for large detuning ($g \ll \delta_{qr}$) and allows us to approach the case of strong coupling between the qubit and the oscillator. Close to resonance the degenerate states have to be treated non-perturbatively.

Figure \ref{Qubit-osc_levelstructure}(a) corresponds to  the weak-coupling (non-resonant) case where the levels of the two subsystems are only weakly perturbed by the coupling, adding red and blue sideband transitions $\ket{01} \rightarrow \ket{10}$ and $\ket{00} \rightarrow \ket{11}$ to the main zero-photon transition $\ket{00}\rightarrow \ket{10}$.

Figure \ref{Qubit-osc_levelstructure}(b) corresponds to the resonant strong-coupling case when the $|01>$ and $|10>$ states are degenerate, in which case the qubit-oscillator coupling produces two "bonding-antibonding" states in the usual way, and the main line becomes split into two lines ("vacuum Rabi splitting".
Of major importance are the linewidths of the qubit and the oscillator relative to the splittings caused by the interaction. To discriminate between the two qubit states, the oscillator shift must be larger than the average level width.

Since the qubit and the oscillator form a coherent multi-level system, as described before (Section \ref{2q_dynamics}), the time evolution can be written as  $c_1(t) \ket{00} + c_2(t) \ket{01} + c_3(t) \ket{10}+ c_4(t) \ket{11}$, which in general does not reduce to a product of qubit and oscillator states, and therefore represents (time-dependent) entanglement. This provides opportunities for implementing quantum gate operation involving qubits and oscillators.

Generation and control of entangled states can be achieved by using microwave pulses to induce Rabi oscillation between specific transitions of the coupled qubit-oscillator system, and the result can be studied by spectroscopy on suitable transitions or by time-resolved detection of suitable Rabi oscillations.

\subsection{Delft persistent current flux qubit coupled to a quantum oscillator}

The Delft experiment \cite{Chiorescu2004} demonstrates entanglement between a superconducting flux qubit and a SQUID quantum oscillator. The SQUID provides the measurement system for detecting the quantum states (threshold switching detector, Fig. \ref{Chiorescu_circuit}). It is also presents an effective inductance that, in parallel with an external shunt capacitance, acts as a low-frequency harmonic oscillator; the qubit and oscillator frequencies are approximately $h\nu_q = h\nu_{01} \approx 6 \;$ GHz and $h\nu_r \approx 3\;$ GHz, corresponding to
Fig. \ref{Qubit-osc_levelstructure}(a).

In the Delft experiment \cite{Chiorescu2004}, performing spectroscopy on the coupled qubit-oscillator multi-level system reveals the variation of the main and sideband transitions with flux bias $\Phi_{ext}$, as shown in
Fig. \ref{Delft_qubit_osc_levels}.
\begin{figure}
\centerline{\epsfxsize=0.45\textwidth\epsfbox{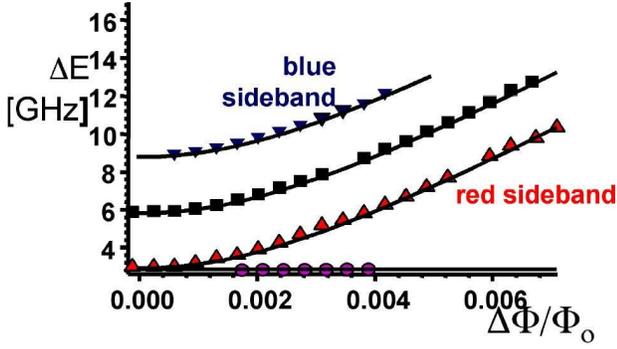}}
\caption{Resonant frequencies indicated by peaks in the SQUID switching probability when a long (300 ns) microwave pulse excites (saturates) the system before the readout pulse. Data are represented as a function of the external flux $\Phi_{ext}$ through the qubit area away from the qubit symmetry point.
The blue $\ket{00} \rightarrow \ket{11}$ and red $\ket{01} \rightarrow \ket{10}$ sidebands are shown by down- and up-triangles, respectively; continuous lines are obtained by adding 2.96 GHz and -2.90 GHz, respectively, to the central continuous line (numerical fit). These values are close to the oscillator resonance frequency $\nu_p$ at 2.91 GHz (solid circles).
{\em Courtesy of J.E. Mooij, TU Delft.}
}
\label{Delft_qubit_osc_levels}
\end{figure}
The presence of visible sideband transitions demonstrates the qubit-oscillator interaction and (weak) level hybridization).

Short microwave pulses can now be used to induce Rabi oscillations between the various qubit-oscillator transitions. In the Delft experiment \cite{Chiorescu2004},
microwave pulses with frequency $\nu_q = \nu_{01} \approx \Delta/h \approx$ 5.9 GHz (qubit symmetry point)  were first used to induce Rabi oscillations with $\sim 25$ ns decay time at  the main qubit transition $\ket{00} \rightarrow \ket{10}$ (and $\ket{01} \rightarrow \ket{11}$) for different values of the  microwave power. The Rabi frequency as function of the amplitude of the microwave voltage demonstrated qubit-oscillator hybridization (avoided crossings) at the oscillator $\nu_p$ ($\nu_r$) and Larmor $\Delta/h$ (!) frequencies.

The dynamics of the coupled qubit-oscillator system can be studied by inducing microwave-driven Rabi oscillation between the blue $\ket{00} \rightarrow \ket{11}$ and red $\ket{01} \rightarrow \ket{10}$ sidebands.
In particular one can study the conditional dynamics in microwave multi-pulse experiments: by coherently (de)populating selected levels via proper timing of Rabi oscillations induced by a first pulse, a second pulse can induce Rabi oscillations on another transition connected to the levels controlled by the first pulse. Alternatively, such Rabi oscillations can instead be blocked by the first excitation.

\begin{figure}
\centerline{\epsfxsize=0.45\textwidth\epsfbox{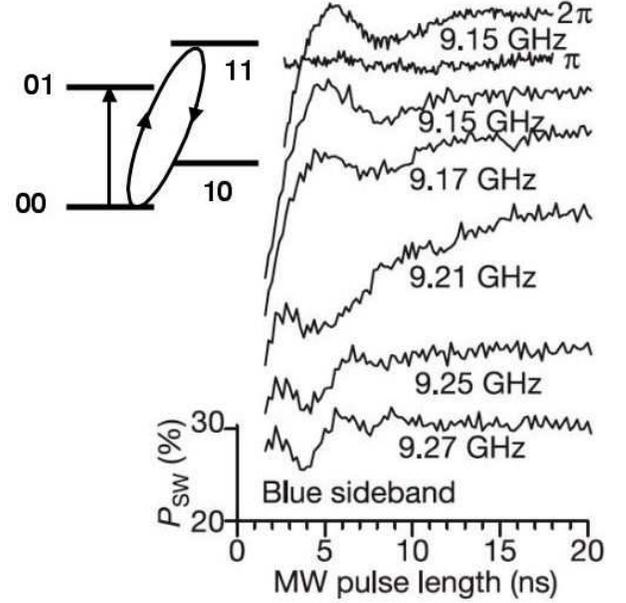}} \caption{
Generation and control of entangled states via $\pi$ and  $2\pi$ Rabi pulses
on the qubit transition $\ket{00} \rightarrow \ket{10}$, followed by Rabi
driving of blue sideband transitions $\ket{00} \rightarrow$. A $\pi$ pulse
excites the system from $\ket{00}$ to $\ket{10}$, which suppresses the blue
sideband transitions $\ket{00} \rightarrow \ket{11}$ (second curve from the
top). On the other hand, with a $2\pi$ pulse the system returns to
$\ket{00}$, which allows Rabi oscillation on the blue sideband. {\em Adapted
from Chiorescu et al. \cite{Chiorescu2004}} } \label{Chiorescu_0011}
\end{figure}
\begin{figure}
\centerline{\epsfxsize=0.50\textwidth\epsfbox{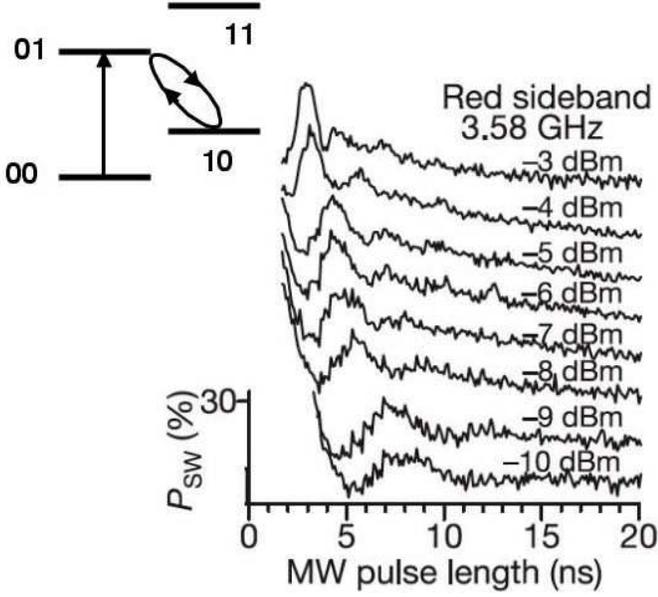}} \caption{
Generation and control of entangled states via $\pi$ and  $2\pi$ Rabi pulses
on the qubit transition $\ket{00} \rightarrow \ket{10}$, followed by Rabi
driving of red sideband transitions $\ket{10} \rightarrow \ket{01}$. A $\pi$
pulse excites the system from $\ket{00}$ to $\ket{10}$, which In the right
panel, a $\pi$ pulse excites the system from $\ket{00}$ to $\ket{10}$, which
populates the $\ket{10}$ state and allows Rabi oscillation on the red
sideband transitions $\ket{01} \rightarrow \ket{10}$. {\em Adapted from
Chiorescu et al. \cite{Chiorescu2004}} } \label{Chiorescu_0110}
\end{figure}
This is shown experimentally in Fig. \ref{Chiorescu_osc33}: by preparing the initial state with initial $\pi$ and $2\pi$ pulses, the sideband Rabi oscillations could be turned off and on again. The corresponding Rabi oscillations are shown in the left part of Fig. \ref{Chiorescu_osc33}(b), demonstrating rapid decay due to the strong damping of the SQUID oscillator (lifetime $\sim$ 3 ns; Q=100-150).

In the previous example, the control pulse (first pulse) was applied to the main $\ket{00} \rightarrow \ket{10}$ transition, controlling the populations of the  "0" and "1" levels. In the next example, microwave control pulse is applied to the $\ket{00} \rightarrow \ket{01}$ transition, inducing Rabi oscillations which populate the first excited state of oscillator. A second microwave pulse (in principle, with different frequency) can then induce Rabi oscillations on the $\ket{01} \rightarrow \ket{10}$ transition (red-sideband). The experimental result \cite{Chiorescu2004} is shown Fig. \ref{Chiorescu_osc4},
\begin{figure}
\centerline{\epsfxsize=0.40\textwidth\epsfbox{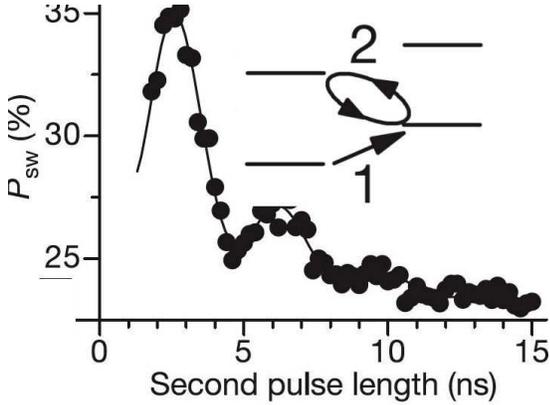}}
\caption{
Generation and control of entangled states and study of decay and lifetimes. Here a Rabi $\pi$  pulse on the oscillator transition $\ket{00} \rightarrow \ket{01}$ populates the state $\ket{01}$, which then allows Rabi oscillation on the red sideband transition $\ket{01} \rightarrow \ket{10}$.
The decay of both the Rabi oscillation and the average probability gives evidence for short oscillator life time ($\sim$ 3 ns; Q=100-150).
{\em Adapted from Chiorescu et al. \cite{Chiorescu2004}}
}
\label{Chiorescu_osc4}
\end{figure}
Clearly, for sufficiently long qubit and oscillator lifetimes, one can prepare entangled Bell states, $\frac{1}{2}(\ket{00} \pm \ket{01})$ and $\frac{1}{2}(\ket{01} \pm \ket{10})$  by applying $\pi/2$ microwave pulses to the $\ket{00} \rightarrow \ket{11}$ and $\ket{01} \rightarrow \ket{10}$ transitions.

\subsection{Yale charge-phase qubit coupled to a strip-line resonator}

In the Yale experiments \cite{Wallraff2004,Schuster2004} a coherent qubit-quantum oscillator system is realized in the form of a Cooper pair box capacitively coupled to  a superconducting stripline resonator (Fig. \ref{Yale_SCB+osc2}) forming one section of a microwave transmission line. The stripline resonator, a finite length (24 mm) of planar wave guide, is a solid-state analogue of the cavity electromagnetic resonator used in quantum optics to study strong atom-photon interaction and entanglement. In the Yale experiments, the qubit is placed in the transverse field in the gap between the resonator strip lines, i.e. inside the microwave cavity. The $E_C$ and $E_J$ parameters are such that the SCT is effectively in the charge-phase region, and it is operated by controlling both the charge and phase (flux) ports.
\begin{figure}
\centerline{\epsfxsize=0.50\textwidth\epsfbox{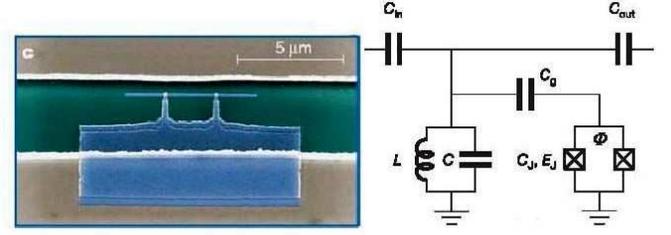}}
\caption{Yale SCB charge-phase qubit coupled to an oscillator in the form of a superconducting microwave stripline resonator.
{\em Adapted from Wallraff et al. \cite{Wallraff2004}}
}
\label{Yale_SCB+osc2}
\end{figure}

The large effective electric dipole moment $d$ of Cooper pair box and the large vacuum electric field $E_0$ in the transmission line cavity lead to a large vacuum Rabi frequency $\nu_{Rabi} = 2dE_0/h$, which allows reaching the strong coupling limit of cavity QED in the circuit.

In the Yale experiment, spectroscopic measurements are performed by driving the qubit with resonant microwave pulses and simultaneously detecting the frequency and intensity-dependent amplitude and phase of probe pulses of microwave radiation sent through a transmission line coupled to the stripline resonator via input/out capacitors ({Fig. \ref{Yale_SCB+osc2}). The oscillator frequency is fixed at $h\nu_r \approx 6\;$ GHz. The qubit level separation, on the other hand, is tunable over a wide frequency range around 6 GHz in two independent ways: (i) by varying the magnetic flux $\phi_e$ through the loop, forming the phase (flux) gate \ref{sectionCPB}, or (ii) by varying the charge $n_g$ via the voltage gate, primarily around the charge degeneracy point, $n_g$=1/2, to minimize the effect of charge fluctuations.

The flux bias is used to tune the qubit transition frequency at $n_g$=1/2 to values larger or smaller than the resonator frequency. The tuning the qubit frequency with the charge gate will provide two distinct cases: the qubit always detuned from the oscillator, and the qubit being degenerate with the resonator at certain values of $n_g$.

A central result \cite{Wallraff2004} is the evidence for the qubit-oscillator hybridization and splitting of the degenerate $\ket{01}$ and $\ket{10}$ states, $\nu_{01} \rightarrow \nu_{\pm}$, as illustrated in Fig. \ref{Qubit-osc_levelstructure} (right). This splitting is often called "vacuum Rabi" splitting because the ocillator is in its ground (vacuum) state.

The $\ket{01}$, $\ket{10}$ level splitting is observed through microwave excitation of the $0 \rightarrow 1$ resonance transition and observing the splitting of the resonance line as the qubit and the oscillator are tuned into resonance (by tuning the qubit frequency to zero qubit-oscillator detuning, $\delta = \nu_q - \nu_r$ = 0).

Another central result \cite{Schuster2004} is the evidence for a long qubit dephasing time of $T_2 >$ 200 ns under optimal conditions: qubit parked at the charge degeneracy point, and weak driving field (low photon occupation number in the resonator cavity). In the experiment \cite{Wallraff2004} the qubit-oscillator detuning is large, the qubit resonance transition $\nu_q =\nu_{01}$ is scanned by microwave radiation, and the dispersive shift of the resonator frequency $\nu_r$ seen by a microwave probe beam is used to determine the occupation of the qubit levels. Scanning the qubit $\ket{00} \rightarrow \ket{10}$ resonance line profile allowed to determine the lineshape and linewith as a function of microwave power, giving the best value of  $T_2 >$ 200 ns. Moreover, observation of the postion of the resonance as a function of microwave power allowed the determination of the ac-Stark shift, i.e. the Rabi frequency as a function of the photon occupation (electric field) of the cavity. The measurement induces an ac-Stark shift of 0.6 MHz per photon in the qubit level separation. Fluctuations in the photon number (shot noise) induce level fluctuations in the qubit leading to dephasing which is the characteristic back-action of the measurement. A cross-over from Lorentzian to Gaussian resonance line shape with increasing measurement power is observed and explained by dependence of the resonance linewidth on the cavity occupation number,  exceeding the linewidth due to intrinsic decoherence at high rf-power.

\subsection{Comparison of the Delft and Yale approaches}

In comparison, the Delft experiment \cite{Chiorescu2004} corresponds to
the case of very large detuning ($\nu_q$ = 6 GHz, $\nu_r$ = 3 GHz; $\delta
= \nu_q - \nu_r \sim \nu_r$; Fig. \ref{Qubit-osc_levelstructure} (left)),
so that one will observe a main resonance line $\ket{00} \rightarrow
\ket{10}$ and two sidebands, $\ket{00} \rightarrow \ket{11}$ (blue) and
$\ket{01} \rightarrow \ket{10}$ (red). Decreasing the detuning $\delta$,
the blue and red sidebands will move away to higher resp. lower
frequencies, and one will arrive at the case of zero detuning and
qubit-oscillator degeneracy. If the qubit-oscillator coupling is larger
than the average linewidth, one will then observe a splitting of the main
line (Fig. \ref{Qubit-osc_levelstructure} (right)).

\section{Experimental manipulation of coupled two-qubit systems}

\subsection{Capacitively coupled charge qubits}

An AFM picture of the NEC SCB 2-qubit system \cite{Yamamoto2003} is shown in Fig. \ref{NEC_2qubit} and the corresponding circuit JJ circuit in Fig. \ref{figCcoupling} (note that in the NEC circuit the left SCT is replaced by a simple SCB).

\begin{figure}[t]
\centerline{\epsfxsize=0.60\textwidth\epsfbox{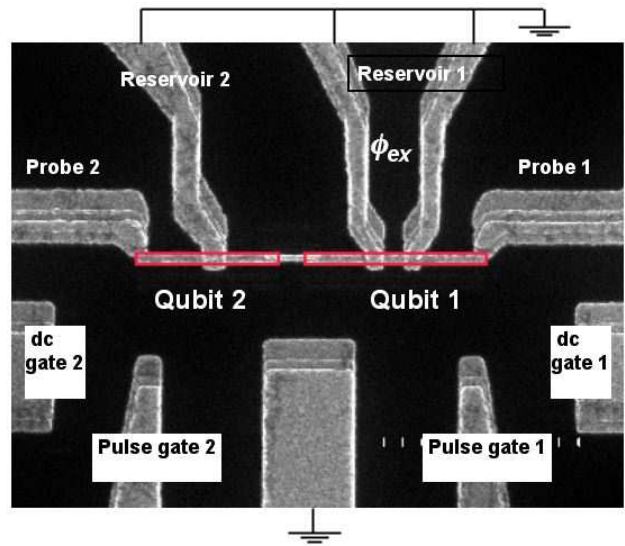}}
\caption{ The NEC 2-qubit system: two capacitively coupled charge qubits. The left qubit is a single Cooper pair box (SCB) and the right qubit is an SCT with flux-tunable Josephson energy.
{\em Courtesy of J.S. Tsai, NEC, Tsukuba, Japan.}
}
\label{NEC_2qubit}
\end{figure}

Two coupled qubits constitute a 4-level system. The Hamiltonian for the NEC system of two capacitively coupled charge qubits (SCBs) was analyzed in the Appendix. The four energy eigenvalues $E_{1,2,3,4}(n_{g1},n_{g2})$ are plotted in Fig. \ref{NEC_energy} as a functions of the gate charges. At each point in the parameter space, an arbitrary two-qubit state can be written as a superposition of the four two-qubit energy eigenstates.
\begin{figure}[t]
\centerline{\epsfxsize=0.48\textwidth\epsfbox{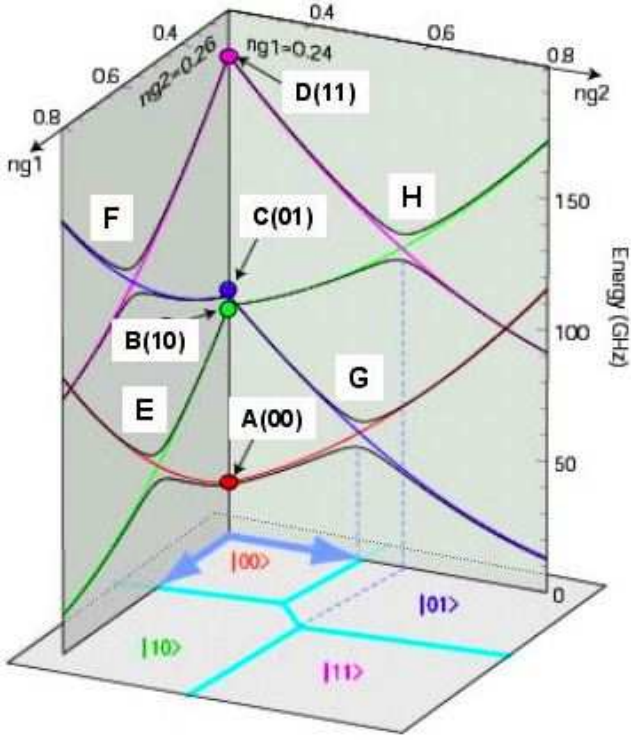}}
\caption{The NEC 2-qubit system: Energy-level structure as a function of the gate charges $n_{g1}$ and $n_{g2}$, independently controlled by the gate voltages $V_{g1}$ and $V_{g2}$.
{\em Courtesy of J.S. Tsai, NEC, Tsukuba, Japan.}
}
\label{NEC_energy}
\end{figure}

As described in Sections IV and X, the general procedure for operating with dc-pulses is to initialize the system to its ground state $\ket{00}$ at the chosen starting point $(n_{g10}, n_{g20})$ and then suddenly change the Hamiltonian  at time $t=0$ to the gate bias $(n_{g1}, n_{g2})$. If the change is strictly sudden, then the state at $(n_{g1}, n_{g2})$ at time $t=0$ is $\ket{\psi(t=0)} = \ket{00}$, which can be expanded in the energy basis of the Hamiltonian,
\begin{equation}
\ket{\psi(0)} = \ket{00} = c_1\ket{E_1} + c_2\ket{E_2} + c_3 \ket{E_3} + c_4 \ket{E_4} \nonumber \\
\end{equation}
This stationary state then develops in time governed by the constant Hamiltonian as
\begin{eqnarray}
\ket{\psi(t)} = c_1 e^{-iE_1t}\ket{E_1} + c_2 e^{-iE_2t}\ket{E_2} + \nonumber \\
+ c_3 e^{-iE_3t}\ket{E_3} + c_4 e^{-iE_4t}\ket{E_4}
\end{eqnarray}
If one re-expands this state in the charge basis of the starting point A (Fig. \ref{NEC_energy}), then one obtains a system with periodic coefficients ${a_i(t)}$,
\begin{equation}
\ket{\psi(t)} = a_1(t)\ket{00} + a_2(t)\ket{01} + a_3(t) \ket{10} + a_4(t) \ket{11}   \nonumber \\
\end{equation}
developing  in time through all of the charge states.

To perform a two-qubit conditional gate operation, one performs a series of pulses moving the system around in parameter space. Specifically, the NEC scheme is to apply sequential dc-pulses to the charge pulse gates of each of the two qubits in Fig. \ref{NEC_2qubit}. Two cases have been studied so far:

(1) $V_{g1}(t)= V_{g2}(t)$}:
This is the case with common control dc-pulses for studied by Pashkin et al. \cite{Pashkin2003}.
Since $\epsilon_1(t)=\epsilon_2(t)$, the plane of operation is at 45 degrees to the axes in Fig. \ref{NEC_energy}. As a result, Pashkin et al. \cite{Pashkin2003} observed interference effects and beating oscillations between the two qubits, as described by Eq. ({\ref{beatingosc}).

(2) $V_{g1}(t)=0; V_{g2}(t)$, $V_{g1}(t); V_{g2}(t)=0$:
This is the case with separate sequential dc pulses on separate gates recently studied by Yamamoto et al. \cite{Yamamoto2003}.
The scheme is illustrated in Fig. \ref{NEC_energy}: Starting at point A, first the system is pulsed in the $n_{g1}$ direction, putting the system in a superposition of the states at points A and B;  then the system is pulsed in the $n_{g2}$ direction, allowing conditional gate operation due to the different time evolution of the states departing from A or B.

Specifically, an entangling gate of controlled-NOT (CNOT) type was demonstrated by Yamamoto et al. \cite{Yamamoto2003} using the pulse sequences shown in Fig. \ref{NEC_2qubit_CNOT}:
\begin{figure}[t]
\centerline{\epsfxsize=0.50\textwidth\epsfbox{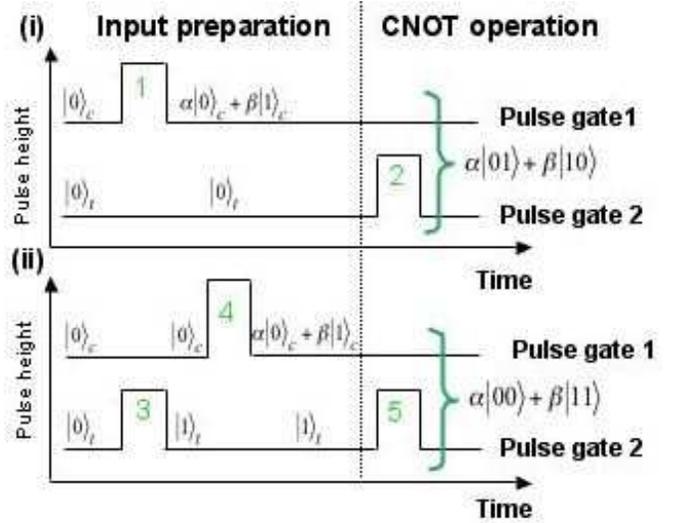}}
\caption{The NEC 2-qubit system: pulse sequences for a CNOT gate (actually, the gate is a $NOT-CNOT$ gate).
{\em Courtesy of J.S. Tsai, NEC, Tsukuba, Japan.}
}
\label{NEC_2qubit_CNOT}
\end{figure}
First one applies a dc-pulse to gate $1$ of the control qubit, moving out (down left) left from $\ket{00}$ (point A in Fig. \ref{NEC_energy}) in the $n_{g1}$ direction to the single-qubit degeneracy point (point E), staying for a certain time,  and then moving back (turning off the pulse), putting the system in a superposition $\alpha\ket{00} + \beta\ket{10}$ (points A and B).
Next one applies a dc-pulse to gate $2$ of the target qubit, moving out (right) in the $n_{g2}$ direction and back. The pulse is designed to reach the first degeneracy point (point G), allowing the state to develop into a superposition of 00 (point A) and 01 (point C) if the control state was 00. If instead the control state was 10 (point B), the dc-pulse on gate 2 will not reach the two-qubit degeneracy point (point H) and the development will be roughly adiabatic, taking the state back to 10 (point B), never reaching 11 (point D).

With timing such ($\pi/2$-pulse) that $00\rightarrow01$ (A $\rightarrow$ G $\rightarrow$ C), the control gate leads to 00+10, and the target gate only modifies the first component ($00 \rightarrow 01$), resulting in 01+10, i.e. one of the Bell states. This is shown in the top panel of Fig. \ref{NEC_2qubit_CNOT}.

If instead one first applies a $\pi/2$-pulse in the $n_{g2}$ direction to the target qubit, inducing $00\rightarrow01$ (C), and then applies a $\pi/2$-pulse in the $n_{g1}$ direction to this state, reaching 01+11, and then again applies a $\pi$/2-pulse to the target qubit $2$, resulting 00+11, one obtains the other Bell state. This is shown in the bottom panel of Fig. \ref{NEC_2qubit_CNOT}.

\subsection{Inductively coupled flux qubits}

Figure \ref{2majer} shows a circuit with two inductively coupled flux qubits forming a four-level quantum system, excited by a single microwave line and surrounded by a single measurement SQUID \cite{MajerPhD2002,Majer2003}.
\begin{figure}[t]
\centerline{\epsfxsize=0.40\textwidth\epsfbox{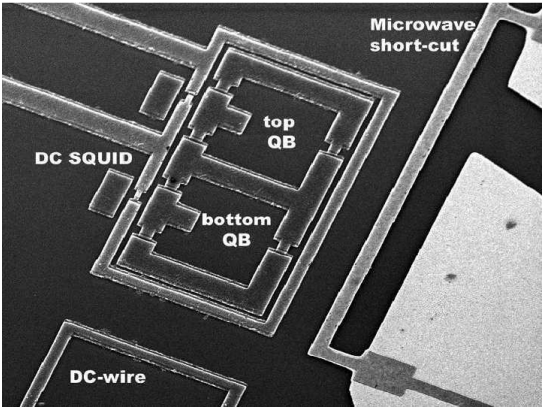}}
\caption{
Scanning microscope image of two inductively coupled qubits surrounded by a DC-SQUID.
{\em Courtesy of J. Majer, TU Delft.}
}
\label{2majer}
\end{figure}
With this circuit Majer et al. \cite{MajerPhD2002,Majer2003} have spectroscopically mapped large portions of the level structure and determined the device parameters entering in the Hamiltonian matrix Eq. (\ref{hamzz}), finding good agreement with the design parameters. Majer et al. found clear manifestations of qubit-qubit interaction and hybridization in the level structure as a function of bias flux. Presently, ter Haar et al. \cite{terHaar2005} are investigating a more strongly coupled system with a JJ in the common leg (cf. Fig. \ref{SCT_JJcoupl}), inducing Rabi oscillations and performing conditional spectroscopy along the lines described in Section XII in connection with the coupled qubit-oscillator system.

A similar device with two flux qubits inside a coupling/measurement SQUID has recently been investigated by et al. \cite{Plourde2004a,Plourde2004b}, so far demonstrating Rabi oscillation of individual qubits. Finally, Izmalkov et al. \cite{Izmalkov2004} have spectroscopically demonstrated effects of qubit-qubit interaction and hybridization for two inductively coupled flux qubits inside a low-frequency tank circuit.

\subsection{Two capacitively coupled JJ qubits}

Capacitive coupling of two JJ qubits (Section IX E, Fig. \ref{JJ_2q}) has recently been investigated by several groups \cite{Berkley2003,McDermott2005}, showing indirect \cite{Berkley2003} and direct \cite{McDermott2005} evidence for qubit entanglement.
Figure \ref{McDermott_2q} shows the circuit used by McDermott et al. \cite{McDermott2005},
\begin{figure}[t]
\centerline{\epsfxsize=0.35\textwidth\epsfbox{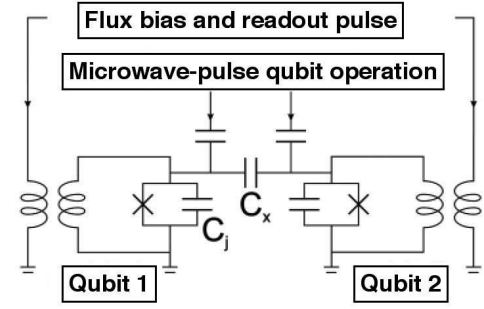}}
\caption{
Circuit scheme for two capacitively coupled current (flux) biased JJ qubits with rf and dc control/readout lines.
{\em Adapted from McDermott et al. \cite{McDermott2005}}
}
\label{McDermott_2q}
\end{figure}
The potential-well and level structure of each JJ qubit under operation and measurement conditions are shown in Fig. \ref{McDermott_1qpot},
\begin{figure}
\centerline{\epsfxsize=0.40\textwidth\epsfbox{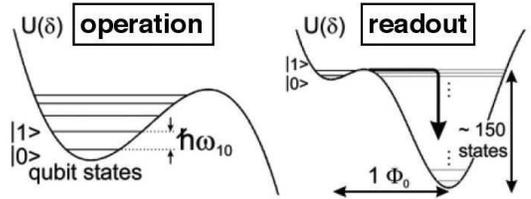}}
\caption{
Potential energies and quantized energy levels the Josephson phase qubit: left, during qubit operation; right, during state measurement, in which case the qubit well is much shallower and state $\ket{1}$ rapidly tunnels to the right hand well.
{\em Adapted from McDermott et al.\cite{McDermott2005}}
}
\label{McDermott_1qpot}
\end{figure}

The 2-qubit circuit behaves as two dipole-coupled pseudo-2-level systems, illustrated in Fig.\ref{Qubit-osc_levelstructure}. With "identical" qubits, the $\ket{01}$ and $\ket{10}$ states are degenerate and become hybridized and split by the interaction. A microwave $\pi$-pulse tuned to the $\ket{0}\rightarrow \ket{1}$ transition of one of the isolated qubits will "suddenly" induce a $\ket{00}\rightarrow \ket{10}$ transition, populating the $\ket{E_+}=\ket{10}+\ket{01}$ and $\ket{E_-}=\ket{10}-\ket{01}$ states with equal weights. This will leave the system oscillating between the two qubits,  between the $\ket{10}$ and $\ket{01}$ states,
$\ket{\psi(t)} = \ket{E_+} + e^{-i\delta E t}\ket{E_-}
= \cos(\delta E t/2)\ket{10} + \sin(\delta E t/2) \ket{01}$
where $\delta E = E_--E_+$.
This means that the two-qubit system oscillates between non-entangled and entangled states.
With one ideal detector for each qubit we can measure the state of each qubit with any prescribed time delay between measurements. With simultaneous measurements we can determine all the probabilities $p_{ij}=|\langle ij\ket{\psi(t)}|^2$, ideally giving
$p_{10} = \frac{1}{2} (1+\cos(\delta E t))$,
$p_{01} = \frac{1}{2} (1-\cos(\delta E t))$, and $p_{00}=p_{11}=0$ in the absence of relaxation, decoherence and perturbations caused by the detectors.

Figure \ref{McDermott_2q_prob} shows the actual experimental results of Ref. \cite{McDermott2005} for the probabilities $p_{ij}(t)$:
\begin{figure}[t]
\centerline{\epsfxsize=0.48\textwidth\epsfbox{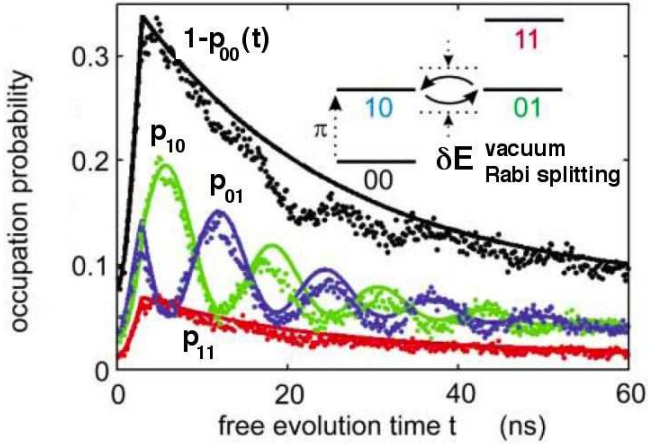}}
\caption{
Interaction of the coupled qubits in the time domain. The qubits are tuned into resonance and the system is suddenly prepared in state (10) by a microwave $\pi$-pulse applied to qubit 1. Simultaneous single-shot measurement of each of the qubits 1 and 2 reads out the probabilities $p_{00}, p_{10}, p_{01}, p_{11}$ for finding the 1+2 system in states 00, 10, 01, and 11, respectively. The full lines represent results of numerical simulations.
{\em Adapted from McDermott et al. \cite{McDermott2005}}
}
\label{McDermott_2q_prob}
\end{figure}
The $p_{10}$ and $p_{01}$ probabilities oscillate out-of-phase, as expected. In addition, the average probabilities and oscillation amplitudes all decay with time. The results are compatible \cite{McDermott2005} with the the finite rise time of the initial $\pi$-pulse (5 ns), the single-qubit readout fidelity (70 percent) and the single-qubit relaxation time $T_1$ (25 ns), and the limited two-qubit readout fidelity, as used in the numerical simulations. Of particular interest is that for simultaneous (within 2 ns) readout of the qubits, the experiments show that readout of one qubit only leads to small perturbation of the other qubit \cite{McDermott2005}, which is promising for multi-qubit applications.

\section{Quantum state engineering with multi-qubit JJ systems}

Due to the recent experimental progress, protocols and algorithms for multi-qubit JJ systems can soon begin to be implemented in order to test the performance of JJ qubit and readout circuits and to study the full dynamics of the quantum systems. The general principles are well known (see e.g. \cite{NielsenChuang2000,Gruska1999}) and have very recently been successfully applied in several other systems to achieve interesting and significant results in well-controlled quantum systems:
ion-trap technology has been used to entangle 4 qubits \cite{Sackett2000}, implement 2-qubit gates and test Bell's inequalities \cite{Schmidt-Kaler2003,Roos2004a}, perform the Deutsch-Josza algorithm \cite{Gulde2003}, achieve teleportation (within the system) \cite{Roos2004b,Riebe2004,Barrett2004}, and perform error correction \cite{Chiaverini2004}; quantum optics has recently demonstrated long-distance teleportation \cite{Walther2004} as well as 4-photon entanglement \cite{Walther2004,Bourennane2004}.
There are presently a considerable theoretical literature on how to implement these and similar protocols and algorithms in JJ circits. Below we will describe a few examples to illustrate what may need to be done experimentally.

\subsection{Bell measurements}

The first essential step is to study the quantum dynamics of a two-qubit circuit and to perform a test of Bell's inequalities by creating entangled two-qubit Bell states (Section \ref{BasicsQC}, Ref.\cite{NielsenChuang2000}) and performing simultaneous projective measurements on the two qubits, similarly to the ion trap experiments \cite{Schmidt-Kaler2003,Roos2004a}. Clearly the experiment will be a test of the quantum properties of the circuit and the measurement process rather than a test of a Bell inequality.

The general principle is to (a) entangle the two qubits, (b) measure the projection along different detector axes ("polarization directions"), (c) perform four independent measurements, and finally (d) analyze the correlations. If the detector axes are fixed, as is the case for the JJ readout (measuring charge {\em or} flux), one can instead rotate each of the qubits.
\begin{figure}[t]
\centerline{\epsfxsize=0.45\textwidth\epsfbox{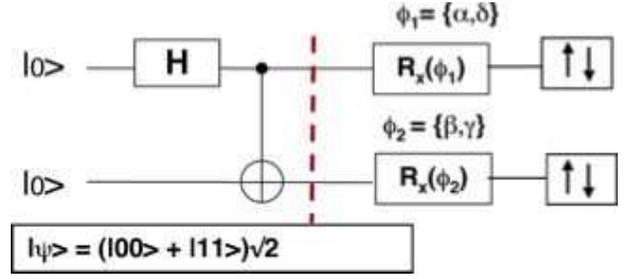}}
\caption{EPR anticorrelation in the singlet state}
\label{Bell_measurement}
\end{figure}

Figure \ref{Bell_measurement} shows the quantum network for creating a Bell state $\ket{\psi}$, perform single qubit rotations $R_x(\phi_1)$, $R_x(\phi_2)$, and finally  perform a projective measurement on each of the qubits along the same common fixed quantization axis. For each setting of the angles, $(\phi_1,\phi_2)$, on then performs a series of measurements obtaining the probabilities $P_{ij}= |\langle{ij}\ket{\psi}|^2 =
\av{\psi\ket{ij}\bra{ij}\psi}$. These can be combined into the results for finding the two qubits in the same state, $P_{same}=P_{00}+P_{11}$, or in different states, $P_{diff}=P_{01}+P_{10}$, and finally into the difference $q(\phi_1,\phi_2) = P_{same} - P_{diff}$. This quantity $q(\phi_1,\phi_2)$ is evaluated in four experiments for two independent settings of the two angles, $ \phi_1=(\alpha,\delta), \;\phi_2=(\beta,\gamma)$, constructing the function
\begin{eqnarray}
B(\alpha,\delta;\beta,\gamma) =  |q(\alpha,\beta) + q(\delta,\beta)| \\ \nonumber
+ |q(\alpha,\gamma) - q(\delta,\gamma)|
\end{eqnarray}

The Clauser-Horne-Shimony-Holt (CSCH) condition \cite{CSCH1969} for violation of classical physics is then $B > 2$ (maximum value $2\sqrt{2}$).
%
%\begin{equation}
%2 \le B \le 2\sqrt{2}
%\end{equation}
%

The application to JJ charge-phase qubits has been discussed by  Refs.\cite{HeZhuWangLi2003,WeiLiuNori2004a}. Experimentally, a first step in this direction has been taken by Martinis et al. \cite{McDermott2005} who directly detected the anticorrelation in the oscillating superposition of Bell states
$\ket{\psi(t)} = \cos(\delta E t/2)\ket{10} + \sin(\delta E t/2) \ket{01}$.

\subsection{Teleportation}

In the simplest form of teleportation an unknown single-qubit quantum state is transferred from one part of the system to another, i.e. from one qubit to another, as illustrated in Fig. \ref{teleport1}.
\begin{figure}[t]
\centerline{\epsfxsize=0.45\textwidth\epsfbox{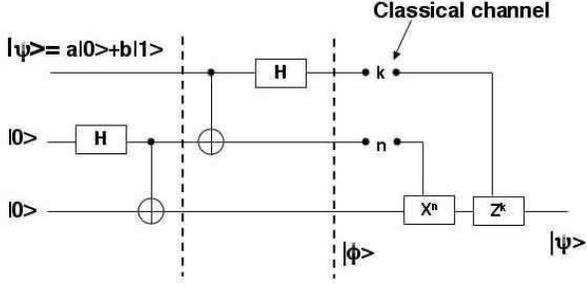}}
\caption{
Teleportation in a 3-qubit system. The unknown qubit ($1$) to be teleported is $\ket{\psi}= a\ket{0}+b\ket{1}$. The result of the measurement is sent by Alice via classical channels to Bob who performs the appropriate unitary transformations to restore the original single-qubit state. As a result of the teleportation, a specific but unknown state has been transferred from one qubit to another.
}
\label{teleport1}
\end{figure}
In Fig. \ref{teleport1}, the initial state is given by
\begin{eqnarray}
(a\ket{0}+b\ket{1})(\ket{00})
\end{eqnarray}
Next, applying CNOT and Hadamard gates entangles qubits $2$ and $3$ into a Bell state,
\begin{eqnarray}
(a\ket{0}+b\ket{1})(1/*2)(\ket{00}+\ket{11})
\end{eqnarray}
which is the resource needed for teleportation (in quantum optics this corresponds to the entangled photon pair shared between Alice and Bob).
One member of the Bell pair (qubit $2$) is now "sent" to Alice and entangled with the unknown qubit to be teleported, creating a 3-qubit entangled state,
\begin{eqnarray}\label{teleport}
\ket{00}(a\ket{0}+ b\ket{1}) \\ \nonumber
+ \;\ket{01}(b\ket{0}+a\ket{1}) \\ \nonumber
+ \;\ket{10}(a\ket{0}-b\ket{1}) \\ \nonumber
+ \;\ket{11}(-b\ket{0}+a\ket{1})
\end{eqnarray}
At this point, qubit $3$ is sent to Bob, meaning that a 3-qubit entangled state is shared between Alice and Bob. Moreover, at this point, in each component of this 3-qubit entangled state in Eq. (\ref{teleport}), the two upper qubits are in eigenstates. This means that a projective measurement of the these two qubits by Alice will collapse the total state to one of the four components. If the result of Alice's measurement is (ij), the first two qubits are in the state $\ket{ij}$ and the 3-qubit state is known, given by the corresponding component in the previous equation.
Any of these 3-qubit products can be transformed by a unitary transformation back to the original state:
\begin{eqnarray}
I (a\ket{0}+ b\ket{1}) = \ket{\psi} \\
\sigma_x (b\ket{0}+ a\ket{1}) = \ket{\psi} \\
\sigma_z (a\ket{0}- b\ket{1}) = \ket{\psi} \\
\sigma_z \sigma_x (-b\ket{0}+ a\ket{1})  = \ket{\psi}
\end{eqnarray}
corresponding to resp. no change, bit flip, phase flip and bit-plus-phase flip of the original unknown state $\ket{\psi}$. Alice's measurement causes {\em instantaneous} collapse, after which Alice by classical means (e.g. e-mail!) can tell Bob which unitary transformation to apply to recover the original single-qubit state.

An alternative approach, without measurement and classical transmission, is shown in Fig. \ref{teleport2},
\begin{figure}[t]
\centerline{\epsfxsize=0.45\textwidth\epsfbox{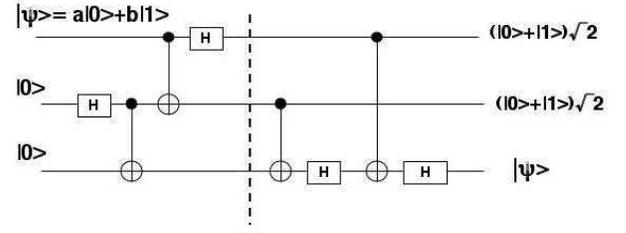}}
\caption{
Teleportation in a 3-qubit system without measurement and classical transmission. The unknown qubit ($1$) to be transferred is again $\ket{\psi}= a\ket{0}+b\ket{1}$. The state of qubits $1$ and $2$ are now used to control CNOT gates to restore the original single-qubit state on qubit $3$. As a result of the teleportation, a specific but unknown state has been transferred from qubit $1$ to qubit $3$, leaving qubits $1$ and $2$ in superposition states.
}
\label{teleport2}
\end{figure}
In this case, the final 3-qubit state is still a disentangled product state with the correct state $\ket{\psi}= a\ket{0}+b\ket{1}$ on qubit $3$,
\begin{eqnarray}
\frac{1}{\sqrt{2}} (\ket{0}+\ket{1}) \frac{1}{\sqrt{2}} (\ket{0}+\ket{1}) (a\ket{0}+b\ket{1})
\end{eqnarray}
while qubits $1$ and $2$ are now in superposition states. If desired, these states can be rotated back to $\ket{00}$ by single qubit gates.

The teleportation protocol has been implemented in ion trap experiments \cite{Riebe2004,Barrett2004} (and of course in quantum optics \cite{Ursin2004}). A proposal for a setup for implementing a teleportation scheme in JJ circuitry is shown in Fig. \ref{chain}, describing a 3-qubit chain of charge-phase qubits coupled by current-biased large JJ oscillators, in principle allowing controllable nearest-neighbour qubit couplings \cite{Lantz2004,Wallquist2004}:
\begin{figure}[t]
\centerline{\epsfxsize=0.40\textwidth\epsfbox{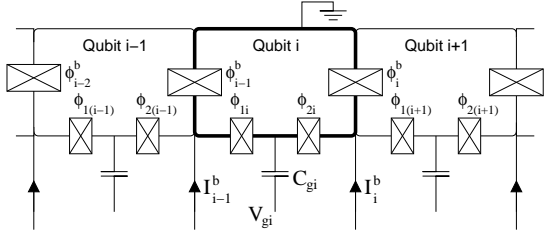}}
\caption{
Network of loop-shaped SCT charge qubits, coupled by large Josephson junctions.
The interaction of the qubits ($i$) and ($i+1$) is controlled by the current bias $I_{bi}$
Individual qubits are controlled by voltage gates, $V_{gi}$.
Single-qubit readout is performed by applying an ac
current pulse to a particular JJ readout junction (not shown), or using an RF-SET capacitively coupled to the island $[27]$.
}
\label{chain}
\end{figure}
In the absence of bias currents, to first order the qubits are non-interacting and isolated from each other. The basic two-qubit gates are achieved by switching on the bias-current-controlled qubit-qubit interaction for a given time. CNOT gates can be achieved in combination with single-qubit Hadamard and phase gates \cite{Lantz2004,Wallquist2004,Wallquist2005}. By applying the sequence of gates shown in Fig. \ref{teleport2} (or an equivalent sequence), the unknown state will be physically moved from the left end to the right end of the chain.
Moreover, by applying coupling pulses simultaneously to several qubits, one can in principle create multi-qubit entangled states in fewer operations than with sequential two-qubit gates \cite{Niskanen2003b,Vartiainen2004a} as well as perform operations in parallel on different parts of the system.

Extending the system to a five-qubit chain one can in a similar way transfer an entangled two-qubit state from the left to the right end of the chain.

\subsection{Qubit buses and entanglement transfer}

With entangled "flying qubits" like photons, quantum correlations can be shared in a spatially highly extended states. However, with solid-state circuitry the issue becomes how to transfer entanglement among spatially fixed qubits.

The standard answer is to apply entangling two-qubit gates between distant qubits. The qubit-qubit interaction can be direct, e.g. dipole-dipole-type interaction between distant qubits, or mediated by excitations in the system. The transfer can be mediated via virtual or real excitations in a passive polarizable medium, a "system bus", or via a protocol for coupling qubits and bus oscillators.

A classic example of protocol-controlled oscillator-mediated coupling is the Cirac-Zoller gate \cite{ZiracZoller1995} between two ions sequentially entangled via exchange coupling with the lowest vibrational mode of the ions in the trap. Another example is the Molmer-Sorensen gate \cite{MolmerSorensen1999,SorensenMolmer1999,SorensenMolmer2000} which creates qubit-qubit interaction via virtual excitation of ion-trap modes. Similar concepts for controlling entanglement have recently been theoretically investigated in applications to JJ-qubit-oscillator circuits \cite{Blais2003,Plastina2001,Plastina2002,Plastina2003,Paternostro2004a,Paternostro2004b,ZhuWangYang2003,YouNori2003,DeChiara2003}

An different approach is to allow the bus ("spin-chain") states to develop in time governed by the fixed bus Hamiltonian and to tailor the interactions and the initital conditions such that useful dynamics and information transfer is achieved \cite{Bose2003,Christandl2004a,Christandl2004b,Albanese2004}.
Also these concepts have been applied to JJ-circuits in a number of recent theoretical studies \cite{DeChiara2004a,DeChiara2004b,Romito2004,Montangero2003,Montangero2004,Facchi2004a,Paternostro2004c}.

\subsection{Qubit encoding and quantum error correction}

Quantum error correction (QEC) (see e.g. Refs.\cite{NielsenChuang2000,Shor1995,Knill2001,Steane1996,Steane1997,Steane2003a,Steane2003b,Steane2004a}) will most certainly be necessary for successful operation of solid state quantum information devices in order to fight decoherence. The algorithmic approach to QEC follows the principles of classical error correction, by encoding bits to create redundancy, and by devising procedures for identifying and correcting the errors based on specific models for the errors.

The quantum circuit in Fig. \ref{QEC_5q} illustrates QEC in terms of five-qubit teleportation with bit errors, illustrating restoration of the state including error correction by measurement-controlled unitary transformation.
\begin{figure}[t]
\centerline{\epsfxsize=0.45\textwidth\epsfbox{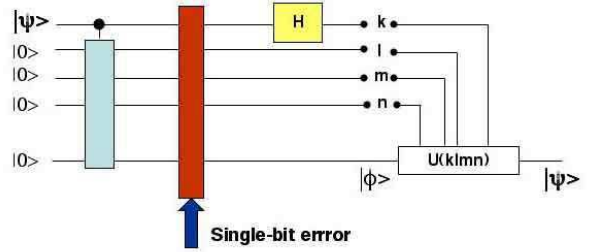}}
\caption{
Teleportation with error correction in a 5-qubit system without measurement.
}
\label{QEC_5q}
\end{figure}

The quantum circuit in Fig. \ref{3qQEC} demonstrates some essential steps of QEC in the case of one logical qubit encoded in three physical qubits:
\begin{figure}[t]
\centerline{\epsfxsize=0.40\textwidth\epsfbox{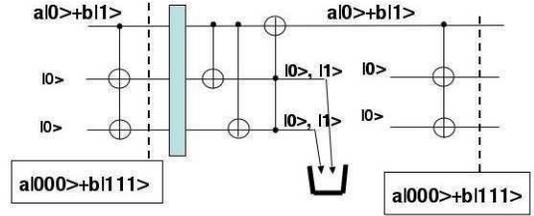}}
\caption{
Coding, decoding and correcting a bit flip error in a 3-qubit logic qubit. The first gate on the left represents two sequential CNOT gates. The last gate befor the garbage can is a control-control-NOT (Toffoli) gate.
}
\label{3qQEC}
\end{figure}
The first step is to encode the physical qubit in logical qubit basis states $\ket{000}$ and $\ket{111}$ by applying two CNOT gates to create a 3-qubit entangled state:
\begin{eqnarray}
(a\ket{0}+b\ket{1}) \;\ket{00} \rightarrow  (a\ket{000}+b\ket{111})
\end{eqnarray}
Next, there may occur a bit-flip error in one of the qubits:
\begin{eqnarray}
a\ket{000}+b\ket{111} \;\; (no \;bit \;flip) \\
a\ket{100}+b\ket{011} \;\; (bit \;flip \;in \;qubit \;1) \\
a\ket{010}+b\ket{101} \;\; (bit \;flip \;in \;qubit \;2) \\
a\ket{001}+b\ket{110} \;\; (bit \;flip \;in \;qubit \;3)
\end{eqnarray}
The next step applies a number of disentangling CNOT gates to check for the type of error. For the four possible states above we obtain:
\begin{eqnarray}
a\ket{000}+b\ket{111} \rightarrow (a\ket{0}+b\ket{1}) \;\ket{00} \\
a\ket{100}+b\ket{011} \rightarrow (a\ket{1}+b\ket{0}) \;\ket{11} \\
a\ket{000}+b\ket{111} \rightarrow (a\ket{0}+b\ket{1}) \;\ket{10} \\
a\ket{000}+b\ket{111} \rightarrow (a\ket{0}+b\ket{1}) \;\ket{01}
\end{eqnarray}
At this stage, qubits $2$ and $3$ have become independent eigenstates, just as in teleportation. The corresponding eigenvalues are called {\em syndrome}, and indicate which corrective operations should be implemented. Remarkably, in two of the above bit-flipped cases, with error syndromes ((01) and (10), qubit $1$ is now correct, the error residing in qubits $2$ or $3$ in the workspace ("ancillas"). The only case where a transformation is needed is when the error syndrom is (11), which corresponds to a bit flip in qubit $1$, to be corrected with a CCNOT (control-control-NOT, or  Toffoli, gate), controlled by the truth table of an AND gate. In Fig. \ref{3qQEC} we have used a compact single-gate notation for CCNOT, while in reality it must be implemented through a sequence of eight two-qubit gates \cite{NielsenChuang2000} (there is no direct three-qubit interaction in the Hamiltonian). At this stage, the 3-qubit state is completely disentangled into a product state.

The final step consists in re-encoding the physical qubit $1$ to the logical qubit $a\ket{000}+b\ket{111}$. However, although the physical qubit $1$ is always correct at this stage, the workspace is not. This can be handled in a number of ways. Figure \ref{3qQEC} dumps the "hot" qubits in the garbage can (i.e. leaves them, or forces them, to relax), and re-encodes with fresh qubits from a larger workspace.
\begin{figure}[t]
\centerline{\epsfxsize=0.40\textwidth\epsfbox{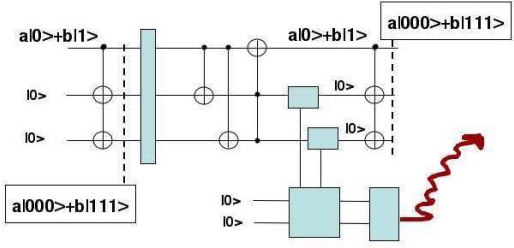}}
\caption{
Coding, decoding and correcting a bit flip error in 3-qubit logic qubit.
}
\label{3qQEC2}
\end{figure}
Alternatively, Fig. \ref{3qQEC2} describes a variation where the old qubits $2$ and $3$ are re-initialized by entanglement with a measurement device which then dissipates the heat from the bitflip and the error correction procedure.

In both of these cases, as described, we need a total workspace with five qubits. In principle, however, one can do with three qubits if we can rapidly re-initialize qubits $2$ and $3$ without inducing new errors. Recently an error correcting procedure with cooling of the system without measurement has been proposed \cite{Sarovar2005}.

Phase flips (sign changes), e.g. $a\ket{0}+b\ket{1} \rightarrow a\ket{0}-b\ket{1}$ can be handled by similar 3-qubit encoding, decoding and correction.
Combining these two approaches gives the 9-qubit Shor code \cite{Shor1995} for correcting also combined bit and phase flips, e.g. $a\ket{0}+b\ket{1} \rightarrow a\ket{1}-b\ket{0}$. The minimum code to achieve the same thing is a 5-qubit code \cite{Knill2001} and there are more efficient codes with 7 qubits \cite{Steane1997,Steane2003a,Steane2003b}.

A related approach to fighting decoherence is to encode the quantum information in noiseless subsystems, so called Decoherence Free Subspaces (DFS) \cite{IoffeNature2002,FeigelmanPRL2004,Celeghini2000,LidarWhaley2003}

A different approach is the so called "bang-bang" dynamic correction method \cite{Viola1999,FaoroViola2003,ShnirmanMakhlin2003,Falci2004a,Facchi2004b,Facchi2005}, basically corresponding to very frequent application of the spin-echo technique.
This is related to the quantum Zeno effect (see e.g. Ref.\cite{Alicki2005}, describing situations where the quantum system via very strong interaction with the environment is forced into a subspace from where it cannot decay.

In this Section we have restricted the discussion to a few different protocols for quantum state engineering, representing basic steps in algorithms for solving specific computational problems. For a discussion of quantum algorithms and computational complexity we refer to the books by Nielsen and Chuang \cite{NielsenChuang2000} and Gruska \cite{Gruska1999}, and to the original papers. We would however like to mention a few papers discussing how to implement a few well-know algorithms in JJ circuitry. The basics of quantum gates in JJ-circuits can be found in e.g. Refs.\cite{MakhlinRMP2001,Makhlin2000b,Siewert2001b}. The Deutsch-Josza (DJ) and related algorithms have been discussed by Siewert and Fazio \cite{Siewert2001b} describing a 3-qubit DJ implementation with three charge-phase qubits connected in a ring via phase-coupling JJs with variable Josephson coupling energy (SQUIDS), and by Schuch and Siewert analyzing a 4-qubit implementation \cite{Schuch2002}. For the Grover search algorithm there seems to be nothing published on the implementation in JJ circuitry. Concerning Shor's factorization algorithm there is a recent paper by Vartiainen et al. \cite{Vartiainen2004b}.
There is also recent work on optimization of two-qubit gates \cite{Zhang2004}, and relations between error correction and entanglement \cite{DeChiara2003}.

Finally there are a number of papers connecting JJ-ciruits wih geometric phases and holononomic quantum computing \cite{Falci2003c,Faoro2003,Falci2000,Cholascinski2004}} and on adiabatic computation \cite{Averin1998} and Cooper pair pumps \cite{MakhlinMirlin2001,Aunola2003}. For a recent paper discussing the universality of adiabatic quantum computing, see Aharonov et al. \cite{Aharonov2004}.

%%%%%%%%%%%%%%%%%%%%%%%%%%%%%%%%%%%%%%%%%%%%%%%%%%%%%

\section{Conclusion and perspectives}

Within 5 years, engineered JJ quantum systems with 5-10 qubits will most likely begin seriously to test the scalability of solid state QI processors.

For this to happen, a few decisive initial steps and breakthroughs are needed and expected: The first essential step is to develop JJ-hardware with long coherence time to study the quantum dynamics of a two-qubit circuit and to perform a "test" of Bell's inequalities (or rather the JJ-ciruitry) by creating entangled two-qubit Bell states and performing simultaneous projective measurements on the two qubits.

A first breakthrough would be to perform a significant number of single- and two-qubit gates on a 3-qubit cluster to entangle three qubits. Combined with simultaneous projective readout of individual qubits, not disturbing unmeasured qubits, this would form a basis for the first solid-state experiments with teleportation, quantum error correction (QEC), and elementary quantum algorithms. This will provide a platform for scaling up the system to 10 qubits.

This may not look very impressive but nevertheless would be an achievement far beyond expectations only a decade back. The NMR successes, e.g. running Shor-type algorithms using a molecule with 7 qubits \cite{NMR_Shor}, are based on technologies developed during 50 years using natural systems with naturally long coherence times. Similarly, semiconductor technologies have developed for 50 years to reach today's scale and performance of classical computers. It is therefore to be expected that QI technologies will need several decades to develop truly significant potential. Moreover, in the same way as for the classical technologies, QI technologies will most probably develop slowly step by step, "qubit by qubit", which in itself will be an exponential development.

Moreover, in future scalable information processors, different physical realizations and technologies might be combined into hybrid systems to achieve fast processing in one system and long coherence and long-time information storage in another system. In this way, solid state technologies might be combined with ion trap physics to build large microtrap systems \cite{Steane2004b}, which in turn might be coupled to superconducting Josephson junctions processors via microwave transmission lines \cite{Tian2004b}.

%%%%%%%%%%%%%%%%%%%%%%%%%%%%%%%%%%%%%%%%%%%%%%%%%%%%%

\begin{acknowledgments}
This work has been supported by European Commission through the IST-SQUBIT and SQUBIT-2 projects, by the Swedish Research Council, the Swedish Foundation for Strategic Research and the Royal Academy of Sciences.
\end{acknowledgments}

%%%%%%%%%%%%%%%%%%%%%%%%%%%%%%%%%%%%%%%%%%%%%%%%%%%%%

\appendix
\section* {Glossary}

\noindent
{\bf Adiabatic evolution} - Development of a quantum system without
transitions among the quantum levels.

\noindent
{\bf Algorithm} - Finite sequence of logical operations, which produces
a solution for a given problem.

\noindent
{\bf Level crossing} - Degeneracy of quantum levels appearing at a
certain value of a controlling system parameter (e.g. gate voltage, bias
flux, etc.).

\noindent
{\bf Anticrossing } -  Lifting of degeneracy (level crossing) of quantum
levels during variation of the system parameters when an interaction is
switched on

\noindent
{\bf Average measurement} - Measurement of an expectation value of a
dynamic variable in a certain state

\noindent
{\bf Bloch sphere} - Geometrical representation of the manifold of
quantum states of a two-level system as points on the unit sphere.

\noindent
{\bf Bloch vector} - Normalized state vector of a two-level system represented by a radial unit vector of the Bloch sphere.

\noindent
{\bf Charging energy} - Electrostatic energy of a capacitor charged with
a single electron (e) (or a single Cooper pair (2e)).

\noindent
{\bf Charge qubit} - Superconducting qubit based on a a single Cooper pair box (SCB), whose computational basis consists of the two charge states of the
superconducting island.

\noindent
{\bf Charge-phase qubit} -  Superconducting qubit based on a single Cooper pair box (SCB) whose charging energy is of the order of the Josephson energy of tunnel junctions

\noindent
{\bf CNOT gate} -  Controlled-NOT gate: two-qubit gate which changes or does not change the state of a target qubit depending on the state of a controlling qubit.

\noindent
{\bf Cooper pair} - Bound state of two electrons (2e), the elementary charge carrier in superconducting equilibrium state.

\noindent
{\bf Coulomb blockade} - Suppression of current through a tunnel
junction or small metallic island due to large charging energy
associated with a passage of a single electron.

\noindent
{\bf CPHASE gate} -  Controlled-phase gate: two-qubit gate which changes or does not change the phase of a target qubit depending on the state of a controlling qubit.

\noindent
{\bf Decoherence} - Evolution of a quantum system, interacting with its
environment; cannot be described with a unitary operator; consists
of decay of phase coherence (dephasing) and/or changing of level population
(relaxation).

\noindent
{\bf Density matrix} - Characteristics of a quantum system, which
contains full statistical information about the state of the system.

\noindent
{\bf Dephasing} - Decay of phase coherence of a superposition state,
represented by decreasing off-diagonal elements of the density matrix.

\noindent
{\bf Entanglement} - Specific non-local coupling of quantum systems when
the wave function of whole system cannot be presented as a product of
partial wave functions

\noindent
{\bf Flux qubit} - Superconducting qubit based on a SQUID, whose
computational basis consists of the two states of the SQUID having opposite directions of the induced flux.

\noindent
{\bf Hadamard gate} - Transformation of computational basis states of a
single qubit to equally weighted superpositions of the basis states (cat
states).

\noindent
{\bf Holonomic quantum computation} - Using the geometric phases when a quantum system is taken around a closed circuit in the space of control parameters.

\noindent
{\bf Gate operation} - Controlled transformation of the state of one or
several qubits; a basic element of an algorithm.

\noindent
{\bf Josephson effect} -  Non-dissipative current flow between two
superconductors separated by a non-superconducting material (insulator,
normal metal, etc.).

\noindent
{\bf Josephson junction} - Junction of two superconductors, which
exhibits the Josephson effect.

\noindent
{\bf Josephson critical current} - Maximal value of the Josephson
current maintained by a particular junction.

\noindent
{\bf Josephson energy} - Inductive energy of a Josephson junction
proportional to the critical Josephson current.

\noindent
{\bf $\pi$ pulse} - High frequency control pulse with a specific
duration applied to a qubit, producing inversion of the qubit level
populations  ($\pi$ rotation; qubit flip)

\noindent
{\bf $\pi$/2 pulse} - High frequency control pulse with a specific
duration applied to a qubit, typically tipping the Bloch vector from a pole to the equator, or from the equator to a pole, on the Bloch sphere.

\noindent
{\bf Phase gate} - Single qubit gate, transforms a superposition of two
quantum states into another superposition with different relative phase
of the states

\noindent
{\bf Precession} - Dynamic evolution of a two-level system in a
superposition state, i.e. linear combination of energy eigenstates.

\noindent
{\bf QND measurement} - Quantum Non-Demolition Measurement: measurement
of a state of a quantum system, which does not destroy the quantum state
and makes possible repeated measurements of the same state.

\noindent
{\bf QPC} - Quantum Point Contact: a constriction in a conductor with ballistic transport through a small number of conduction channels.

\noindent
{\bf Qubit} - Quantum two-level system; basic element of a quantum
processor.

\noindent
{\bf PCQ} - Persistent Current Qubit: synonymous with flux qubit.

\noindent
{\bf Rabi oscillation} - Dynamics of two-level system under resonant
driving perturbation, consists of periodic oscillation of the level
populations with the frequency proportional to the amplitude of the
perturbation.

\noindent
{\bf Readout} - Measurement of a qubit state.

\noindent
{\bf Relaxation} - Change of population of the energy eigenstates
resulting in approaching the equilibrium population.

\noindent
{\bf SCB} - Single Cooper pair Box: superconducting analog of SEB,
where it is energetically favorable to have only paired electrons on the
island.

\noindent
{\bf SCT} - Single Cooper pair Transistor: a superconducting device containing a small island whose charging energy is controlled by an electrostatic gate electrode to increase or decrease current flowing through the island from one large electrode (source) to another (drain).

\noindent
{\bf SEB} - Single Electron Box: small metallic island connected to a
large electrode via resistive tunnel junction, whose charging energy
hence amount of trapped electrons is controlled by an electrostatic gate.

\noindent
{\bf SET} - Single Electron Transistor: a device containing a small
island whose charging energy is controlled by an electrostatic gate electrode to
increase or decrease current flowing through the island from one large
electrode (source) to another (drain).

\noindent
{\bf rf-SET} - SET driven by an rf signal, is used as an ultra sensitive
electrometer by monitoring a linear response function of the SET, which
is sensitive to the electrostatic gate potential.

\noindent
{\bf SQUID} - Superconducting Quantum Interferometer Device: a device
consisting of a one or more Josephson junctions included in a superconducting loop.

\noindent
{\bf dc-SQUID} - SQUID containing two Josephson junctions.

\noindent
{\bf rf-SQUID} - SQUID containing one Josephson junction.

\noindent
{\bf Single-shot measurement} - A measurement which gives an "up/down" answer in one single detection event.

\end{document}